# Expression of Interest

# for a

# Comprehensive New Detector at RHIC II

August 18, 2004

P. Steinberg, T. Ullrich (Brookhaven National Laboratory)

M. Calderon (Indiana University)

J. Rak (Iowa State University)

S. Margetis (Kent State University)

M. Lisa, D. Magestro (Ohio State University)

R. Lacey (State University of New York, Stony Brook)

G. Paic (UNAM Mexico)

T. Nayak (VECC Calcutta)

R. Bellwied, C. Pruneau, A. Rose, S. Voloshin (Wayne State University)

and

H. Caines, A. Chikanian, E. Finch, J.W. Harris, M. Lamont, C. Markert,

J. Sandweiss, N. Smirnov (Yale University)

# Table of Contents

# 1  Introduction

There are compelling physics questions to be addressed by a new comprehensive detector at a future, high-luminosity RHIC II collider. These form the basis for this Expression of Interest.

- **What precisely are the properties of the strongly-coupled Quark Gluon Plasma (sQGP)?** Can a more weakly interacting QGP state be formed and investigated at RHIC?

- **How do particles acquire mass and what is the effect of chiral symmetry restoration on hadronization in a dense medium?** What is the chiral structure of the QCD vacuum and its influence on and contributions of different QCD vacuum states to the masses of particles?

- **Is there another phase of matter at low Bjorken-$x$, *i.e.* the Color Glass Condensate (CGC)?** If present, what are its features and how does it evolve into the QGP? If not, are parton distribution functions understood at low Bjorken-$x$ and can they describe particle production?

- **What are the structure and dynamics inside the proton, including parton spin and orbital angular momentum?** What are the contributions of gluons and the QCD sea to the polarization of the proton. What is the flavor-dependence? Are there tests for new physics beyond the Standard Model from spin measurements at RHIC II (such as parity-violating interactions)?

The RHIC II complex will be the only QCD facility to have the capability of addressing these questions. We propose that a new comprehensive detector system is needed for RHIC II to address these questions adequately and in an effective way. The primary focus of the new detector is to measure and identify hadrons, electrons and muons, and to identify and measure photon and jet energy over a large rapidity range and full azimuth in proton-proton (polarized and unpolarized), proton-nucleus and nucleus-nucleus collisions. This detector utilizes precision tracking and particle identification to large transverse momentum (20 GeV/$c$) in a 1.3 T solenoidal magnetic field with electromagnetic and hadronic calorimetry and muon identification over $-3.5 < \eta < 3.5$. There is additional coverage forward, outside the magnet. An in-depth experimental program utilizing the unique features of this detector system and that of RHIC II is necessary to address these compelling physics questions in an era with heavy ions in the Large Hadron Collider (LHC) and will be introduced in this document. Furthermore, this physics is complementary to the LHC ion program and symbiotic with any future electron-ion collider (eRHIC) program.

# 2 Physics Overview

## 2.1 Properties of the Strongly-Coupled Quark Gluon Plasma

There has been considerable recent discussion about the properties of the matter created at RHIC.[1] Whether or not it is deconfined is still debatable. However, measurements show that it is highly interacting, that the quark and gluon degrees of freedom are necessary to describe the experimental results, and that purely hadronic descriptions have failed. A strongly-coupled Quark Gluon Plasma (sQGP) appears to have been formed.[1] This was unanticipated in that a weakly-coupled QGP was the expectation from theory and lattice calculations. Although lattice calculations show that even at LHC energies the Stefan Boltzmann limit for an ideal gas cannot be reached, the coupling at RHIC seems surprisingly high and most features of the produced matter can be described with ideal fluid dynamics. It is obvious that the properties of this sQGP must be established precisely and unambiguously. To do so, we must also understand whether or not a more weakly interacting QGP state can be formed at RHIC, under what conditions, and if not, why not. This requires establishing experimentally the conditions and properties of the high density stage of RHIC collisions. The initial temperature, density and entropy are needed, as well as the subsequent evolution of the system (also the thermodynamic variables and state of the system), and ultimately the equation(s) of state describing the system at various stages. The question of whether the system reaches a deconfined state remains to be determined at RHIC. It is anticipated that this will be studied initially in the next few years at RHIC. The new comprehensive detector, with its lepton pair and triggering capabilities over large acceptance, will address with precision the issue of deconfinement by determining the temperatures at which the quarkonium states dissociate[2] through measurements of the yields of the J/$\psi$, $\psi$', $\Upsilon$(1S), $\Upsilon$(2S) and $\Upsilon$(3S) states. From simple binding arguments the temperature dependence of the melting sequence is expected to be $T(\psi') < T(\Upsilon(3S)) < T(J/\psi) < T(\Upsilon(2S)) < T(\Upsilon(1S))$. Comparing the measured ratios of the quarkonium states with model calculations and in the future refined lattice calculations will establish the dissociation temperatures of quarkonium states, and the temperatures and conditions where the system becomes deconfined.[i]

Detailed properties of the high density state will be established using $\gamma$–jet[3] and jet-jet tomography[4] (utilizing identified leading particles and topological jet identification where possible) as a function of collision variables (*e.g.* nucleus-nucleus impact parameter and geometry, orientation relative to the reaction plane, transverse momentum, rapidity, particle type and flavor-dependence of the jet). This is depicted in Figure 1. Furthermore, in jet and leading particle measurements, the fragmentation function of identified particles will be measured and compared to those measured in this device in p+p interactions to determine precisely the modification of the fragmentation functions of specific particles as a function of their constituent quark content.[5] This is essential to ascertain the properties of the high density state and to understand the interaction of this state with traversing partons. Thus, particle identification is essential at high transverse momentum ($p_T$) to accomplish these measurements.

---

[i] The initial temperature created in RHIC collisions can be determined from $\gamma\gamma$ or lepton-lepton HBT. These measurements are difficult but feasible with the increased luminosities and the appropriate detector systems.

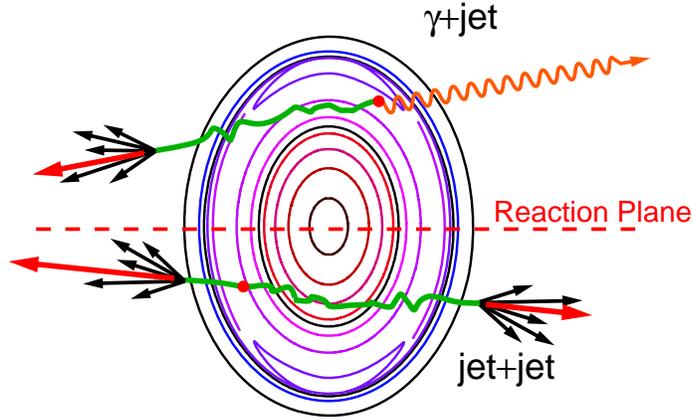

**Figure 1:** High density overlap region as viewed along the direction of the colliding heavy ions. Displayed are contours of equal energy density in a non-central impact parameter Au + Au event from hydrodynamics.[6] The reaction plane of the collision is denoted by the horizontal dashed line. Two examples of hard scatterings are shown, resulting in a photon-parton (top) and a parton-parton pair (bottom). The partons fragment after propagating through the medium.

Propagating heavy quarks are expected to lose much less energy than propagating light quarks since for heavy quarks gluon bremsstrahlung is suppressed at small angles (known as the dead cone effect).[7] Thus, jets with D- or B-mesons as leading particles should be less suppressed than light quark jets. Measurement of identified heavy quark jets and comparison with light quark jets, in the ratios $D/\pi$ and $B/\pi$ at high $p_T$, will determine whether gluon bremsstrahlung is the dominant energy loss mechanism. Likewise, measuring the modification of fragmentation functions of hadrons with heavy constituent quarks will determine the extent and mechanism of energy loss from interactions of propagating quarks with the matter. Similarly, ratios of leading anti-particles to leading particles, such as $\bar{p}/p$ and $K^-/K^+$, at high $p_T$ can be used to distinguish energy loss effects of gluon versus quark jets.[3] Anti-particles are more likely to originate from fragmentation of gluons than from quarks, whereas particles come both from quark and gluon fragmentation. Measurement of the rapidity and rapidity interval dependence of particle correlations should provide some selection on quark versus gluon jets, since the strengths of the $gg$, $qg$ and $q\bar{q}$ parton-scattering subprocesses vary with the rapidity of the trigger particle (or jet) and with the rapidity interval (gap) between the trigger and associated particles (or jets). Furthermore, a program of jet studies as a function of collision geometry, number of jets (2-, 3-jet events) in the final state, and energy for various rapidity combinations of the jets should lead to a precise understanding of the propagation of quarks and gluons in the high density state.

## 2.2 The Constituent Quark Content of Hadrons and the Influence of QCD Vacuum States on Hadron Masses

The Higgs field is responsible for the mass of particles in a chirally symmetric medium. However, when chiral symmetry is broken, the quark condensate contributes a significant part of the mass of the light and strange hadrons. This is depicted in Figure 2. Specific contributions of the various partons (gluons, sea quarks, valence quarks) to the hadron mass can be determined by measuring the fragmentation function of specific hadrons in p+p interactions. Measurement of these fragmentation functions requires particle identification of leading particles in jets at high $p_T$. Such measurements in A+A collisions establish how fragmentation functions are modified by the propagation of the various types of quarks in a dense medium and reflects the contributions of different constituent quarks to the hadron masses in the medium. It would be extremely exciting if these fragmentation functions were to reflect properties of a chirally restored medium, although this connection has yet to be established theoretically. A possible measurement, though, could be the fragmentation of a parton into a short-lived resonance inside and outside the medium. In addition to accounting for the constituent quark masses, the chiral quark condensate is responsible for inducing transitions be-tween left-handed and right-handed quarks, where $q\bar{q}=q_L\bar{q}_R+q_R\bar{q}_L$. Therefore, helicities of (leading) particles in jets (*e.g.* determined by detecting the polarization of leading Λ particles) may provide information on parity violation and chiral symmetry restoration.[8]

It is interesting to determine in heavy ion collisions the extent to which hadron formation proceeds via quark coalescence and whether effects of a chirally restored medium can be identified in the hadron yields or resonance yields, masses or widths. Recent observations[10] of two new heavy+light quark hadron states, which decay to $D_s\pi$ and $D_s^*\pi$ respectively, have led to suggestions of chiral-doubling[11] in heavy-light quark systems. Identification of such states at RHIC would provide information on chiral symmetry restoration in dense matter. Measuring D-meson production will explore these chiral dynamics since the masses of $0^+$ and $0^-$ states should merge as the chiral quark condensate melts. Measuring $D_s/D$ ratios would distinguish between production mechanisms such as pQCD, thermal and chirally symmetric production models. Measuring the mass and width of the ρ in various directions relative to the jet cone, *e.g.* in the near- and away-side jet cones, may be sensitive to and uncover effects of chiral restoration on the ρ.

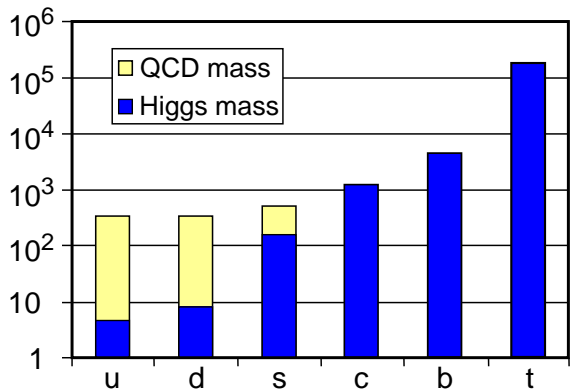

**Figure 2**: The particle mass (in MeV on a logarithmic scale) for the various quarks. Contributions of the Higgs mass and the chiral quark condensate are shown.[9]

## 2.3 Another Phase of Matter – the Color Glass Condensate?

Recent observations of suppression of high $p_T$ hadrons at forward rapidities (η = 3.2) in d + Au interactions[12] at RHIC are consistent with gluon saturation or formation of a color glass condensate (CGC)[13] at low Bjorken $x$ (~ $10^{-3}$). If this is a CGC, its features and how this precursor evolves into a QGP must be understood. Figure 3 shows possible phases of QCD as a function of parton rapidity and transverse momentum. Precise measurements at sufficiently low-$x$ at forward rapidities at RHIC can investigate the existence of the CGC. Correlations between forward and mid-rapidity jets (Mueller-Navelet di-jets), direct photons at forward rapidities, γγ HBT to measure the coherence of the sea-quark source, Drell-Yan at forward rapidities, and measurements and comparisons of heavy meson production in A+A collisions relative to binary collision scaling of p+p interactions at forward rapidities will reveal the existence, properties and evolution of a condensed gluonic phase (CGC) of matter at low $x$. These measurements require precision hadron, photon and jet measurements at large momentum at forward rapidities and associated high $p_T$ hadron, photon and jet measurements near mid-rapidity. The existence of a CGC at low $x$ would require the Large Hadron Collider heavy ion program, which is predominantly at low $x$ due to the large collision energy, to disentangle effects of the CGC and that of a Quark Gluon Plasma in order to understand these states of matter. At RHIC low $x$ (dominant at forward rapidities) and high $x$ (high $p_T$ particles at mid-rapidity) effects can be distinguished easily.

Another approach[14] suggests that conventional leading-twist calculations employing nuclear parton distribution functions rather than a CGC can describe the new data. There are also questions[15] as to whether the average Bjorken-$x$ values are sufficiently small for inclusive forward particle production to probe saturation. If there is no evidence for a CGC at RHIC II, it is still important to determine whether the parton distribution functions determined from deep inelastic scattering quantitatively describe particle production in p+A collisions. Particle correlations in p+A collisions, such as di-hadrons or di-jets produced at large rapidity (forward di-jets or di-hadrons) with small rapidity intervals, test the applicability of leading-twist formulations by probing small $x$ phenomena. If the CGC picture is correct, then the rapidity distribution of recoil hadrons would differ from those of leading-twist approaches.

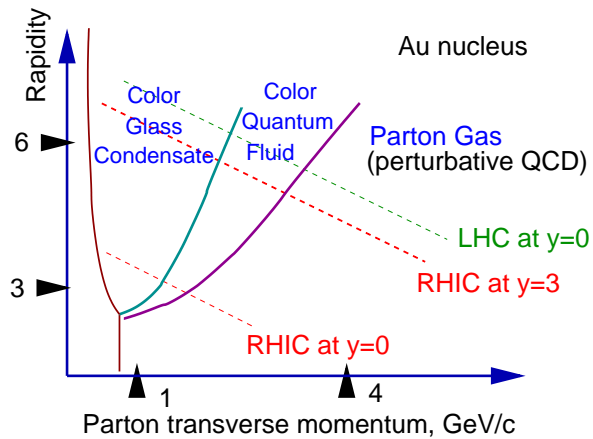

**Figure 3:** Possible phases of QCD as a function of parton rapidity and transverse momentum.[16]

It is essential to measure and understand the parton distributions of the incident nuclei and the initial stages of collisions at RHIC in general. This is particularly important for a description of hard processes and other perturbative QCD aspects of the collision process, and for understanding the evolution to a QGP. Although the quark distribution functions have been measured in deep inelastic scattering, determination of the gluon distributions in nuclei will only begin soon at RHIC. The measurements at large rapidities require that faster partons in one projectile be used to probe soft "low-$x$" partons in the other. Thus, at large rapidities, we probe aspects of the nuclear wave function where parton densities are sufficiently large that saturation phenomena should occur ("nuclear shadowing"). This will require studies of proton-nucleus and nucleus-nucleus collisions, and instrumentation and measurements in the forward direction at RHIC II. All of this is necessary to understand completely the initial conditions, the overall collision process and formation and evolution of a QGP. In addition, it is important to understand the process of energy deposition in the collision process. Extracting the net baryon density vs. rapidity provides information on the transport of quantum numbers from the projectiles to smaller rapidities ("nuclear stopping"). Thus, both from the "global variable" and "parton" perspectives there are compelling reasons to study the forward region through particle identification and tracking at very small angles.

## 2.4 The Structure and Dynamics inside the Proton

Initial determination of the contribution of gluons to the polarization of the proton will be accomplished with the existing RHIC detectors in the current RHIC spin program.[17] Important questions about the spin structure of the proton are best answered with the performance projected for RHIC II and with suitable detectors. Our knowledge of how the quark contribution to the spin of the proton is distributed amongst the different flavors is limited. Recent results from semi-inclusive deep inelastic scattering[18] represent five years of data collection from a state-of-the art experiment. Despite this effort, our knowledge about the separation of valence and sea contributions to the spin of the proton is limited by both statistical precision and systematic uncertainties associated with the fragmentation process. For the spin-averaged parton distribution functions, the contributions from dif-ferent quark flavors and the separation of valence and sea contributions are largely determined from deep inelastic scattering of neutrinos through their charged-current weak interactions. There are no prospects for studying neutrino deep inelastic scattering from polarized targets in the foreseeable future. Electroweak probes can still be employed by production of $W^{\pm}$ in polarized proton collisions at $\sqrt{s}$ = 500 GeV. When the $W^{\pm}$ are produced with sufficient rapidity, the large parity violating asymmetry $A_L$ is directly proportional to the quark or antiquark polarization. This proportionality is most easy to see in a leading-order calculation. Full treatments of next-to-leading order corrections have been made[19], and it has been found that the sensitivity to contributions from different quark (anti-quark) flavors remains.

Determination of the helicity asymmetry distribution for the antiquarks is expected to provide important insight into the non-perturbative structure of the proton. The excess of $\bar{d}$ over $\bar{u}$ at small Bjorken $x$ observed in Drell-Yan experiments[20] has been explained by a pion cloud model of the proton and by the Chiral Soliton Model[21]. The latter predicts a difference $\Delta\bar{u} - \Delta\bar{d}$ that is even larger than the difference of the unpolarized quark distributions. The recent semi-inclusive deep inelastic scattering data[18] cannot tell if this asymmetry is present. Robust measurements of this quantity via parity violation in $W^{\pm}$ can answer this question. The polarization of strangeness in the sea can be probed by charm-tagged $W^{\pm}$ production. A large acceptance detector and the best performance of RHIC II for polarized proton collisions at $\sqrt{s}$ = 500 GeV is required for such a measurement.

Additional probes of gluon polarization are possible through measurements of double longitudinal spin asymmetries ($A_{LL}$) for heavy quarks produced in p+p collisions. Heavy quark production is dominated by gluon-gluon fusion. The contribution of competing production mechanisms for heavy flavors through quark-antiquark annihilation is small at RHIC, since that requires an anti-quark in the initial state. Thus, the gluons in the proton can be accessed directly through measurements of heavy quarks. Furthermore, charm and bottom production probe the gluon density in the proton at different momentum fractions (Bjorken-$x$) and scales ($Q^2$). Measurements of the production of heavy quarkonia also probes gluons in the proton and will only be possible at the higher RHIC II luminosities. [Understanding the unpolarized production of heavy quarkonium states is therefore also important to the spin physics program as well as the heavy ion program.] Heavy-flavor production can be selected through various leptonic decay channels over a large kinematic range in the proposed detector. In addition, $B \rightarrow J/\psi + X$ tagging the $J/\psi$ through displaced electron and muon vertices provides a means of identifying open beauty production.

The unpolarized aspect of hadron production of heavy flavors has recently attracted significant attention. Although open charm production is fairly well understood, beauty production exhibits large discrepancies between theory[22] and recent data from HERA[23] and LEP.[24] This has led to an extensive discussion of physics beyond the Standard Model (SM) as an explanation for this discrepancy. In the minimal supersymmetric extension of the SM, gluinos can decay into a standard model bottom quark and a lighter supersymmetric sbottom quark. This would increase the yield in bottom production and thus provide a mechanism to explain the apparent discrepancy between data and theory. RHIC could play an important role in understanding this discrepancy through its ability to investigate energy- and spin-dependent charm and bottom production.

Studies of the collisions of transversely polarized protons are of equal importance as those with longitudinal polarization. The transversity densities of quarks and anti-quarks will be accessible at RHIC, but can be probed with greater precision at RHIC II. The chiral-odd character of the transversity structure function makes it inaccessible to inclusive deep inelastic scattering, since electroweak (and QCD) vertices preserve quark chirality. But, transversity is a leading-twist parton distribution function similar in importance to the spin-averaged and helicity asymmetry distributions. Lattice QCD is nearly at the point where robust calculations of moments of structure functions are within reach. The indications from the lattice are that the transverse polarization of quarks in a transversely polarized nucleon is large. Measurements of semi-inclusive deep inelastic scattering are underway that aim to probe transversity through the use of a spin-dependent, chiral-odd fragmentation function (Collins function) to analyze the transverse polarization of the current quark. Initial measurements of the azimuthal asymmetry of particle yields correlated with the polarization direction for one beam (analyzing power) are also underway with the existing RHIC detectors. They may provide sensitivity to transversity through the Collins function, but the contributions from other dynamics that can also give rise to non-zero $A_N$ must first be disentangled. The increased luminosity of RHIC II and the broad acceptance of the proposed new detector will make possible double-transverse spin asymmetry ($A_{TT}$) measurements for high-$p_T$ inclusive jets and the Drell-Yan process. These observables are directly sensitive to the transversity distribution without contributions from differing dynamics.

The Sivers function is a correlation of the form $S_T \cdot (P \times K^\perp)$ between the proton transverse polarization vector ($S_T$), its momentum (P), and the transverse parton momentum relative to the proton direction ($K^\perp$). Non-zero values of the Sivers function arise from the orbital motion of partons within the proton. In collisions of unpolarized protons with transversely polarized protons the single transverse-spin asymmetry in the relative intra-jet azimuthal angle (*i.e.* non-collinearity) of dijets or alternatively the measurement of the relative azimuthal angle between photons and away-side jets are sensitive to the Sivers function.[25] It is also of interest to measure the analyzing power for Drell-Yan production. The gluon interactions that permit this time-reversal-odd distribution function to be non-zero lead to an expected change in sign in comparison of Sivers-type transverse spin effects in semi-inclusive deep inelastic scattering and polarized proton Drell-Yan production. Measurements of the analyzing power for Drell-Yan production will be possible with the projected luminosities for RHIC II and the large acceptance of the proposed new detector.

There are interesting predictions for physics beyond the Standard Model (SM) whose existence may be discovered using spin measurements at RHIC II.[26] These include new parity-violating interactions that would lead to significant modifications of SM predictions. Here parity violation arises within the SM for quark-quark scattering through the interference of gluon- and $Z^0$-exchange. Observation of a parity-violating single-spin asymmetry in inclusive single-jet production at RHIC would signify quark compositeness[27], as depicted by the solid curves labeled $\varepsilon\eta = \pm 1$ in Figure 4. Since the magnitude of deviations from the SM prediction (solid curve labeled SM in Figure 4) is extremely small and increases with transverse jet energy, such measurements are very difficult requiring the highest possible luminosities and data rates, and large coverage for jet measurements to the highest possible transverse energies.

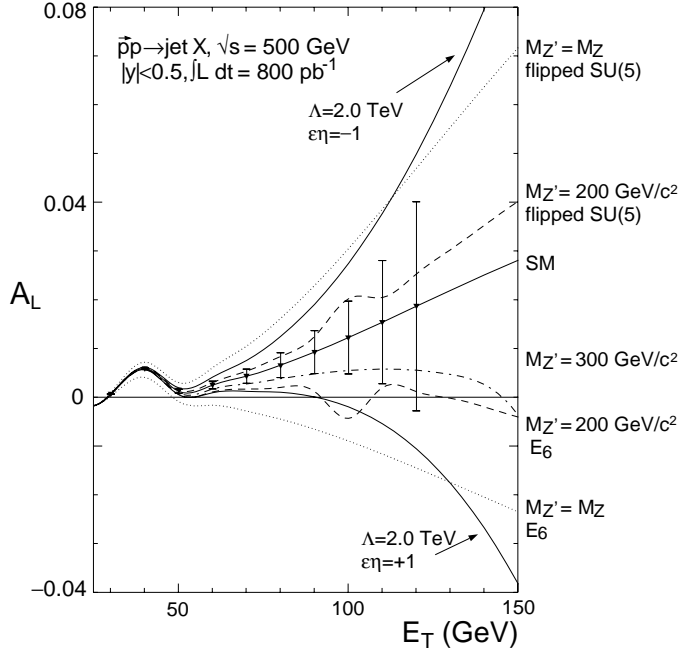

**Figure 4:** Single spin asymmetry for inclusive jet production as a function of jet transverse energy. The standard model prediction is shown (labeled SM) along with predictions for quark compositeness (solid curves labeled $\varepsilon\eta = \pm 1$) and leptophobic model predictions for an extra heavy vector boson that couples directly to quarks (dashed and dotted curves labeled by mass of the new vector boson).

A dedicated comprehensive p+p detector at a high luminosity RHIC II facility is an effective way to explore the structure and dynamics of the proton beyond the present RHIC spin program capabilities. This would allow a focus on rare processes, such as polarization of the QCD sea and parity violating interaction in the Standard Model, that are complementary to a future dedicated eRHIC experiment. RHIC is in a unique position to explore a certain aspect of physics beyond the SM, as it can explore a region of phase space that is unconstrained by current and future experimental efforts at other collider facilities. The capabilities of the proposed (semi-hermetic) detector system certainly benefit the study of physics beyond the standard model at RHIC II.

## 2.5 Detector Requirements

To be able to accomplish the physics listed above the new detector system will contain full electromagnetic and hadronic calorimetry coverage (for measurements of jets and photons, triggering and correlations), high resolution tracking in a large integral magnetic field times track-length ($\int B \cdot dl$), particle identification up to large transverse momenta ($> 20$ GeV/$c$) including flavor dependence of leading particles and detailed fragmentation studies, and high rate data-acquisition and triggering capabilities to utilize the high luminosity effectively for low cross section measurements and photon-, particle-, and jet-correlations.

# 3  Jet Physics

## 3.1  Introduction

The new comprehensive detector will utilize high $p_T$ particles and jets to probe the quark-gluon plasma (QGP) at RHIC, study its properties and gain a better understanding of high density Quantum Chromodynamics (QCD). Proton-proton and proton-nucleus data will be measured in the same detector, as necessary, to distinguish final state effects from nuclear modifications of the parton distribution functions (PDF). The RHIC energy regime appears to be ideal for these studies. The higher energy regime of the Large Hadron Collider (LHC) will provide increased particle and jet yields at high $p_T$ and low $x$. However, recent measurements in the forward direction at RHIC show indications of possible gluon shadowing in the initial state at such low $x$. Thus the mid-rapidity region at the LHC may be dominated by initial state saturation making deconvolution of final state effects, such as parton energy loss and jet quenching, difficult. While measurements in the forward region at RHIC can provide information on gluon shadowing, and possibly saturation or a color-glass condensate, the midrapidity region at RHIC can be used to study predominantly final state effects. Therefore, at RHIC, measurements can be performed over a specific part of phase space (*e.g.* forward- or mid-rapidities) to select the $x$ region of the dominant process of interest.

Precision studies of jets at RHIC will require identification of all particles in a jet to the highest possible $p_T$. The latest reasonable statistics estimates for our new detector show that at RHIC-II this corresponds to jet energies of 40-50 GeV and fragmentation products (identified particles) out to 20-30 GeV/$c$. The present study aims at identifying photons, jet energy profiles, and leading particles at high $p_T$ to distinguish between jets from light, strange, heavy quark and gluon fragmentation. We will attempt to isolate three jet events in order to provide a clean sample of jets from gluon fragmentation. Likewise, the associated baryon, meson, particle, and anti-particle content of all jets will be measured. Conducting these measurements in pp, dA, and AA collisions will provide a wealth of information on jet fragmentation in the vacuum and in the medium in order to understand the flavor dependence of hadronization.

## 3.2  γ-Tagging of Jets

Photons do not re-interact with the medium, and thus can be used to determine the parton energy of the original hard scattering. The efficient reconstruction of high momentum photons, which can be employed for jet tagging out to a γ $p_T$ of 20 GeV/$c$ with good statistics at RHIC-II, is unique to this device. However, the fact that a high $p_T$ photon can be measured and correlated to an away side jet does not lead to a physics result in itself. Instead defining the jet energy of the away side based on energy conservation between the γ and the jet (*i.e.* a γ-tag) is in principle a clean jet energy trigger, assuming one can apply isolation cuts to diminish the background from fragmentation photons (see below) or one applies a sufficiently high γ momentum threshold. The main physics goals of γ-tagged jets, though, still remains the same as in jet studies without a tag. In other words, in order to extract actual physics from a γ-tagged jet the statistics of single particle and pair correlations in the away side jet will set the scale for actual measurements. Some numbers are given in the following chapter.

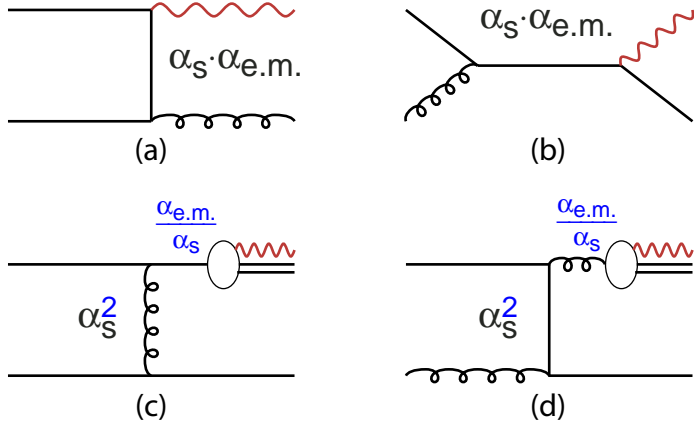

**Figure 5:** LO examples of direct (a+b) and fragmentation (c+d) contributions to the γ+jet cross section.

An additional feature of γ-tagged jets is that the production mechanism of the γ determines the original parton for the away side jet fragmentation. There are only two processes that produce γ-jet combinations, annihilation (Figure 5a) and Compton scattering (Figure 5b). The Compton process requires a gluon in the initial state, while annihilation requires an antiquark in the initial state. Thus, neither production diagram is readily available in a proton-proton interaction. The sea antiquark required for the annihilation process has a softer parton distribution function (lower $p_T$) than the gluon, whereas the gluon for the Compton process is generally harder. Therefore a larger percentage of recoil gluon jets will appear at lower $p_T$ (from annihilation) whereas higher $p_T$ will be dominated by recoil quark jets (from Compton), which can be used to distinguish between quark and gluon jets on the basis of their overall recoil jet energy on the away-side of a photon in proton-proton interactions. However, this effect is predicted to be very dependent upon the incident hadron reaction, as seen in Figure 6. In the case of a pion-proton reaction the distinction between gluon and quark jets as a function of the photon $p_T$ is very strong, whereas for a proton-proton reaction, which will constitute most of the initial collisions at RHIC, the quark jets (from Compton scattering) will always dominate, therefore the $p_T$ dependence of the relative contribution of the two processes cannot easily be explored in order to distinguish between quark jets and gluon jets on the away-side as was done by OPAL. We will try to use the γ-tag to distinguish gluon and quark jets, but the effectiveness of such a tag will have to be determined experimentally and no claims are made here on the basis of such a tag. Still, the γ-tag defines the energy of the jet and in that sense is of utmost importance.

There is, however, a serious background process, namely fragmentation in which the photon originates from the fragmentation of a final state parton as depicted in Figure 5c+d. Despite the fact that the Compton and annihilation subprocesses are of order $\alpha_s^2$, the fragmentation contribution is present already in LO since the parton-to-photon fragmentation functions are effectively of order $\alpha/\alpha_s$ in perturbative QCD[28], where α denotes the e.m. fine structure constant. This process constitutes a considerable background source in which the photon does not carry the full original away side jet energy but only $z \cdot E_\gamma$ where $z$ is determined by the not well known photon fragmentation function. This process complicates the interpretation of γ-tagged jets considerably. So far in γ+jet studies at high-energy colliders the photon is experimentally required to be 'isolated' in order to suppress this background. This is usually achieved by demanding that the amount of hadronic energy allowed in the jet cone around the photon direction is limited to a small fraction of the photon energy (usually 10%). While this isolation procedure works successfully in pp collisions its usefulness in Au+Au can be questioned. The fragmentation background can only be successfully removed at large energies where (a) its contribution is decreasing in NLO and (b) isolation cuts, even in Au+Au, become more effective. The latter requires a hadronic calorimeter, which is unique to our RHIC-II detector proposal. In addition the superior detector resolution in the electromagnetic calorimeter and the high rate capability of all of our detector components allow us to extend the γ-jet tagging capabilities out to a γ $p_T$ of 20 GeV/$c$. This means particle composition studies of a jet in the high multiplicity environment are no longer limited to a statistical approach but rather can be measured in detail with high statistics as a function of single leading particle flavor or intra-jet identified particle correlations. The whole issue of particle identified fragmentation and its relevance to particle production is therefore the main physics thrust of the γ-tagged jet studies as well. Detailed estimates of the achievable statistics are given at the end of this section.

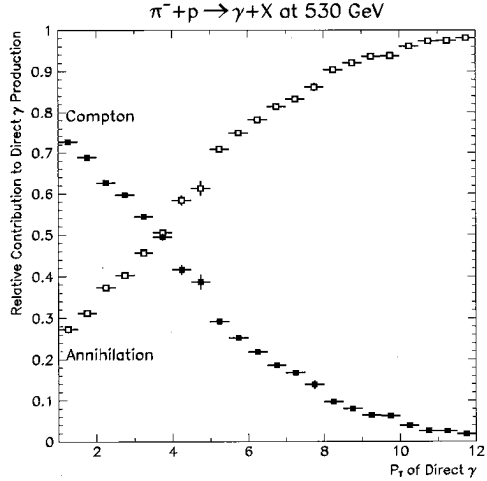
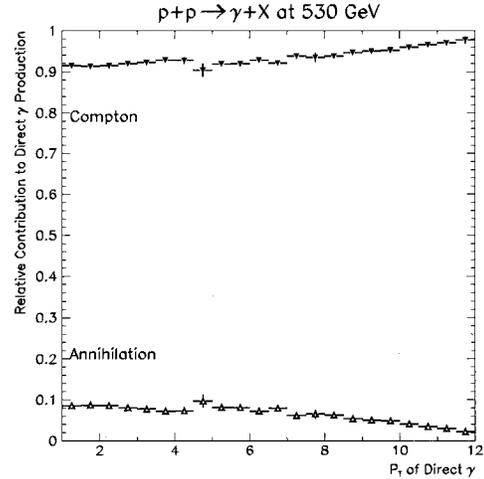

**Figure 6a:** Relative contributions of Compton and annihilation sub-processes in direct photon events vs. $p_T$ of the direct photon. The figure shows that at high values of direct photon $p_T$ the annihilation sub-process dominates in $\pi^- p \to \gamma + X$. These results are HERWIG predictions.

**Figure 6b:** Relative contributions of Compton and annihilation sub-process on direct photon events vs. $p_T$ of the direct photon. The figure shows that for $pp \to \gamma + X$ the Compton diagram dominates. These results are HERWIG predictions.

It should be emphasized that the flavor dependence in γ-jet, γ-leading hadron, di-hadron, and di-jet correlations will be studied as a function of *x* and orientation relative to the reaction plane (established by lower momentum particles in each event). This complex set of correlation data will be necessary for detailed determination of the energy loss mechanism and properties of the QGP.

## 3.3 Jet Fragmentation Studies in pp, pA, and AA Collisions

A unique capability of the proposed detector is particle identification out to particle momenta exceeding 20 GeV/$c$ over a pseudo-rapidity range of $|\eta| < 3.5$. This enables us to measure the particle composition of high energy jets in concert with the jet energy profile measured in the calorimeters. The detailed particle composition of jets in elementary collisions has been the study of many high energy experiments which led to the parameterization of the hadronization process in parton-parton collisions via fragmentation functions. Fragmentation functions are not calculable in QCD but are rather an attempt to factorize the probability $D_q^B(x,Q^2)$ of a certain parton (q) leading to a certain hadron (B) as a function of the *x* of the initial parton and the *z* ($p_h/p_{jet}$) of the final hadron. By applying a statistical approach to many inclusive cross section measurements of identified hadrons in high energy collisions, one can determine the relative contribution of valence quarks, sea quarks, and gluons to the formation of specific hadrons. In that sense the fragmentation function is a parameterization to answer the question of how particles acquire mass. In general, the gross fea-tures of fragmentation functions for different single parton to single hadron conversions are similar as a function of $x$ and $z$ (for a summary see *e.g.*, Ref. 29), which leads to the notion of a universal fragmentation function, in particular for light quarks. Figure 7 shows such a universal function and its potential modification due to radiative energy loss (induced gluon bremsstrahlung) in a dense medium. However, detailed studies of inclusive cross sections have shown, via a statistical approach to the fragmentation process, that as a function of $x$, different partons contribute with different weights to specific hadronization processes. For example, Bourelly and Soffer[30] have shown that for octet baryon production there is a very close relation between the parton distribution functions and the fragmentation functions. In particular strange and heavy quarks have a finite contribution to the formation of non-strange baryons, especially at smaller $x$[31], which means that simple parton-hadron duality pictures must be augmented by exact measurements of the jet content from a particular parton fragmentation[32].

As an example, Figure 8 shows the relative contributions of quark fragmentation functions to two final state baryons (p,Λ) according to the statistical model and based on cross sections measured in elementary particle collisions[30]. Please note the considerable contribution of heavy quarks to the proton and Λ fragmentation at low $x$. If the medium modifications in central heavy ion collisions are flavor dependent, varying for different partons, and the relative contribution of partons to the fragmentation function differs as a function of $x$, then the modification of the fragmentation function cannot be universal as depicted in Figure 7 but rather will have a different shape as a function of $x$ and parton species.

The new detector has a unique capacity to measure high $p_T$ particles, jets, and correlations over a large range of $x$ due to its extended pseudo-rapidity coverage. An estimate of the dependence of η on $x$ is given by the $x \sim p_T e^{-\eta}$. More detailed calculations show however that the $x$-coverage depends on the $Q^2$ scale to be probed. An example for $x$-ranges at fixed RHIC pseudo-rapidities is given in Figure 9.

Based on these calculations we expect the new RHIC-II detector to probe the perturbative regime in the range of $x = 0.1$ to $10^{-4}$ and thus extend the present knowledge of fragmentation in the unquenched regime of pp collisions for various types of partons over a wide range of $x$-values, including heavy quarks and potentially gluons via the γ-jet tag. Particle identification in the jet helps to understand jet formation mechanisms in proton-proton collisions and will allow us to investigate modifications of the fragmentation functions in nucleus-nucleus collisions. Nucleus-nucleus data will be compared to proton-nucleus data taken at the same energy and with the same detector configuration and at different energies (EMC and HERMES) in order to distinguish effects of the QGP and of the cold nuclear medium.

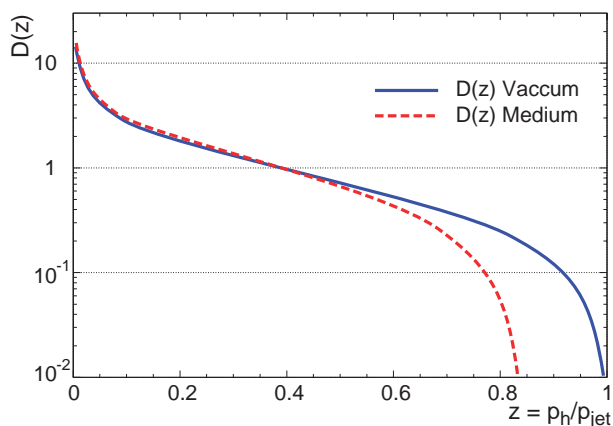

**Figure 7:** Universal modification of universal fragmentation function based on the assumption of induced gluon bremstrahlung as the main radiative energy loss in the medium[33].

If differences in parton contributions to the hadronization can be further corroborated experimentally, one would expect the assumption of a universal modification, due to radiative energy loss, to the universal fragmentation function to be an oversimplification. This idea would then have to be modified for each particle as each parton species would be expected to suffer different amounts of energy loss in the dense opaque medium. In general it is already anticipated that gluons, light and heavy quarks interact differently. This will be discussed below.

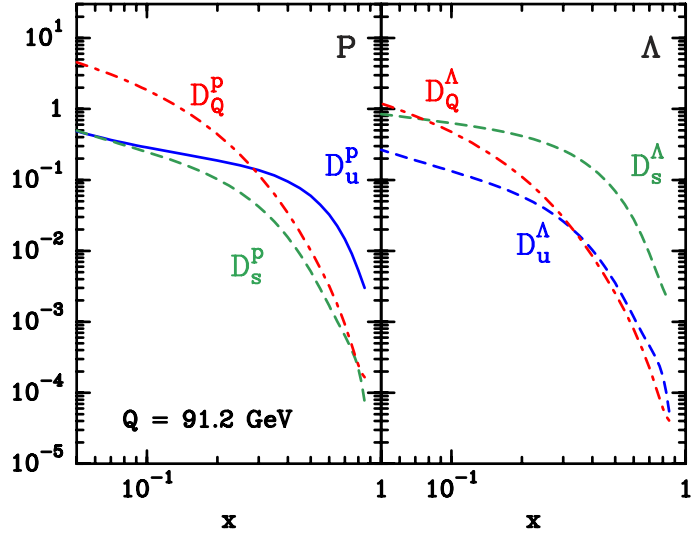

**Figure 8:** The quark octet baryons fragmentation function $D^B_q(x, Q^2)$ and $D^B_q(x, Q^2)$ ($B = p, \Lambda$, $q = u, d, s$ and $Q = c, b, t$), as a function of $x$ at $Q = 91.2$ GeV.

**Gluon quenching:** For parton momenta much higher than the constituent quark mass the energy loss is expected to be universal for quarks and when scaled by the color (Casimir) factors ($C_F/C_A$) identical to the gluon energy loss. The color factor, though, will increase the multiplicity in the un-quenched gluon jet[34] and the energy loss of the gluon in the quenching scenario. In proton-proton interactions the away-side $p_T$ distributions can be used to select quark versus gluon jets to further study their particle content, while in nucleus-nucleus interactions the parton energy loss complicates this selection criterion considerably. However, in nucleus-nucleus interactions the ratios of leading anti-particles to leading particles, such as $\bar{p}/p$ and $K^+/K^-$, at high $p_T$ can be used to distinguish energy loss effects of gluon versus quark jets, because anti-particles are more likely to be produced from fragmentation of hard-scattered gluons than from quarks, whereas particles are produced in both, quark and gluon fragmentation. At high $p_T$ perturbative–QCD predicts that these antiparticle–to-particle ratios should decrease with increasing momentum. Furthermore, a difference in the anti-particle to particle ratios as a function of $p_T$ pro-vides information on the relative energy loss of quark jets relative to gluon jets and can be compared to results from proton-proton and proton-nucleus interactions to determine the energy loss difference between fast quarks and gluons traversing the medium in the nucleus-nucleus case. A prediction by X.N. Wang is shown in Figure 10.

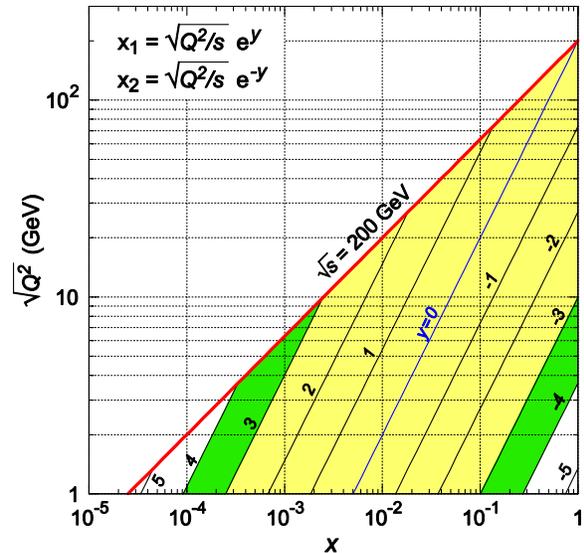

**Figure 9:** Comparison of $x$-ranges covered by RHIC as a function of rapidity and $Q^2$.

Another possibility to study gluon jets is to trigger on three jet events, in which case the lowest energy jet, which is emitted transversely to the remaining di-jet axis has to be a gluon jet. Simulations to investigate whether this relatively low energy jet can be identified in heavy ion collisions at RHIC and whether the production cross section of such three-jet events is sufficiently high at RHIC-II are underway.

**Light quark quenching:** Although the s-quark quenching is not expected to show large deviations from that of the u- and d-quark, one might be able to measure detailed differences in the intermediate $p_T$ range where the constituent quark mass cannot be neglected. These detailed measurements will certainly help to establish a more complete picture of hadronization of the abundantly produced particles, although the intermediate $p_T$ range is likely to be also populated by other production mechanisms, such as thermal parton recombination.

One additional light quark effect that is not easily quantifiable is the apparent contribution of non-valence quarks to the universal fragmentation function. Certainly the parton distribution function of sea-quarks is significantly softer than the valence quark distribution, but for the radiative energy loss the color factor scaling cannot be applied. In any case, the sea-quark contribution to the fragmentation process is of different weight at different values of $x$, and it contributes more strongly to the low $z$ part of the fragmentation function. Therefore the modification of this contribution to the fragmentation in an opaque medium cannot be generalized and has to be measured for each particle as a function of $z$ and $x$. Certainly studies of the nuclear modification factor $R_{AA}$ for identified leading particles and as a function of the multiplicity of the jet will provide more detailed information on how the fragmentation functions are modified in nucleus-nucleus. In the context of identified leading particles, it is interesting to note that the production rate of strange baryons per jet in elementary reactions is nearly independent of jet energy in contrast to the strong dependence measured for the total charged multiplicity, as shown by OPAL[31] in Figure 11. OPAL also finds that in simulations the relative abundance of strange mesons and baryons is dependent on the flavor of the jet (see Figure 11b). It will be important to measure how or indeed if medium modifications in relativistic nucleus-nucleus collisions affect these results.

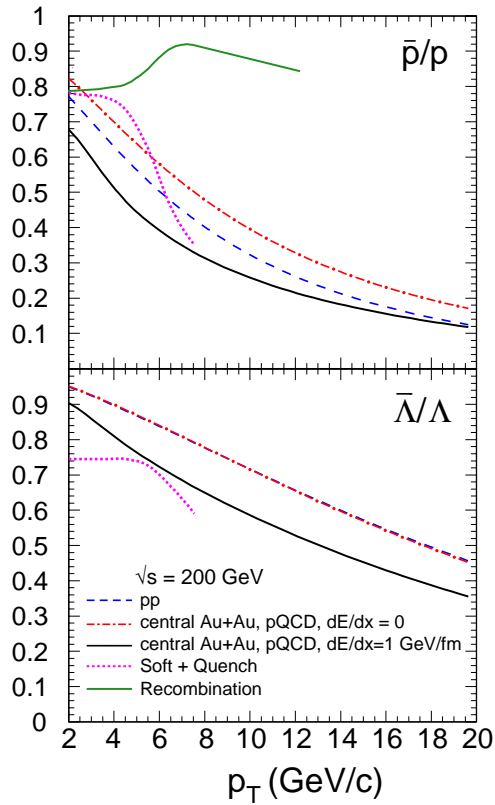

**Figure 10:** The ratio of $\bar{p}$ to p (upper panel) and $\bar{\Lambda}$ to $\Lambda$ (lower panel) spectra as a function of $p_T$ in pp (dashed) and central Au+Au collisions at $\sqrt{s}$ = 200 GeV without energy loss (dot-dashed) and with an energy loss $dE_q/dx$ = 1 GeV/fm (solid). The mean free-path $\lambda_q$=1 fm. Gluons are assumed to lose twice as much energy as quarks.

**Heavy quark quenching:** For the heavy quarks (c and b) the dead-cone effect, described by Dokshitzer and Kharzeev[35], predicts that heavy quarks suffer considerably less energy loss in the medium than light quarks since for heavy quarks gluon bremsstrahlung at small angles is suppressed. Hence, jets with a D or B-meson as leading particles should be less suppressed than light quark jets. If the magnitude of the quenching effect were found to be similar for light and heavy quark jets, then gluon bremsstrahlung could not be the dominant mechanism. Thus, measurements of identified heavy quark jets and comparison with light quark jets, *e.g.* via ratios of D/π and B/π, will help identify the energy loss mechanism primarily responsible for jet quenching. In order to study heavy flavor jet quenching, displaced vertices will be used to identify and trigger on heavy flavor decays. High $p_T$ electrons in coincidence with a leading hadron, both emanating from a vertex that is displaced from the primary nucleus-nucleus reaction vertex, will provide a trigger for heavy flavor decays. Examples of other specific decay modes of interest for a displaced vertex, heavy flavor trigger are $B^0 \to J/\psi + K^0_s$, and D and $D^* \to K\pi$ and $K\pi\pi$.

**Alternatives to gluon bremsstrahlung:** There are indications already in proton-nucleus (*i.e.* cold nuclear matter) measurements that the flavor composition and the overall multiplicity in a jet is changed from proton-proton to proton-nucleus interactions due to the coalescence of partons with quarks in the medium[36,37]. This leads to a modification of the fragmentation function already in proton-nucleus interactions. The effect can be explained in the context of rescaling models, which implement partial deconfinement inside nuclei by modifying the fragmentation function perturbatively[38]. This model is an alternative to the gluon bremsstrahlung models[39,40]. It is interesting to note that in a rescaling model the level of deconfinement can be adjusted[8], and therefore the $R_{AA}$ measurements in nucleus-nucleus collisions at RHIC II should be compared to a total deconfinement calculation. A recent paper[41] also pointed out that the photon contribution to the radiative energy loss might not be negligible, which will lead to important γ-measurements that can only be fully achieved with the new detector capabilities.

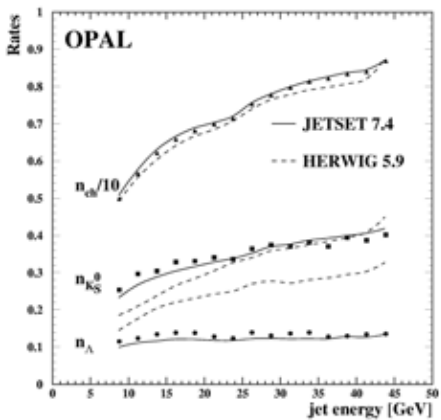
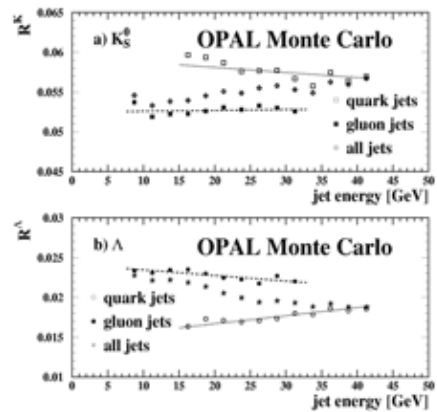

**Figure 11a:** Production rates per jet of charged particles, $K^0_s$ mesons and Λ baryons, $n_{ch}/10$, $n_K$, and $n_\Lambda$ as a function of the jet energy compared with predictions of the models JETSET 7.4 and HERWIG 5.9. The charged particle rates are scaled down by a factor of 10.

**Figure 11b:** Relative production rates of (a) $K^0_s$ and (b) Λ in JETSET 7.4 events for pure quark and gluons jets as a function of the jet energy. Straight lines are fitted to the points. The statistical errors are smaller then the size of the symbols. The zeros of the vertical axes have been suppressed.

**Energy loss azimuthal distribution:** Early jet measurements showed that jet production in hadron-nucleus reactions is enhanced beyond a simple linear A scaling factor for moderate $\sqrt{s}$[42], whereas for higher incident energies the scaling approaches a linear A dependence[43,44]. The enhanced A dependence of the jet and di-jet cross sections at $\sqrt{s}$ comparable to RHIC energies is largely attributed to jet-like contributions from the underlying background which do not scale linearly with A. These jet-like contributions can be measured in a direction perpendicular to the jet cone and then compared to the jet content itself[12], first in pp collisions to determine a reference and then in AA collisions, as this dependence might change in the opaque medium if the gluon radiation is not collinear. In other words, does the gluon radiation stay confined in the cone or does it also affect the particle production pedestal outside of the cone? Furthermore, does it change the flavor composition of the particle production outside the jet cone? A first detailed theoretical study of jet fragmentation softening versus jet suppression was recently performed by Lokhtin and Snigirev for the case of nucleus-nucleus collisions at RHIC and LHC[45]. The softening of the fragmentation function with no substantial jet rate suppression would indicate that small angle gluon radiation (*i.e.* in the jet cone) dominates the medium-induced partonic energy loss, whereas an increasing contribution of wide-angle gluon radiation or collisional loss would lead to jet rate suppression. Again, this question can only be addressed experimentally by measuring the complete jet and the 'out-of-cone' energy loss via the particle identified momentum spectrum of the underlying event and the jet itself. In order to distinguish the effect of gluon radiation outside of the jet from other particle production mechanisms the jet energy profile must be measured together with the jet particle composition, which can only be accomplished in the new detector.

**String fragmentation versus recombination in jets:** Previous jet experiments using elementary collisions attempted to distinguish the different particle production mechanisms in a jet cone: the standard quark-diquark picture (direct production through string fragmentation) and the so-called popcorn mechanism (quark coalescence or recombination of independently produced quarks). DELPHI[46] has shown in rapidity correlation studies for baryons that the baryon-antibaryon pair generated in a simple fragmentation process always stays within the rapidity window of the jet if produced through the direct process of string fragmentation. However, if one applies a recombination or quark coalescence scenario to independent quarks in the QGP, the rapidity correlation of the baryon-antibaryon pair should be weaker.

Differences in the rapidity correlations between $p\bar{p}, \Lambda\bar{\Lambda}, \Xi\bar{\Xi}$, and $\Omega\bar{\Omega}$ and mixed pairs could be used to better understand the various contributions in the jet cone. Also the relative contribution of baryons to mesons in a jet is expected to be different depending on the production mechanism. Furthermore, asymmetries in the production rates between $D^+$ and $D^-$ mesons in proton-proton reactions (the so-called leading particle effect[47,48]) can only be understood in the context of a quark recombination of two partons in the initial hadron[49,50] or independently produced partons[51].

By extending these types of measurements from elementary pp-collisions to complex AA-collisions, we can first measure baseline correlations and yields of identified particles in order to sort out the relative production mechanism contributions in pp and then distinguish partonic energy loss effects, which should, for example, affect the baryon-antibaryon correlations. The picture of hard parton recombination can then be further extended to include thermal partons as well which has lately been investigated by some theorists (*e.g.* Ref. 52).

**Measurements unique to the new RHIC-II detector:** The main advantage of the new detector over any existing or upgraded RHIC device is that it combines the relevant features of a large and symmetric coverage for tracking, PID, and calorimetry over a large momentum range with extremely high data acquisition rates and superior triggering capabilities. The PID coverage in terms of coordinate and momentum space is unparalleled, and the inclusion of hadronic calorimetry allows, for the first time at RHIC, unambiguous jet energy and particle profile, high $p_T$ particle correlation, and γ-jet measurements over a large range in $(x,Q^2)$.

The key measurements in the context of jet studies at RHIC-II will remain those of jet suppression in the opaque medium via single particle yield, flow comparisons between pp and AA, and many particle correlations to distinguish between jet structures and bulk matter. It is a feature of this detector, that these measurements are no longer limited to the leading jet particle but rather include the full jet profile in terms of energy, particle multiplicities and species. The capability to break down the jet content particle by particle from the pion to the B-meson over a $p_T$ range from 500 MeV/$c$ to above 20 GeV/$c$ will enable for the first time one to address the fundamental issues of hadronization and particle production in pp and AA collisions on a highly detailed level. As an example of the wealth of particle species fragmenting from a parton, Figure 12 shows the fragment particle yields per event in the new detector as a function of the fragment transverse momentum for jets with an incident parton $p_T$ > 10 GeV/$c$ in √s = 200 GeV proton-proton collisions at RHIC-II. In more forward directions this distribution is limited by the kinematics, therefore we also show the momentum distributions of various fragmentation particles at mid- and forward-rapidities in Figure 13.

The unparalleled ability of the comprehensive new detector design in particle and energy identified jet studies in pp and AA will allow our field to bridge the gap between high energy and nuclear physics in a way that enables us to answer the core question of elementary particle production in nature. The rather awkward attempts to describe hadronization through factorization via fragmentation functions in the vacuum will be complemented by measurements in different phases of matter with different coupling constants. This will put serious constraints on the validity of the factorization approach and our ideas of hadronization mechanisms and it will therefore help to answer the fundamental question of hadron production and chiral symmetry breaking. In particular the measurement of the particle-identified content of jets from relatively moderate $p_T$ out to high $p_T$ is unique to this detector. For example, γ-jet measurements are possible starting from a γ momentum of 7 GeV/$c$ out to 20 GeV/$c$ with sufficient statistics. The upgraded RHIC detectors can neither match the low $p_T$ cutoff due to resolution constraints nor the high $p_T$ cutoff due to statistics limitations. Table 1 shows our statistics for γ-jets assuming a 30 nb$^{-1}$ run (14 weeks at RHIC-II design luminosity with 50% dead time).

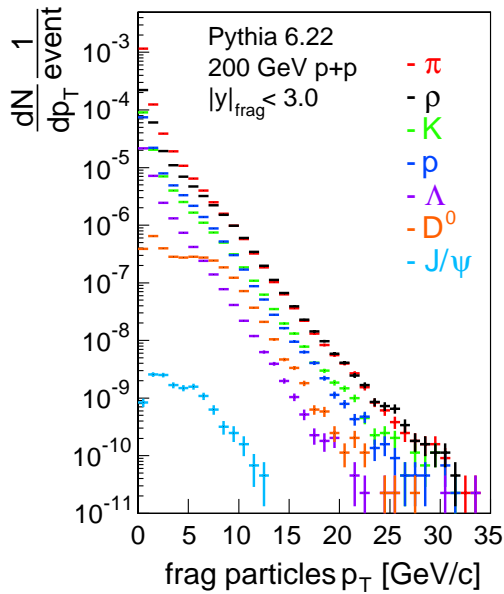

**Figure 12:** Fragmentation particle spectra per proton-proton collision from an incoming parton with $p_T$ > 10 GeV/$c$ in |η| < 3 as a function of the transverse momentum.

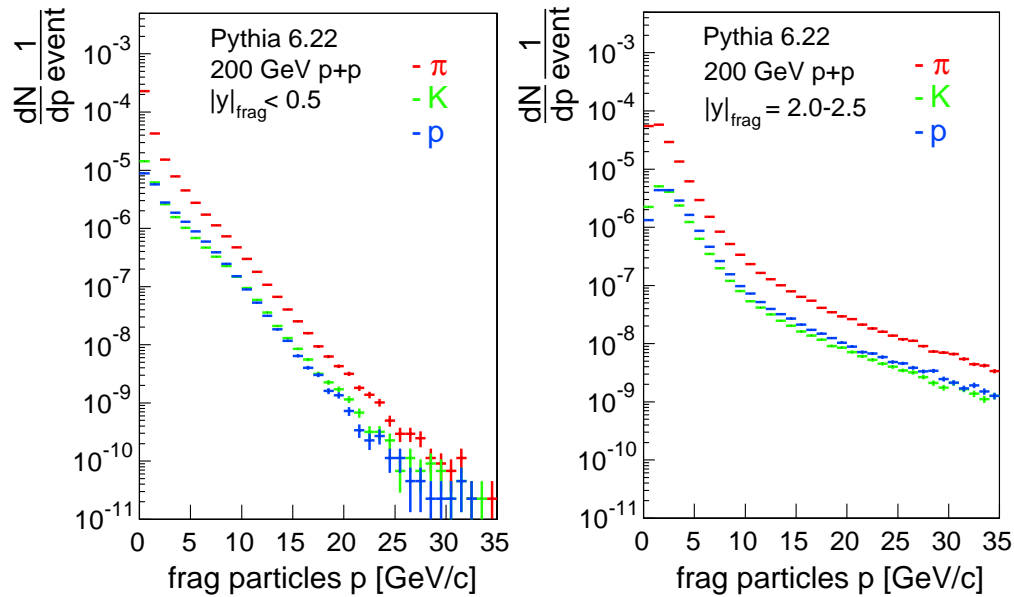

**Figure 13:** Fragmentation particle multiplicities per proton-proton collision from an incoming parton with $p_T > 10$ GeV/$c$ in central and forward rapidities as a function of the total momentum of the fragmentation particle.

| Transverse energy of $\gamma$ | Number of events |
|---|---|
| 20 GeV | 33,000 |
| 30 GeV | 1,500 |

**Table 1:** $\gamma$-jet rates in Au-Au collisions.

One unique measurement of the RHIC era was the determination of jet properties in the medium via azimuthal jet particle correlations. Early measurements by STAR indicated that the correlations inside the jet cone and between same-side and away-side jets in a di-jet event can be measured through two particle correlations. The dramatic disappearance of the away-side jet in the opaque medium produced in central Au-Au collisions is widely seen as proof for partonic jet quenching in a QGP-like medium. These measurements still lack the $p_T$ reach that is required to distinguish modification of the fragmentation process from other production mechanisms at more moderate $p_T$, such as recombination of thermal and shower partons. It is very important to (i) extend the $p_T$ reach of two-particle correlations to a $p_T$-range where fragmentation can be safely tested, and (ii) extend the measurements to high $p_T$ correlations of more than just charged hadrons (*i.e.* identified hadrons). The differences in correlations between different particle species (*e.g.* baryons vs. mesons, particles vs. anti-particles, and light quarks vs. heavy quark particles) will add to the understanding of the fragmentation process and its modification in the QGP phase. As an example Table 2 shows the anticipated numbers for certain single particle yields and two-particle correlations in a specific $p_T$ window achievable with our detector. Neither of the upgraded RHIC detectors will be able to achieve these measurements, since the increase in PID coverage from PHENIX and STAR to this detector is a factor 72 and a factor 3, respectively. Furthermore, the STAR PID coverage does not extend beyond 5 GeV/$c$, PHENIX cuts off at around 10 GeV/$c$, while the coverage of the new detector extends to beyond 20 GeV/$c$, as shown in Figure 14.

| di-jet leading particle yields | $p_T$-cutoff and PID | number of particles |
|---|---|---|
| | > 10 GeV/$c$ protons | 15,000,000 |
| | > 20 GeV/$c$ protons | 30,000 |
| γ-jet leading particle yields | $p_T$-cutoff and PID | number of particles |
| | > 10 GeV/$c$ protons | 5,000 |
| | > 15 GeV/$c$ protons | 100 |
| two-particle jet correlations | $p_T$-cutoff and PID | number of pairs |
| | > 4 GeV/$c$ Lambda pairs | 50,000 |
| | > 10 GeV/$c$ proton pairs | 5,000 |

**Table 2:** Leading particle yields and correlation pair yields in di-jets and γ-jets in Au-Au.

An interesting extension of the PID capabilities and superior tracking resolution to high $p_T$ is to distinguish jet and bulk matter contributions in the intermediate $p_T$ range by measuring the $\langle p_T \rangle$ and multiplicity fluctuations in the event. It is expected that the thermal bulk matter will fluctuate with a significant difference to that expected from the explicitly non-thermal unquenched jet contribution. The quenched jet contribution might partially thermalize but the directivity of the jet could be preserved in parton recombination[22].

**In summary:** Measuring particle identified fragmentation functions as a function of the hadron momentum fraction $z$ at different $x$ scales in pp and AA collisions, *i.e.* in the vacuum and in an opaque medium that behaves like an ideal fluid, will enable us to determine not only the exact parton contribution to each produced hadron, but also the energy and $x$ dependence of the process that leads to massive hadrons. In that sense these measurements address one of the fundamental, unanswered questions of physics, namely how particles acquire their mass. These detailed studies are seriously limited in the LHC detectors and the upgraded RHIC detectors, due to the lack of the combined detector capabilities necessary for these measurements, namely PID for particle momenta up to 20 GeV/$c$ and calorimetry continuously from central to forward rapidities.

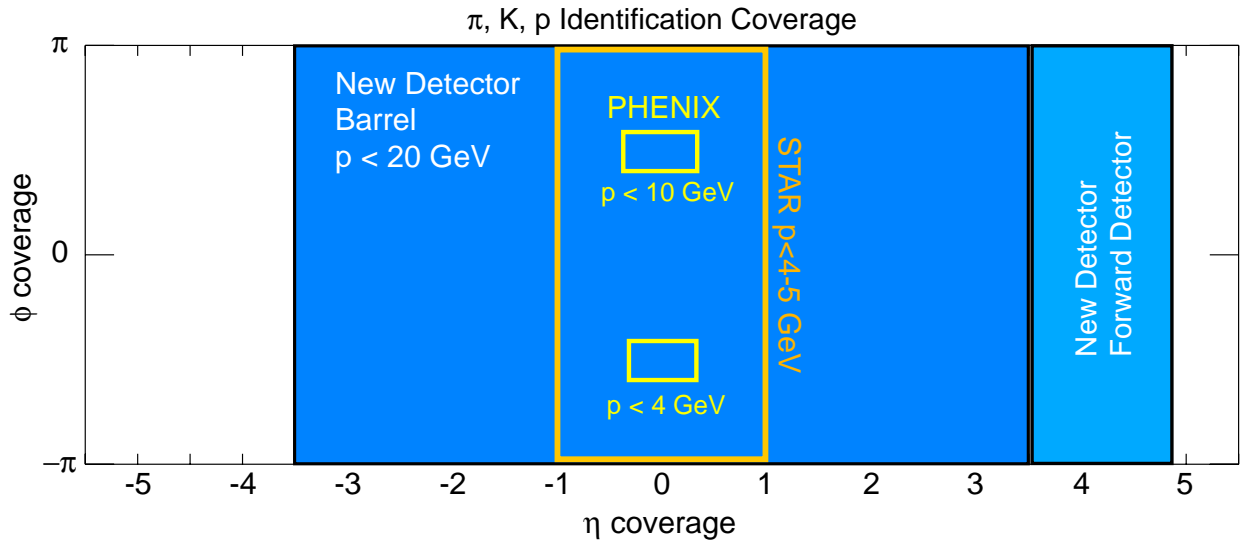

**Figure 14:** Comparison of particle identification capabilities in the new detector and the upgraded RHIC-I detectors as a function of pseudorapidity coverage and ϕ. The boxes show STAR and PHENIX, the blue shaded area shows the new detector coverage. The maximum transverse momentum for which particle identification is possible is indicated by the values in the boxes.

## 3.4 Chiral Symmetry Restoration via Resonances in and out of Jet Cones

In the QCD vacuum chiral symmetry is spontaneously broken by the formation of the quark condensate, thereby generating the major part of the mass of most hadrons and inducing a large mass splitting of chiral partner states in the hadronic spectrum, *e.g.* π-σ, ρ(770)-a1(1260) or N(940)-N(1535). The (approximate) restoration of chiral symmetry at high temperatures, as predicted by lattice QCD[53], implies that, at the phase transition, chiral partners become degenerate. Therefore, medium modifications of hadron masses and widths are generally considered as valuable signals for chiral restoration, especially for short-lived resonances decaying at least partially in the produced hot and dense medium. So far, the most promising observable for medium modified hadron properties are dilepton spectra which directly probe the spectral functions of vector mesons, most notably the ρ(770) due to its large electromagnetic decay rate[54,55]. Thus, it is of prime importance to gain information on medium effects on the chiral partner of the ρ, the $a_1$(1260), *e.g.*, through its π–γ decay channel[56]. RHIC measurements were not yet able to gather evidence for chiral symmetry restoration either due to a decoupling of the chiral and the deconfinement transition or a lack of viable candidate measurements. The new detector offers a unique expansion of these studies by potentially allowing the simultaneous reconstruction of the ρ and the $a_1$(1260). Besides the standard π–ρ decay channel of the $a_1$, the detector's unique capability in low momentum γ detection enables us to explore the π–γ channel for the $a_1$. The γ capabilities are expanded via the inclusion of additional γ converters also needed for the γ–γ HBT. Whether the very broad unquenched $a_1$ resonance can be detected in central AA events with a high combinatorial background is under investigation, but studies indicate that in pp and peripheral events RHIC-II offers the unique opportunity to measure in medium and out of medium masses of the ρ and the $a_1$ in the same event through the simultaneous measurement of quenched and unquenched jets. An unquenched jet in central Au-Au collisions is likely to be emitted from the surface of the medium and therefore any resonance contained in an unquenched jet will be produced out of medium, whereas the quenched away side jet will lose most of its energy to the medium and likely thermalize within the matter. Therefore thermal ρ's and $a_1$'s, either from the thermalized bulk or a quenched jet, will exhibit medium modifications possibly indicative of chiral restoration. A measurement of the ρ/$a_1$ ratio based on two particle azimuthal correlations in dijets in central Au-Au collisions will be another good candidate for measuring chiral symmetry restoration, due to the approximate degeneracy of the ρ and $a_1$ mass distributions at/close to the critical temperature. As it is essential to scrutinize theoretical models, measurements of medium modifications of the other light vector mesons, φ and ω, will also be conducted. To assess the relative yields of in-medium and free decays, as well as additional medium effects on the hadronic decay products of a resonance, the new detector will allow for a comparative study of di-lepton and di-hadron decay channels for the ρ and the φ, which will also enable investigation of changes in the pertinent branching ratios. In order to ensure likely decay of the resonance in the medium and simultaneously sufficient transverse momenta to detect the leptonic or hadronic decay products in the detector's high field environment we plan to trigger on resonances that are emitted transverse to the di-jet thrust axis.

## 3.5 Parity and CP Violation Studies

It has been proposed that heavy-ion collisions may create meta-stable regions of space in which parity and CP symmetries are spontaneously violated in strong interactions[57]. These CP-odd meta-stable regions may exist in hot QCD despite the fact that the parameter for the effective strength of the CP-violating term in the QCD Lagrangian is zero to very high precision in nature. The regions should create observable effects in the produced hadrons, a unique measurement as parity and CP violation have never been seen in the strong interaction. Their measurement would thus have implications not only in the understanding of the 'Strong CP problem' (why do strong interactions respect parity and CP invariances?), but also because it is closely related to a core issue of symmetry restoration (in this case, the axial U(1) symmetry) in the excited, high temperature state possibly created in heavy-ion collisions. Since the effect is relatively small and changes sign every event, event-by-event correlation methods must be employed to search for a signal. Several such methods have been suggested, of which many involve looking for certain asymmetries in charged particle production. The most recently suggested methods along these lines[58] is practicable with current RHIC detectors, but it is our current understanding that this method will have a difficult time distinguishing true CP violation from other non-CP violating effects that lead to similar charge correlations. Such methods will of course continue to be explored.

A much less ambiguous signal could be seen by observing longitudinal helicity correlations of produced hyperons. CP violating regions affect the net helicity of quarks and anti-quarks. Such helicity flips are as likely to affect the strange quark as the light u and d quarks. The spin of the $\Lambda$ is dominated by its strange quark and hence there should exist correlations in the helicities of $\Lambda$s (here $\Lambda$ means $\bar{\Lambda} + \Lambda$) produced in these domains. Since $\Lambda$s decay through the parity violating weak decay this effect should be observable by measuring the $\Lambda$ helicity correlations event-by-event. The $\Lambda$ helicity may be determined, to a good approximation, as up or down by measuring the decay proton's direction in the decay center of mass frame relative to the $\Lambda$ line of flight in the collision frame. For each event the ratio $R = N_{up}/(N_{up} + N_{down})$ is determined; if CP-odd domains exist, non-statistical fluctuations in R should be observed. While not background free, it is believed that backgrounds can be controlled sufficiently such that should non-statistical deviations be observed, CP violation could be determined with very high confidence. The main physics source of background is from $\Xi$ feed-down which creates longitudinally polarized $\Lambda$s in the $\Xi$ rest frame. Fluctuations in the event-to-event $\Xi$ yield could create a fake CP violation signal as described above. However, taking a pessimistic view that the $\Xi$ decay asymmetry, $\alpha = 1$ (not the observed value of 0.45) and that all $\Lambda$s are from $\Xi$ decays, no measurable signal was observed in 200 million simulated events. This is so because the $\Lambda$ is polarized in the $\Xi$ rest frame and the effect is washed out when boosted into the $\Lambda$ rest frame. Another possibility is to only look for $\Lambda$ helicity correlations of pairs of $\Lambda$s identified as resulting from a "jet". By triggering on a high $p_T$ particle and looking at the asymmetry ratio R for $\Lambda$ on the away side you can hope to enhance the signal by increasing the probability that both $\Lambda$s were created in the CP violating bubble. However, while this increases the probability of the $\Lambda$s being created close to each other, our initial estimates show that the event statistics needed are greatly increased due to the need to create, and identify, high $p_T$ trigger particles.

Whether these metastable states actually have observable effects depends on the probability of formation, their size, number density and lifetime. There currently exists only rough theoretical guidance as to what signal strength can be expected from this method[59]. Initial (and likely optimistic) simulations, assuming that 1/10[th] of the produced strange quarks result in a primordial $\Lambda$, that the topological charge (Q) distribution is a Gaussian of width 10, that each unit of Q produces a helicity flip of one quark, and that 30% of these flips affects strange quarks, show that even with 50% efficiency for identifying produced $\Lambda$s, a meaningful $3\sigma$ measurement will take a minimum of 100 million central events. The number of events cannot be reduced by triggering; thus high luminosity and data rates are clearly essential (as is excellent vertexing capability for hyperonic lifetimes) for such a measurement to be performed. These high rate and reconstruction efficiencies are only possible at RHIC II.

# 4 Quarkonium Program

## 4.1 Theoretical Motivation

In 1986 Matsui and Satz proposed in a landmark paper[60] the study of *J/Ψ* production in AA collisions as a signature of a QGP. They predicted a suppression of *J/Ψ* due to the color screening of the static potential between heavy quarks. This idea was later (2000) refined by Kharzeev and Satz [61] after the octet model for quarkonium production was established. In this approach, the color octet state ($c\bar{c}g$) can be broken up by hard gluons that are not available in a hadron gas but are abundant in a QGP.

With the prospect of a possible measurement of bottonium at RHIC and LHC theoretical efforts began also to include the ϒ states[62,63]. Earlier, most studies focused on the measurements of charmonium states, mainly because of their relatively large production cross-section when compared to bottonium states. However, bottonium spectroscopy in nucleus-nucleus collisions has various advantages compared to charmonium spectroscopy. Bottonium mesons are massive (~10 GeV/$c^2$) and their decay leptons have relatively large momenta and are thus easy to distinguish from electrons originating from most background sources. The combinatorial background in this mass range is extremely small and multiple scattering is of little concern. While the interpretation of charmonium suppression is made more difficult by the rather large cross-section for absorption by co-moving matter, the situation for bottonium is considerably easier. Calculations of the absorption of directly produced ϒ by hadronic co-movers show that this effect is essentially negligible[62].

As early as 2001, studies of the heavy quark potential on the lattice became available that quantitatively predicted a sequential suppression of the various quarkonium states[64,65]. They indicated that the ϒ(2S) state dissolves at temperatures and conditions almost identical to those for the *J/Ψ*, while the ϒ(1S) state stays unsuppressed up to temperatures of more than twice the critical temperature $T_C$. Such high temperatures are not likely to be achieved at RHIC energies, making the ϒ(1S) state a possible standard candle with which to compare. On the other hand, because of its low binding energy the ϒ(3S) is predicted to dissolve at temperatures below $T_C$. Therefore, a measurement of the various bottonium states can shed light on the production (via ϒ(1S)) and suppression mechanisms (ϒ(2S) and ϒ(3S)) of heavy quark bound states, avoiding many difficulties inherent in charmonium measurements. Only recently[66] new lattice calculations have revealed the importance of the entropy term in the calculation of the free energy *F*, which is commonly interpreted as the heavy quark potential at finite *T*. The conclusion from these results is that the charmonium ground state (*J/Ψ*) persists in the QGP as a well defined resonance with no significant change in its "zero temperature" masses at least up to $T \sim 1.5\ T_C$. The *J/Ψ* gradually disappears for $T > 1.5\ T_C$ and is gone at 3 $T_C$ (see Figure 15). These results, however, must be treated with caution. So far the calculations[67,68] performed by several groups are inconsistent, and all ignore the width of the quarkonium states but predict only position and yield. It is still rather likely that future calculations, once the treatment of resonance widths are properly taken into account, will show that the *J/Ψ* will dissolve also at RHIC[69]. The current expectation, based on the different binding energies and masses of the various quarkonium states, predicts a systematic suppression pattern, and thus allows measurements to "bracket" the temperature in the initial stage. As for now the expectations at RHIC are: $T_{\text{diss}}(\Psi') < T_{\text{diss}}(\Upsilon(3S)) < T_{\text{diss}}(J/\Psi) \approx T_{\text{diss}}(\Upsilon(2S)) \leq T_C$ (RHIC) $< T_{\text{diss}}(\Upsilon(1S))$. While this might be modified somewhat by future theoretical refinements, what will remain true from these new lattice results is the importance of a comprehensive study of **all** experimentally accessible quarkonium states rather than focusing on one, the J/Ψ, alone. A systematic study of heavy quarkonium spectroscopy, with a full determination of the suppression pattern across the various states, remains the most direct probe of deconfinement. It is also a signature that most closely acts as a thermometer of the hot initial state, which, with future improved lattice calculations, can be directly compared to QCD.

At RHIC, the initial nucleon-nucleon collisions may not be the only source of charmonium production. Regeneration of quarkonium, either in the plasma phase[70,71] or in the hadron phase[72], could counter the effects of suppression, ultimately leading to enhanced quarkonium production. In the plasma phase there are two basic approaches: statistical and dynamical coalescence. Both approaches depend on being able to measure quarkonium relative to the total $Q\bar{Q}$ production. The first calculation in the statistical approach assumed an equilibrated fireball in a grand canonical ensemble[70]. This effect, however, is considered to be small at RHIC because of the still moderate abundance of heavy $Q\bar{Q}$ pairs. Dynamical coalescence assumes that some of the produced $Q\bar{Q}$ pairs can also form quarkonia which would not be produced otherwise. These models predict enhancements of up to a factor of 2-3 at RHIC[73]. Quarkonium production in the hadron gas is expected to be negligible at RHIC for the J/Ψ while larger enhancements are predicted for the Ψ' which depends strongly on the cross-section of J/Ψ with π and ρ. It is important to note that any secondary production will be at lower center-of-mass energies than the initial production from nucleus-nucleus interactions. Thus the production kinematics will be different, leading to narrower rapidity and $p_T$ distributions. Secondary quarkonia could be separated from primary quarkonia, subject to suppression, by appropriate kinematical cuts. Such cuts will also be useful to identify (rare) J/Ψ's from B-meson decays. Most striking is the dependence on $\sqrt{s}$ of the ratio of produced J/Ψ relative to the number of $c\bar{c}$ pairs, depicted in Figure 16. A measurement of an excitation function of this ratio over a range of $\sqrt{s}$ = 30-200 GeV could help to disentangle the suppression from a possible enhancing mechanism such as recombination/coalescence. Such a measurement, however, is extremely demanding in terms of statistics since heavy quarks and quarkonia will need to be measured with good statistics over a wide range of beam energies.

In general the *production* of quarkonia is to a large extent not understood. The ability to extract information about the plasma from the features in AA collisions relies on our understanding of production in pp and pA interactions. The suggested importance of the color-octet mechanism, initially widely accepted, is now under scrutiny in light of recent measurements at CDF, BaBar and Belle (*e.g.* double $c\bar{c}$ production). A polarization measurement, *i.e.*, the analysis of the angular distribution of the quarkonium decay products $d\sigma/d\cos\theta^* \sim 1 + \alpha \cos^2\theta^*$ in its rest frame, is a crucial test for the color-octet mechanism. Since the octet production matrix elements of Non-Relativistic QCD[74] (NRQCD) lead to a polarization pattern different from the color singlet model, a polarization measurement can provide us with significant information on quarkonium production. For example, quarkonia at large $p_T$ are predicted to be almost fully transversely polarized, *i.e.*, α ~ 1, as a result of production via gluon fragmentation while at smaller $p_T$ < 5 GeV/*c*, the quarkonia are predicted to be produced essentially unpolarized. The observation of this polarization pattern would test the underlying theory. Recently the measurement of quarkonium polarization was also suggested as a possible signature for QGP formation[75]. These measurements require a large acceptance detector and good statistics.

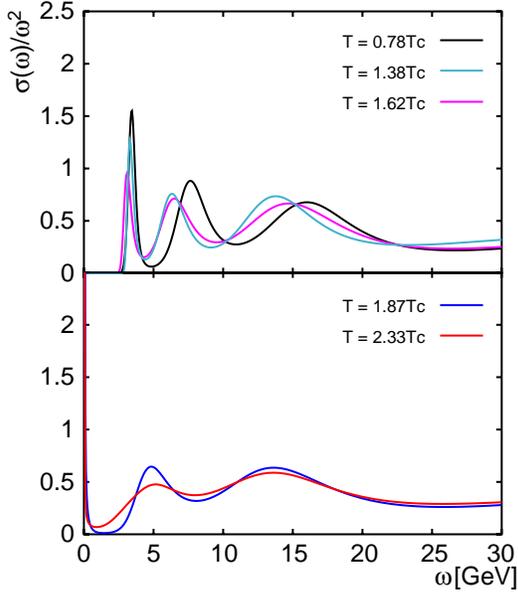
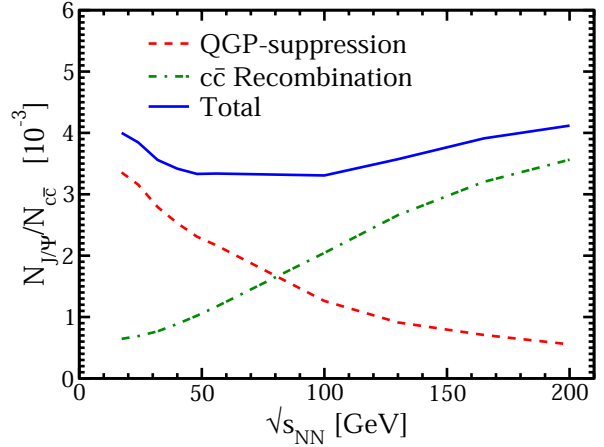

**Figure 15:** Thermal vector spectral functions (J/Ψ) from Maximum Entropy Method (MEM) analyses of meson correlation functions calculated in quenched QCD on anisotropic lattices[66]. The "melting" of the J/Ψ occurs gradually with increasing $T$. Early temperatures at RHIC are expected to be around $T = 2\,T_C$.

**Figure 16:** Excitation function of the ratio of produced J/Ψ to the number of $c\bar{c}$ pairs in central heavy-ion collisions. Calculations by R. Rapp[76].

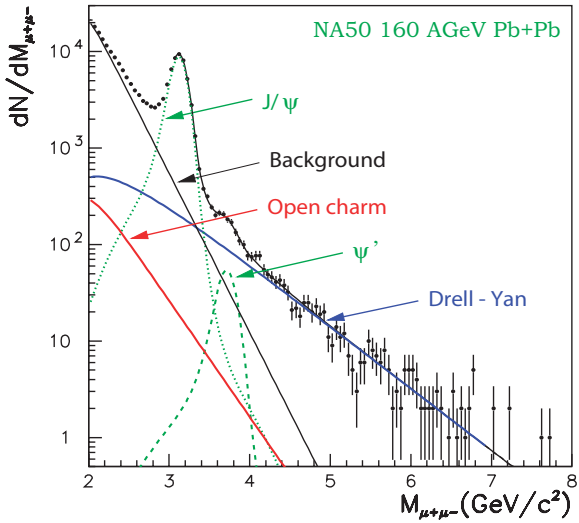
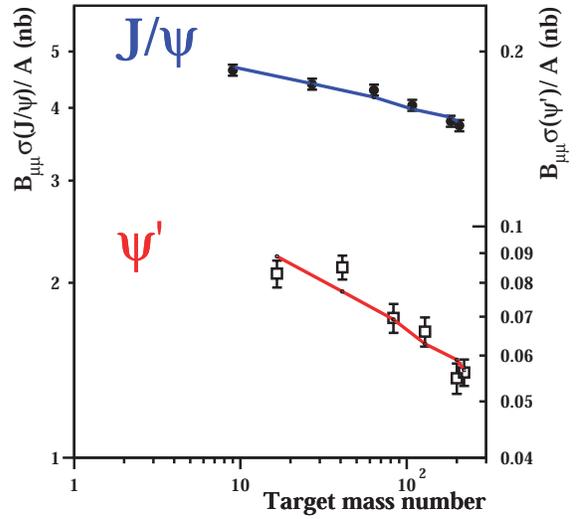

**Figure 17:** Invariant muon pair mass spectrum from 160 AGeV Pb+Pb collisions from NA50[83].

**Figure 18:** NA50 measurements of J/Ψ and Ψ' absorption in cold nuclear matter using 450 AGeV pA collisions. The respective absorption cross-sections are $\sigma_{abs}(J/\Psi)$ = 4.2±0.5 mb and $\sigma_{abs}(\Psi')$ = 9.6±1.6 mb. These measurements are performed at mid-rapidity only ($x_F \approx 0$).

At relatively large transverse momentum, typical quarkonium suppression effects, such as color screening, become negligible. Instead, the $q\bar{q}$ pair can be seen as a hard probe that interacts with the medium. In particular, any color octet $q\bar{q}$ can suffer jet quenching. The relative abundance of charmonium resonances can provide an experimental handle on studying such phenomena as each resonance may have a different octet contribution. We must exercise caution, however, as a variety of competing charmonium production models exist. In parallel with any nucleus-nucleus studies, it is therefore important to investigate and compare production mechanisms in proton-proton and proton-nucleus interactions, all at low (central region) and high $x_F$ (forward region)[77,78]. Dead cone or other effects can be important for heavy quark systems[79]. A measurement of the quarkonium nuclear modification factor at high-$p_T$ can provide a unique experimental probe for studying energy loss and color diffusion[80].

## 4.2 Experimental Status

The prospects of a "clean" QGP signature triggered an extensive experimental program at the CERN/SPS. HELIOS-III[81], NA38[82] (which became NA50[83] and then NA60[84]) conducted detailed measurements of the dimuon invariant mass spectrum at mid-rapidity. Despite early enthusiasm and enormous statistics (see Figure 17) the picture that evolved is still rather ambiguous. The measurements at SPS must also be understood in light of the many results on quarkonium production in pA collisions from fixed target experiments at Fermilab's Tevatron (*e.g.* E866). The status at the SPS can be summarized as follows:

- The $J/\Psi$ and $\Psi'$ have a substantial absorption cross-section (4.2 and 9.6 mb at $x_F \sim 0$) in normal nuclear matter as derived from pA studies[85] (see Figure 18). Studies of the A dependence (Cronin exponent) were conducted in pA collisions at Fermilab[86] over a large range of $x_F$ but at larger energies. A very strong $x_F$ dependence was observed for $x_F > 0.2$ as depicted in Figure 18.
- The $J/\Psi$ is substantially suppressed in semi-central and central Pb+Pb collisions[83] beyond that expected from absorption in cold nuclear matter as depicted in Figure 20. The suppression is centrality dependent.
- Feed-down contributions from higher $\chi_c$ states ($\chi_c \to J/\Psi + \gamma$) are considered to be large (20-40%)[87] but so far have not been measured at SPS energies. See decay scheme in Figure 21.
- Alternate models[88] are able to describe the observed $J/\Psi$ suppression by assuming absorption of the bound state by the co-moving matter as depicted in Figure 22. Charm production is still unknown at the SPS and leads to uncertainty in any statement of suppression (a dedicated experiment, NA60, has just completed its first ion run).
- There are no measurements at forward rapidities at SPS energies to complement the existing measurements.

The picture emerging from SPS studies is inconclusive, and the many missing pieces of vital information have resulted in quarkonium suppression being regarded as an interesting study, but not a conclusive one. On the other hand, the vast experience gained at the SPS can be used to improve the measurement. The main lesson learned is that a simple measurement of $J/\Psi$ in AA collisions as a function of centrality is *insufficient* to draw substantial conclusions. Rather, a systematic and detailed study of all related aspects, *i.e.*, a comprehensive quarkonium program, is necessary.

The experimental situation in pp interactions is considerably better[89], especially with the beginning of Run II at the Tevatron. There is a tremendous flow of new results and excellent measurements on quarkonia from CDF and D0. This includes $J/\Psi$, $\Psi'$, $\Upsilon(1S)$, $\Upsilon(2S)$, and $\Upsilon(3S)$ cross-sections from $p_T = 0$ up to 18 GeV/$c$ for $|y| < 1.8$ as well as the $\chi_c$ feed-down contributions (see for example Figure 23 and Figure 24). Polarization measurements of all quarkonium states, including the bottonium states, are expected soon. This, however, does not necessarily contribute to the interpretation of future measurements at RHIC since the production mechanisms at Tevatron energies are different from those at the lower RHIC energies (flavor creation vs. flavor excitation and shower/fragmentation)[90].

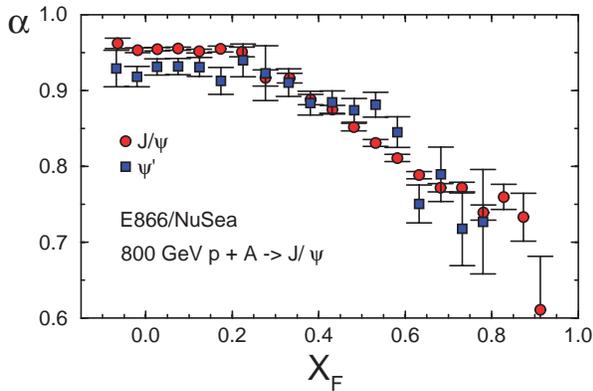

**Figure 19:** Measurement of $J/\Psi$ and $\Psi'$ absorption in 800 GeV pA collisions over a large range of $x_F$ by E866/NuSea[86].

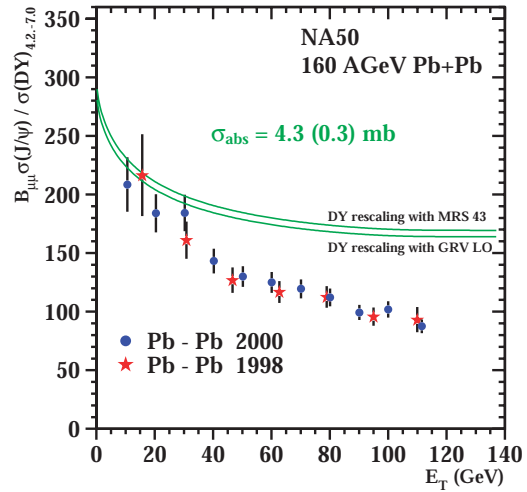

**Figure 20:** $J/\Psi$ absorption in 160 AGeV Pb+Pb collisions from NA50[85]. The green curve depicts expectations when taking only nuclear absorption into account. Suppression beyond nuclear absorption is large (factor 2).

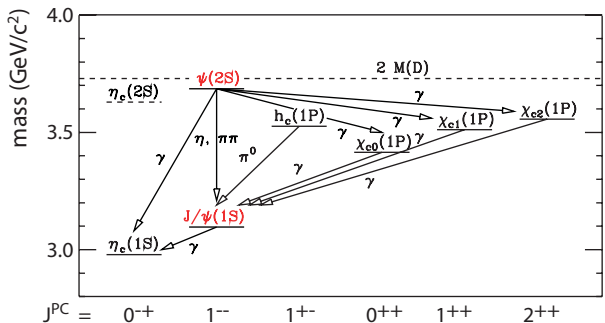

**Figure 21:** Decay scheme of charmonium states.

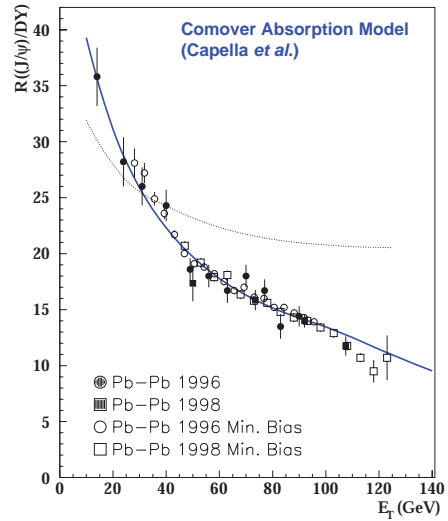

**Figure 22:** Ratio of $J/\Psi$ over Drell-Yan, i.e., the measure for $J/\Psi$ suppression, versus $E_T$ from NA50. The blue line is a prediction from a comover absorption model[88].

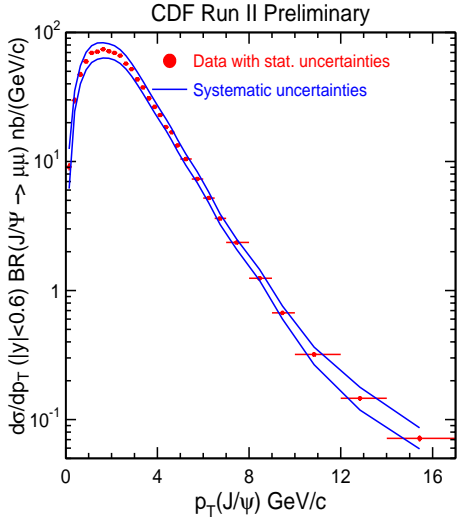
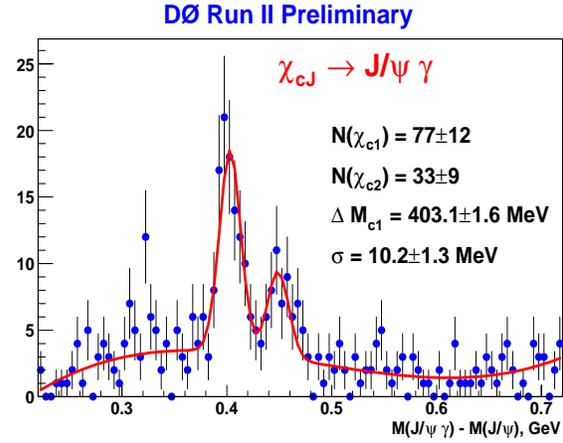

**Figure 23**: Preliminary $p_T$ spectra of $J/\Psi$ at $\sqrt{s} = 1.96$ TeV from CDF[89] for $p_T = 0$ up to 18 GeV/$c$.

**Figure 24:** Invariant $J/\Psi\,\gamma$ mass spectrum as measured by D0[89] at $\sqrt{s} = 1.96$ TeV. The peaks correspond to the two states $\chi_{c1}$ and $\chi_{c2}$. Both are a large source of feed-down into $J/\Psi$ but are extremely difficult to reconstruct.

Measurements at RHIC have so far been inconclusive and lack considerable statistics. This situation will improve once the processing and analysis of data taken during Run 4 is completed although it should be noted that the statistics accumulated at Fermilab and SPS in similar measurements are far larger than that achievable at RHIC prior to RHIC-II. The situation at RHIC does not appear to be very favorable for a detailed spectroscopic study:

- PHENIX
    - Small acceptance for $Q\bar{Q} \to e^+e^-$ in central arms
    - Moderate acceptance for $Q\bar{Q} \to \mu^+\mu^-$ in muon arms
    - Insufficient momentum resolution to resolve the $\Upsilon$ states in the muon arms
    - Acceptance too small for polarization measurements
    - Acceptance too small for measurements of $\chi_c$ states
    - Direct measurement of charm mesons difficult to impossible
- STAR
    - Moderate acceptance for $Q\bar{Q} \to e^+e^-$ in central barrel
    - Lack of trigger capabilities for $J/\Psi \to e^+e^-$
    - Moderate DAQ rate
    - Acceptance too small for measurements of $\chi_c$ states
    - Acceptance too small for polarization measurements
    - Momentum resolution insufficient to resolve $\Psi$ and $\Upsilon$ states.
- PHOBOS/BRAHMS
    - No measurement possible

While RHIC-II will provide ample luminosity to perform, in principle, all required measurements, STAR and PHENIX will require substantial upgrades to overcome some of the shortcomings listed above. Even with the proposed upgrades only a fraction of what is required to fully understand production, possible enhancement, and suppression mechanisms will be exploitable.

## 4.3 Requirements for a Quarkonium Program

Charmonium physics in heavy ion collisions is as compelling as it was in '86 when first proposed by Matsui and Satz[60]. In order to fully study and understand all aspects of quarkonium physics in heavy collisions and thus exploit its features in probing the QGP a new detector is required that must fulfill diverse requirements. The following table relates physics motivations, probes and the subsequent detector requirements:

| Topic | Probes and Studies | Requirements |
|---|---|---|
| Baseline measurement | $J/\Psi$, $\Psi'$, $\Upsilon(1S)$, $\Upsilon(2S)$, $\Upsilon(3S)$ in AA, pA, and pp as a function of: <br> • centrality <br> • $p_T$ <br> • rapidity $y$ <br> • $x_F$ <br> • $\sqrt{s}$ <br> • beam mass A | High rates and large acceptance to record sufficient statistics in available (limited) beam-time. <br> High momentum resolution to resolve the $\Upsilon$ states. |
| Nuclear effects (shadowing, absorption) | Study pp and pA collisions: <br> • Measure $x_1$, $x_2$, $x_F$ dependence, measure A dependence (Cronin) | Large acceptance (incl. forward coverage) extending to large $x_F$ and low $x_{BJ}$. |
| Distinguish suppression vs. recombination | • Charm production: $\sigma(p_T, y)$ <br> • $v_2$ of $J/\Psi$ <br> • $p_T$ dependence of suppression | High resolution vertex detectors (charm). <br> Azimuthally symmetric detectors ($\Delta\phi = 2\pi$) for correlation and elliptic flow measurements. |
| Contribution from feed-down | • Measure $\chi_c$ (challenge: soft $\gamma$ at large $\eta$) at least in pA | Photon detection capabilities at mid- and forward-rapidities (see Figure 28). <br> High rates, good energy and momentum resolution to enhance S/B ratio for the $\chi_c$. |
| Quarkonium production | • Quarkonium polarization at least in pp and pA | Large acceptance to reach large $\cos\theta^*$ (see Figure 27) |

From this simple table alone it becomes evident that a comprehensive picture can only emerge from an extensive program of RHIC-II running over several years. Certain measurements need to be repeated for various beam energies and beam species. Only RHIC-II luminosities will allow an exploitation of the full spectrum of physics possibilities. For systematic quarkonium spectroscopy, a detector with *substantially* larger coverage than presently available with excellent resolution to resolve the various $\Psi$ and $\Upsilon$ states, good electron and muon identification, reasonable photon ID for the $\chi_c$ states in the forward direction, and triggering and data recording capabilities at the highest possible rates will allow these measurements to be completed with sufficient precision and statistics.

## 4.4 Quarkonium Measurements with a New Comprehensive Detector

The "golden" quarkonium decay modes for the proposed new detector are $Q\bar{Q} \to e^+e^-$ and $Q\bar{Q} \to \mu^+\mu^-$. In what follows we assume the detector layout as discussed in section 7 but with a reduced acceptance to account for edge effects. The following parameters are relevant for a quarkonium measurement:

- High precision tracking in $|\eta| < 3$, $\Delta\phi = 2\pi$
    - resolution sufficient to resolve the $\Upsilon(1S)$, $\Upsilon(2S)$, and $\Upsilon(3S)$ states (see Figure 50)
- $\mu$ identification in $|\eta| < 3$, $\Delta\phi = 2\pi$ via muon detectors *after* the tracking devices
- Electron ID using electromagnetic calorimetry and PID in $|\eta| < 3$, $\Delta\phi = 2\pi$
- Photon detection via electromagnetic calorimetry in $|\eta| < 3$ (4), $\Delta\phi = 2\pi$

The $J/\Psi$ production cross-section in pp collisions at $\sqrt{s} = 200$ GeV can only be approximated due to the lack of constraining data above $\sqrt{s} = 60$ GeV. However, the predictions depend only weakly on the choice of the underlying parton distribution functions[91]. The cross-section for the $\Psi'$ can then be estimated in the color evaporation[91] model using the $J/\Psi$ cross section. In our estimates we have excluded the feed-down contribution from the $\chi_c$ states. The cross-section of the $\Upsilon$ states in pp collisions is rather well constrained by available data above and below RHIC energies. There is little information on the radiative decays of the $\chi_b$ available yet so the measured cross-sections are not corrected for these feed-down effects. Table 3 lists the PDG values for mass and branching ratio into $e^+e^-$ ($\mu^+\mu^-$) of the various quarkonium states and summarizes our cross-section estimates at mid-rapidity ($y = 0$) for pp at $\sqrt{s} = 200$ GeV. In order to scale from pp to *minimum* bias A+B collisions we use $\sigma_{AB} = \sigma_{pp} (AB)^\alpha$ where A and B are the mass numbers of the two colliding nuclei and $\alpha$ is an exponent that reflects effects in (cold) nuclear matter which must be determined experimentally. We use $\alpha = 0.96$ for the $J/\Psi$ and 0.94 for the $\Psi'$ as measured by E866[92] and $\alpha = 0.9$ for the $\Upsilon$ states as observed by E772[93]. Since the yield is dominated by the region $x_F = 0$ we did not take the $x_F$ dependence of $\alpha$ into account. This scaling method neglects all (unknown) effects in the hot and dense matter such as suppression and/or recombination.

As mentioned above these factors strictly apply only for minimum bias collisions. For centrality selected events the detailed geometry of the collision must be taken into account. We, however, do not intend to use any centrality trigger but instead use dedicated quarkonia triggers in minimum bias collisions in order to collect data from a wide range of centralities.

|        | State | Mass (GeV/$c^2$) | BR($e^+e^-$) | $d\sigma/dy|_{y=0}$ | BR·$d\sigma/dy|_{y=0}$ |
|--------|-------|------------------|--------------|---------------------|------------------------|
| J/$\Psi$ | 1S    | 3.097            | 6.0%         | 0.63 µb             | 38 nb                  |
| $\Psi'$  | 2S    | 3.686            | 0.9%         | 0.15 µb             | 1.4 nb                 |
| $\Upsilon$ | 1S    | 9.460            | 2.5%         | 2.5 nb              | 63 pb                  |
| $\Upsilon'$ | 2S    | 10.023           | 1.2%         | 0.9 nb              | 11 pb                  |
| $\Upsilon''$ | 3S    | 10.355           | 1.8%         | 0.6 nb              | 12 pb                  |

**Table 3:** Quarkonia mass, measured branching ratio into $e^+e^-$ (close or identical to $\mu^+\mu^-$), and estimated cross-section in pp at $\sqrt{s} = 200$ GeV.

We used PYTHIA 6.22 to generate $J/\Psi$ and $\Upsilon$ in 200 GeV pp in order to study the acceptance of the new comprehensive detector. Since the geometric acceptance and decay kinematics for decays into $e^+e^-$ and $\mu^+\mu^-$ are identical, in what follows we depict only the results for the $e^+e^-$ channel.

Figure 25 shows the geometric acceptance for $J/\Psi$ (left) and $\Upsilon$ (right) as a function of pseudo-rapidity $\eta$. Full azimuthal coverage ($\Delta\phi = 2\pi$) is assumed. The various curves correspond to different momentum cuts on the decay leptons. The upper horizontal double-arrow depicts the acceptance of the new detector. The upper limit of the acceptance of the PHENIX central arm ($\Delta\phi = \pi/2$), $|\eta| < 0.35$) for lepton momenta $p > 200$ MeV/$c$ is indicated by the vertical line at $\eta=0.35$. Note that the central arm only detects electrons and covers half the full azimuth. The acceptance corresponding to the two PHENIX muon arms ($1.2 < \eta < 2.4$. and $-2.2 < \eta < -1.2$, $p > 2$ GeV/$c$) is indicated by the arrow in both plots. The overall geometric acceptance for $J/\Psi$ is ~50% for decay leptons with $p > 1.5$ GeV/$c$ and ~80% for the $\Upsilon$ even for lower momenta cuts of 3 GeV/$c$. As seen in the plots the coverage $|\eta| < 3$ is ideally suited to capture almost all of the quarkonia decays.

A very important variable for the study of quarkonia production and absorption in nuclear matter is $x_F$. Figure 26 shows the geometric acceptance for $J/\Psi$ (left) and $\Upsilon$ (right) as a function of $x_F$. The various curves correspond to different momentum cuts on the decay leptons. Solid curves are for $|\eta| < 3$, dashed curves are for $|\eta| < 4$ assuming that the barrel acceptance is augmented by forward detectors. As one can clearly see the reach in $x_F$ can be considerably expanded by electron identification and tracking capabilities in the forward region $|\eta| > 3$. Measurements at large $x_F$ (low $x$) open a new window to study quarkonium production and suppression, a region with maximum nuclear absorption and significant shadowing. The green curves depict the acceptance of the two PHENIX muon arms.

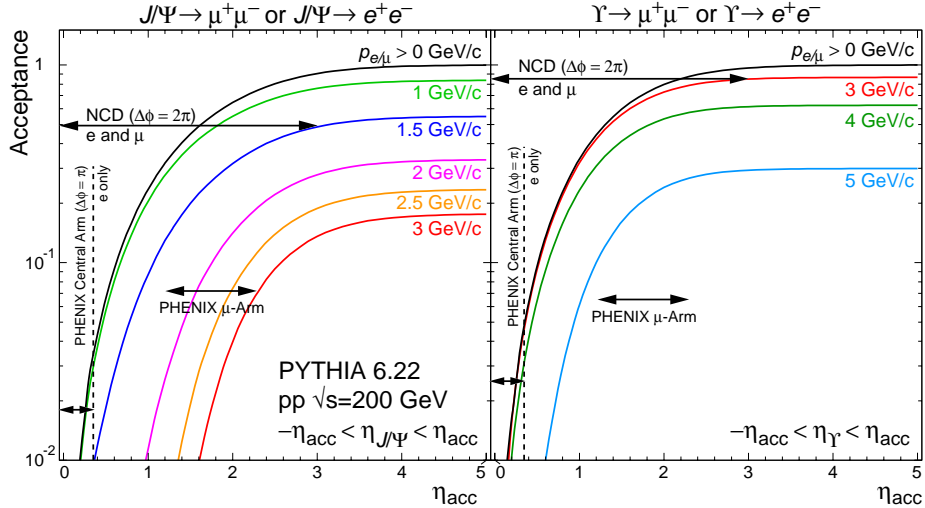

**Figure 25:** Geometric acceptance for $J/\Psi$ (left) and $\Upsilon$ (right) as a function of pseudo-rapidity $\eta$. Full azimuthal coverage ($\Delta\phi = 2\pi$) is assumed. The various curves correspond to different momentum cuts on the decay leptons. The acceptance of the new comprehensive detector ($|\eta| < 3$, $\Delta\phi = 2\pi$) is indicated by the upper double-arrow. The upper limit of the acceptance of the PHENIX central arm ($|\eta| < 0.35$) for lepton momenta p > 200 MeV/$c$ is indicated by the vertical line at $\eta=0.35$. Note that the central arm only detects electrons and covers half the full azimuth. The acceptance corresponding to the 2 PHENIX muon arms ($1.2 < \eta < 2.4$. and $-2.2 < \eta < -1.2$, $p > 2$ GeV/$c$) is indicated by the horizontal double-arrow in both plots. PYTHIA 6.22 was used to generate the $J/\Psi$ and $\Upsilon$ distributions.

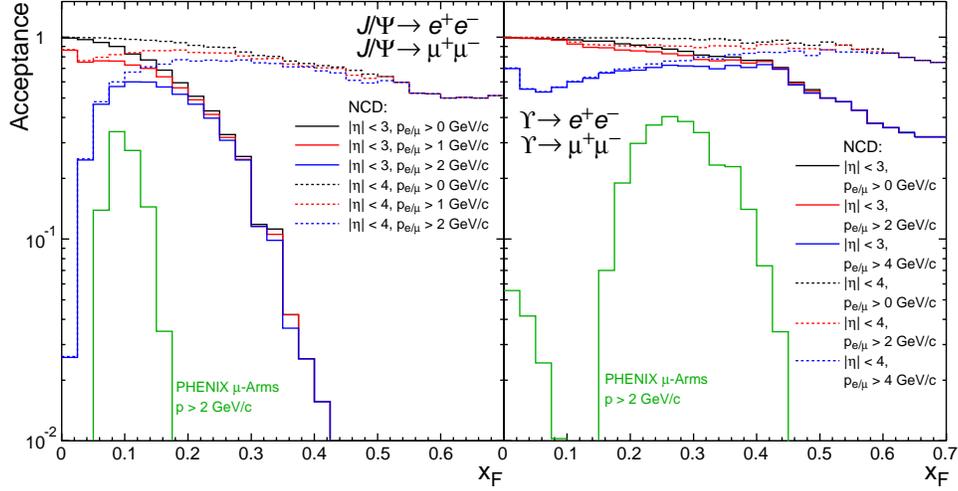

**Figure 26:** Geometric acceptance for $J/\Psi$ (left) and $\Upsilon$ (right) as a function of $x_F$. Full azimuthal coverage ($\Delta\phi = 2\pi$) is assumed. The various curves correspond to different momentum cuts on the decay leptons. Solid curves are for $|\eta| < 3$, dashed curves are for $|\eta| < 4$ assuming that the barrel acceptance is augmented by forward detectors (see text). The green curves depict the acceptance of the two PHENIX muon arms. PYTHIA 6.22 was used to generate the $J/\Psi$ and $\Upsilon$ distributions.

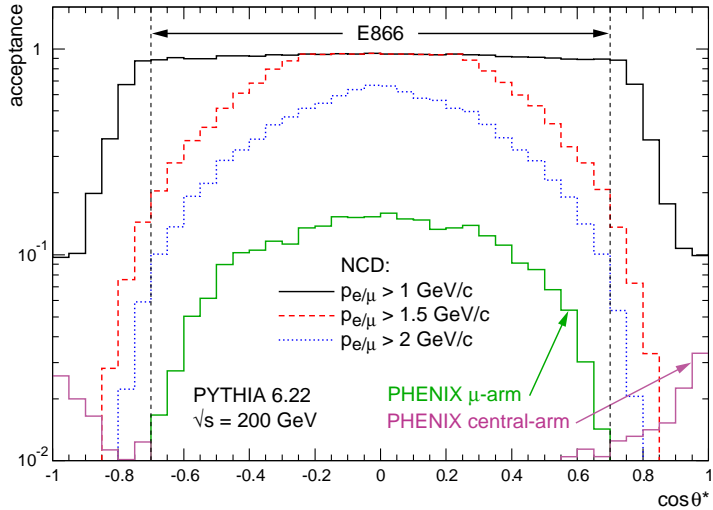

**Figure 27:** Geometric acceptance for $J/\Psi$ as a function of $\cos\theta^*$. The various curves correspond to different momentum cuts on the decay leptons. The vertical dashed lines depict the $\cos\theta^*$ coverage of E866 which is one of the few experiments that have conducted a successful quarkonium polarization measurement[94].

As discussed earlier a polarization measurement of the $J/\Psi$ could yield valuable information on the quarkonium production mechanism. The key observable is the differential cross-section $d\sigma/d\cos\theta^* \sim 1 + \alpha \cos^2\theta^*$. It is obvious that a precise determination of $\alpha$ is best conducted at large $\cos\theta^*$. Figure 27 shows the geometric acceptance for $J/\Psi$ as a function of $\cos\theta^*$. The various curves correspond to different momentum cuts on the decay leptons. The vertical dashed lines depict the $\cos\theta^*$ coverage of E866 which is one of the few experiments that conducted a successful quarkonium polarization measurement[94]. As can be seen the coverage of the proposed detector design is well suited to conduct a world-class polarization measurement.

It is crucial for the interpretation of the quarkonia yields to account for the feed-down contributions from the $\chi_c$ states. They are bound loosely and dissolve at considerably lower temperatures than, for example, the $J/\Psi$. This results in a distortion of the suppression measurement. The measurement of $\chi_c$ is therefore mandatory. Unfortunately this is one of the hardest measurements to conduct as demonstrated in Figure 28. The $\chi_c$ decays into $J/\Psi$ and a photon. The latter has a rather low momentum but can only be cleanly detected at energies above 2 GeV; an even higher cut is likely required in central Au+Au collisions. However, by applying this cut the decay kinematics pushes the photons out to forward rapidities as depicted in the right-hand plot of Figure 28. Still, with the new detector design a considerable fraction can be detected within the barrel $|\eta| < 3$ while electromagnetic calorimetry in the forward region $3 < \eta < 4$ will expand the acceptance sufficiently to capture the full decay phase space. This is a measurement that to-date cannot be conducted at RHIC; in fact it has so far not been performed in any heavy-ion experiment. Since few measurements on $\chi_c$ production are available we used PYTHIA 6.22 to estimate the inclusive cross-section in pp collisions. The $\chi_c$ A-dependence is unknown since no measurement has been conducted so far. It is expected[95] that the $\alpha$ value used to scale from pp to AA collisions is close to that of the $J/\Psi$ and $\Psi'$. Current model calculations[95] predict that the higher absorption cross-section of the $\chi_c$ is compensated by its larger formation time. In our estimates we assume $\alpha = 0.93$ independent of $x_F$.

Due to the lack of detailed detector simulation and reconstruction code we estimated the reconstruction and trigger efficiencies for quarkonia measurements based on experiences and tests conducted in STAR and known SLD performances that were scaled to RHIC Au+Au conditions (central HIJING events). In the proposed new detector the hadron rejection is additionally enhanced by an overall particle ID performance out to high momenta. Muon detection is limited to $p_\mu > 2$ GeV/$c$ and we use an overall trigger and reconstruction efficiency for muon pairs of 0.6 independent of $p_T$. The electron pair efficiency is assumed to vary between 0.4–0.6 dependent on the electron momenta. For the $\chi_c$ states we fold in an additional $\gamma$ reconstruction efficiency of 0.5-0.9 dependent on the energy. As always, efficiency and background rejection are strongly correlated. Stricter track

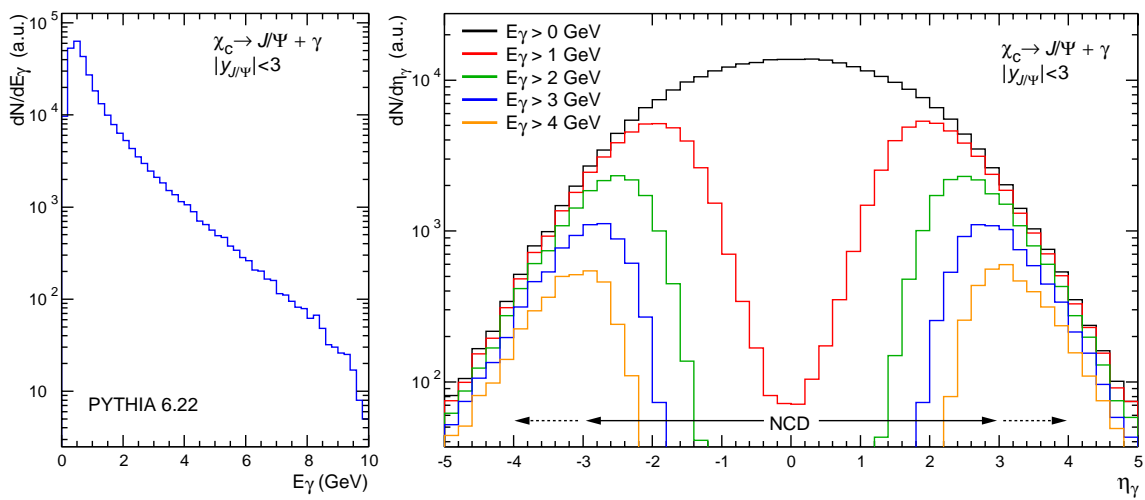

**Figure 28:** Left plot: energy spectrum of the decay photon in $\chi_c \to J/\Psi\,\gamma$ for the case where the $J/\Psi$ is detected in $|\eta| < 3$. Right plot: Rapidity distribution of the decay photons. The various lines depict the spectra for different $E_\gamma$ cuts. Simulations were done using PYTHIA 6.22.

quality and PID cuts decrease the yield while improving the signal-to-background ratio.

The yields for quarkonia in Table 4 and Table 5 are based on an integrated luminosity of 30 nb$^{-1}$. This corresponds to a 14 weeks RHIC-II runtime with a combined detector and machine efficiency of 50%. The assumed average luminosity per fill is $7 \cdot 10^{26}$ cm$^{-2}$ s$^{-1}$ (7 mb$^{-1}$ s$^{-1}$).

|  | $e^+e^-$ channel | | | $\mu^+\mu^-$ channel | |
|---|---|---|---|---|---|
|  | $p_e$ > 1.5 GeV/$c$ | 2 GeV/$c$ | 3 GeV/$c$ | $p_\mu$ > 2 GeV/$c$ | 3 GeV/$c$ |
| J/Ψ | $21 \cdot 10^6$ | $18 \cdot 10^6$ |  | $18 \cdot 10^6$ |  |
| Ψ' | $0.6 \cdot 10^6$ | $0.5 \cdot 10^6$ |  | $0.5 \cdot 10^6$ |  |
| ϒ |  |  | $32 \cdot 10^3$ |  | $32 \cdot 10^3$ |
| ϒ' |  |  | $6 \cdot 10^3$ |  | $6 \cdot 10^3$ |
| ϒ'' |  |  | $6 \cdot 10^3$ |  | $6 \cdot 10^3$ |

**Table 4:** Estimated quarkonia yields for a 14 weeks Au+Au run at RHIC-II assuming an integrated luminosity of 30 nb$^{-1}$. Numbers for various cuts on the lepton momenta are listed for each decay channel. For details see text.

|  | $E_\gamma$ > 2 GeV | | $E_\gamma$ > 4 GeV | |
|---|---|---|---|---|
|  | $|\eta_\gamma|$ < 3 | $|\eta_\gamma|$ < 4 | $|\eta_\gamma|$ < 3 | $|\eta_\gamma|$ < 4 |
| χ$_c$ | $68 \cdot 10^4$ | $90 \cdot 10^4$ | $11 \cdot 10^4$ | $28 \cdot 10^4$ |

**Table 5:** Estimated yields for the sum of all χ$_c$ states (χ$_{c0}$, χ$_{c1}$, χ$_{c2}$) for different lower cuts on the γ energy and acceptance. The numbers refer to a detection of the J/Ψ in both the e$^+$e$^-$ and μ$^+$μ$^-$ channels. The cut on the lepton momentum is $p_{lepton}$ > 2 GeV/$c$ for muons and electrons. For details see text.

All quarkonia yields presented in the table are sufficient to perform a detailed study of the $p_T$, $y$, $x_F$, and cos θ* dependences of quarkonia production with abundant statistics. Even a reduction of a factor of 10 (~10 days running) will yield enough statistics to study $p_T$ and $y$ charmonium spectra, thus improving the efficiency of energy and species scans dramatically. Polarization (cos θ*), absorption ($x_F$), and χ$_c$ measurements are more demanding and, according to these studies, only feasible with a new large acceptance high rate detector. The combined measurement in the dielectron and dimuon channel and the large acceptance of the proposed detector allows for a "world-class" study of quarkonium production and medium-induced effects that are not possible with the proposed upgrades of the current RHIC detectors.

# 5 Forward Physics in a New RHIC II Detector

So far, much of the physics studied at RHIC has focused on the production of particles at mid-rapidity. These are particles with far less longitudinal energy than transverse, presumably produced via parton-parton scatterings, which traverse the produced medium. These are the particles which act as "probes" of the system, since presumably little of their energy is bound up with the overall collision dynamics. Thus, they are the most sensitive to QCD- and QGP-related dynamical effects relevant to the dynamics after the liberation of partons from the incoming nucleon wave functions.

When one moves away from midrapidity (for example, full dN/dη distributions are shown in Figure 29) new dynamical regimes are opened up as partons take progressively less of the initial momentum, *i.e.*, low-$x$ physics becomes more relevant. These ultra-soft partons have interactions which are coherent over a larger space-time range, making them sensitive to collective effects in the initial-state wave function. These effects are usually discussed in terms of several important physics effects:

**Color glass condensate**: The initial state parton distributions may be dominated by semi-classical parton saturation effects, which lead to long-range rapidity correlations, even among the produced particles.

**Hydrodynamics**: Hydrodynamic evolution translates pressure gradients into expansion and asymmetries in the final state hadrons. It is an open question whether the hydro limit is reached away from the regions of highest particle density.

**Baryon density**: The stopping of the initial-state baryons, for which the precise mechanism is unknown but data at various energies is now available, leads to a non-uniform baryon density stretching across the full rapidity range. This is also the domain of "leading particles" for which experimental data do not tell a unique story across various systems.

These various physics effects are probably in competition for several interesting classes of observ-ables which are only just starting to be explored at RHIC.

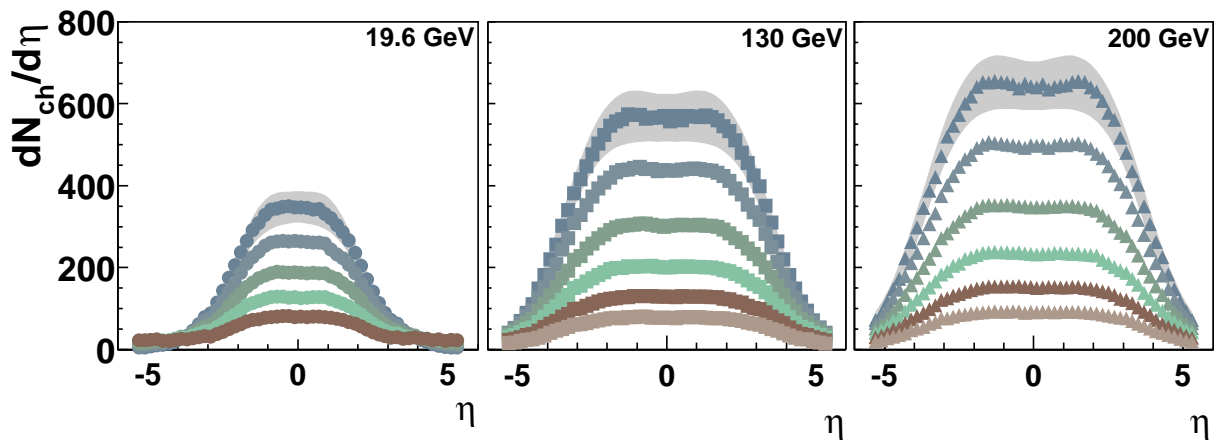

**Figure 29:** Pseudorapidity distributions of inclusive charged particles in Au+Au collisions for three energies (from PHOBOS).

**Particle yields at forward rapidities and large $p_T$:** Interesting single-particle observables in the forward region include "limiting fragmentation" in the yields of inclusive charged particles and the suppression of high-$p_T$ particles in the forward direction. The precise nature of this effect requires detailed studies of identified particles out to kinematic regions (high $p_T$ and high rapidity) so-far unavailable in the far-forward region at RHIC.

**Dynamical correlations:** Most of the forward physics covered so far by detectors at RHIC have been single-particle observables. However, it has become clear that multi-particle correlations are crucial for extracting the dynamical properties of the system. Jet-like correlations and their disappearance have been fundamental for the current understanding of energy loss in the produced medium. The study of Mueller-Navelet dijets[106] is a useful tool to study the transition from perturbative dynamics to the onset of Color Glass dynamics. Finally, hydrodynamics by itself is a comprehensive approach to mapping space-time correlations into momentum-energy correlations, which also have a characteristic phase-space structure. However, the rapidity dependence of $v_2$ has proven to be a challenge for hydrodynamic calculations so far. Thus it is clear that these physics topics require a large-acceptance multiparticle spectrometer.

## 5.1 Color Glass Condensate

One of the more intriguing theoretical concepts to arise in the wake of the RHIC results has been the "Color Glass Condensate". Similar to a "state of matter" consisting of the interactions of real particles, it is a state of matter characterized by a many-particle wave function of interacting virtual quarks and gluons. The density of partons is so high that a description in terms of classical fields appears to make more sense and such a description is now being implemented in phenomenological models and applied to RHIC data. The dominant feature is the presence of a momentum scale generated by the transverse parton density, and quantum effects in particular momentum regimes that lead to breakdowns in the factorization theorems of pQCD. Many inclusive observables have been modeled, with varying degrees of success, using the CGC approach: ranging from particle multiplicities[96] to spectra at high $p_T$[97]. There have even been attempts to elucidate shadowing behavior in p+A collisions using the CGC methodology.

CGC phenomena are expected to dominate the low-$x$ region, which in the case of hadron-hadron collisions is at large rapidity, where a parton from one projectile with momentum fraction $x_1$ scatters off a parton in the other projectile with momentum fraction $x_2$. The kinematics of 2→1 scattering processes (*e.g.* used in studies of Drell-Yan or $J/\Psi$ production[98]) suggests that $x_1 \sim m_T e^y$ while $x_2 \sim m_T e^{-y}$, where $m_T$ and $y$ refer to the transverse mass and rapidity of the produced gluon. In other words, these are scattering processes with large asymmetries, *e.g.* a large-$x$ valence quark scattering off an ultra-low-$x$ gluon. The CGC methodology offers a systematic way of addressing the coherent initial-state wave function of the proton and nucleus at low-$x$, where the density is so high that it generates a dimensionful scale $Q_s^2$, which reflects the transverse parton density of the wave function ($xG_A(x)/A$).

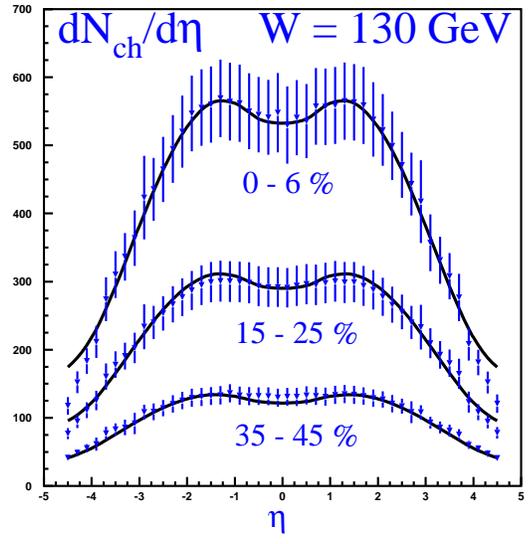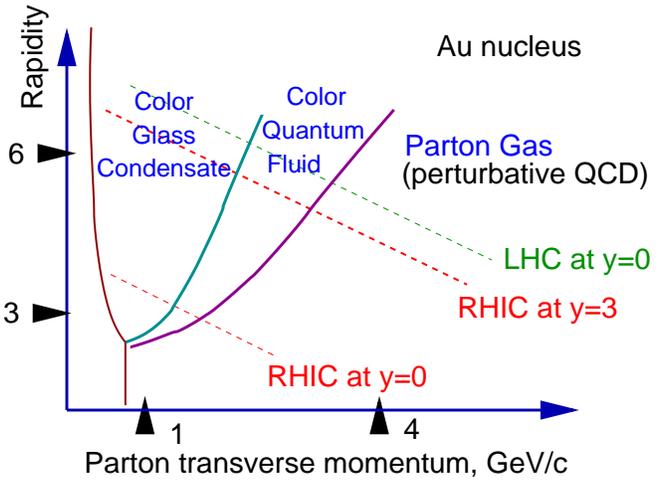

**Figure 30:** Pseudorapidity density of inclusive charged particles compared with a CGC calculation.

**Figure 31:** "Phase Diagram" of nuclear matter in $y\text{-}q_T$ space showing the range where color glass condensate is thought to exist.

Nuclei enhance this scale typically by factors of $A^{1/3}$ (*i.e.* $Q^2_s(A) \sim A^{1/3} Q^2_s$), almost like a multiple scattering scenario. The appeal of the CGC approach is that because this typical scale may be in the perturbative regime (*i.e.* $Q^2_s(A) > \Lambda_{QCD}$), it can be handled semi-perturbatively. Thus the CGC may well provide an explanation of "nuclear shadowing," where the nucleons in a nucleus appear to have fewer low-$x$ partons than in free-space. This is an appealing alternative to traditional discussions of shadowing, which to date are typically just parameterizations of DIS data on nuclei[99].

The canonical demonstration of the CGC approach applied to "soft" physics, *e.g.* the inclusive particle multiplicities measured over $4\pi$, is shown in Figure 30. This suggests that the final state particle production is dominated by the initial-state gluon distribution, with only minimal modification from hadronization – an effect known as "local parton-hadron duality"[100] (LPHD) and well used in jet physics from $e^+e^-$ to Tevatron data. However, these studies are dominated by a limited region in $Q^2$, requiring more differential studies out to higher $p_T$.

Higher-$p_T$ particle production at forward rapidities in d+Au and Au+Au may fall into a region called the "Color Quantum Fluid"[97] shown in the "phase diagram" in Figure 31. In this region, non-trivial correlations may persist into a region normally dominated by pQCD evolution, due to the generation of a larger scale $Q_s^2/\Lambda_{QCD}$. While such correlations were not seen at mid-rapidity in d+Au collisions, recent BRAHMS data, shown in Figure 32, points to this effect being active at forward rapidities[12]. They have measured the ratio of hadron spectra in the forward direction, normalized by proton-proton data scaled up by the number of binary collisions. These ratios already show a strong suppression of particle production in the forward region, which becomes stronger with increasing rapidity and centrality[101]. This is a natural expectation of the CGC approach; these are so far the most compelling data that point to its relevance in nuclear collisions.

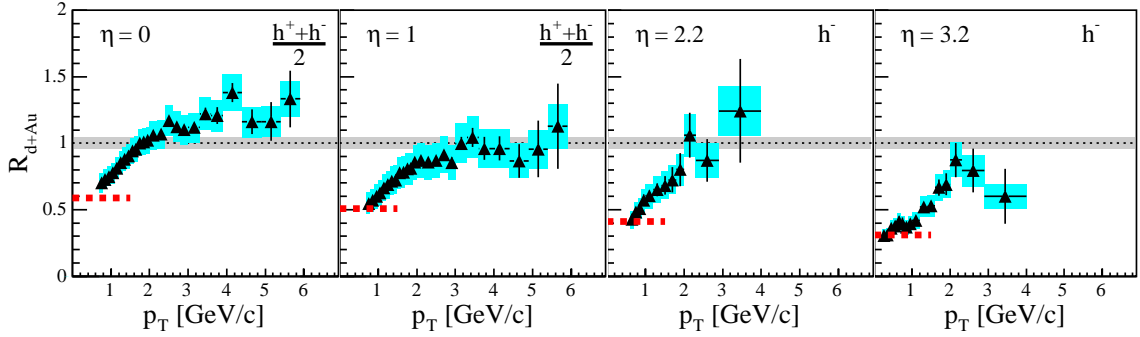

**Figure 32:** BRAHMS data[12] on the nuclear suppression factor $R_{dAu}$ as a function of pseudorapidity.

Since the CGC wave function is ultimately a coherent superposition of gluon states, it implies non-trivial quantum-mechanical correlations between partons even at different rapidities. Provided LPHD holds, they should also be visible in the final state. All of this points to the need for multi-particle measures to assess the CGC description of the proton and nuclear wave functions. To date, only a few multiparticle measurements have been suggested, but they are starting to increase with the availability of the newer single-particle data and the interest they have generated.

Since the CGC is formed on very short time scales ($\tau \sim \eta/Q_s \sim 0.2$ fm/$c$ at RHIC), one might expect substantial long-range correlations in particle production, as was proposed by Kovchegov and McLerran[102]. One might also expect a weakening of the HBT correlation if the final state pions trivially reflect the initial wave function, although it is not known whether the conversion from gluon to pion scrambles the phases, such that the correlation ultimately becomes maximal. Another recent suggestion has been to use the photons coherently radiated off the saturated quark states to study the disappearance of HBT correlations in the forward direction[103]. Two-photon HBT correlations have been measured by the WA98 experiment, and have found a system of abundant pions and yields of thermal direct photons in excess of what one would expect using pQCD calculations[104]. Such studies may ultimately require a large acceptance forward detector, using both conversion dielectron pairs as well as direct tagging of photons in an electromagnetic calorimeter, and perhaps in combination[105].

There is also some recent progress in experimental data pertaining to correlations in the forward direction, due to a preliminary study of forward-midrapidity (Mueller-Navelet[106]) correlations by STAR. Azimuthal correlations between tagged $\pi^0$'s from far-forward rapidities ($\eta \sim 3.8$) and charged hadrons near midrapidity ($\eta \sim 0$) have been found both in p+p and d+A collisions, and are shown in Figure 33. The size and strength of the p+p correlations match well with PYTHIA calculations tuned on earlier p+p data and using standard PDFs. The d+A data however is somewhat different, finding a substantial weakening of the two-particle correlation at moderate $p_T$. This is a potential signature of classical coherence induced by the CGC disturbing almost-trivial kinematic correlations that would have been naively predicted by PYTHIA. This clearly deserves additional study over larger acceptance with a wider variety of particle species.

At this point, it is only fair to point out that while parton saturation calculations have achieved consistency with experimental data, they have not made any unique predictions which have been borne out by measurements. The particle multiplicities in A+A might have been estimated by compari-sons with more elementary systems (pp, $e^+e^-$). Furthermore, the predicted high-$p_T$ suppression in d+A collisions due to the Color Quantum Fluid phase was not observed[107]. Moreover, traditional pQCD calculations have been able to reproduce the low-$x$ structure functions at HERA and PYTHIA has made successful predictions of the correlations between forward and mid-rapidity particle described below. Still, it is undeniable that the low multiplicities measured in the final state must reflect some coherence present in the initial state as the multiple collisions of baryons are taking place (something which should also affect the baryon dynamics[108]). Moreover, one of the shortcomings of the RHIC experimental program to date has been precisely the lack of a broad acceptance multiparticle spectrometer in the forward region that would be sensitive to the global features and dynamical correlations induced by the presence of a CGC in the low-$x$ sector of the initial state parton distributions. To address this issue once and for all truly requires a capable new detector with broad forward coverage.

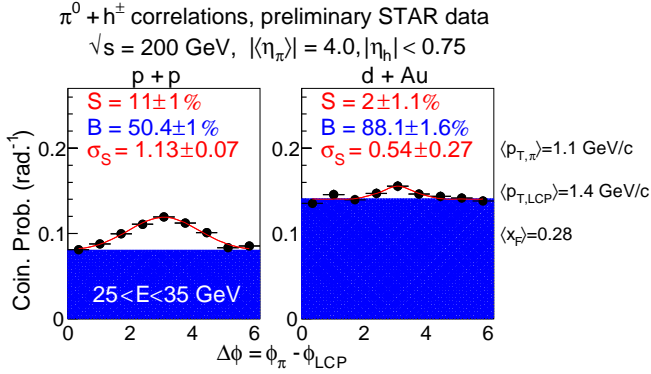
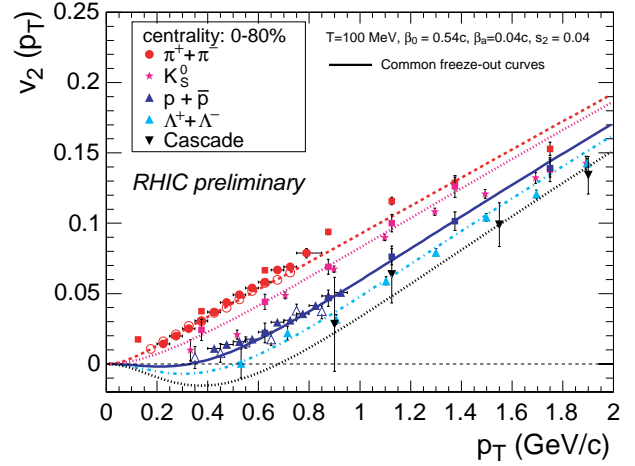

**Figure 33:** Correlation of charged particles at midrapidity with $\pi^0$s emitted in the forward direction (STAR preliminary).

**Figure 34:** Elliptic flow ($v_2$) measured vs. $p_T$ for a selection of identified particle species. The characteristic "splitting" for different particles is reproduced by hydrodynamic-inspired "blast wave" fits.

## 5.2 Hydrodynamic Evolution

One of the dramatic changes in the understanding of particle production going from the moderate SPS energies to the high-energy of the RHIC collider is the success of relativistic hydrodynamics in capturing several important quantitative and qualitative features of soft particle production[109]. Hydrodynamic behavior is seen inclusively by the evident hardening of identified particle spectra at low-$p_T$ due to "radial flow" as well as dramatic opposite-side correlations of the bulk of produced particles, known as "elliptic flow".

Elliptic flow has been measured extensively by the RHIC experiments in limited regions of phase space, mainly at midrapidity[110]. Already here, several qualitative features, like the splitting of the trends in $v_2(p_T)$ for particles of different mass[111] (shown in Figure 34), the saturation of $v_2$ for large $p_T$, and the dramatic scaling of baryon and meson $v_2$ when dividing $v_2$ and $p_T$ by the number of constituent quarks[112], have been measured and confirmed by several RHIC experiments. The large acceptance central detector, with PID out to 20 GeV/$c$, will be an important extension to this existing experimental program.

However, while hydrodynamics makes impressive predictions at moderate $p_T$ for $v_2$ as a function of centrality (collision geometry) and $p_T$ (up to 2 GeV/$c$) several important deviations make it imperative to have a large acceptance spectrometer for exploration of the limits of the hydrodynamic description. It is generally thought that RHIC collisions are sufficiently dense that they saturate the so-called "hydro-limit", which should hold true in the limit of zero mean-free path such that the only scale that enters the dynamics is the geometrical asymmetry itself[113]. However, measurements from STAR and PHOBOS of $v_2$ away from mid-rapidity show that there is a dramatic decrease of the $p_T$-integrated $v_2$ as a function of pseudorapidity[114]. To date, it is not known whether or not this change is driven by any change in hadrochemistry or simply by the monotonically changing particle density. It is also not known whether this change is accompanied by a major change in the trend of $v_2(p_T)$, which is approximately energy independent at mid-rapidity for RHIC energies (preliminary 62.4 GeV through 200 GeV)[115]. However, the limiting fragmentation observed in the forward direction[128] may already be providing some hint, which only further detailed studies at least out to moderate $p_T$ (4 GeV) can elucidate completely.

Another important deviation of data from hydrodynamics is in the realm of HBT correlations. A large range of theoretical calculations predicted that $R_{out}/R_{side}$ at midrapidity would rise to 1.5, mainly due to the assumption that there will be a substantial "QGP stall" during the mixed phase[116]. The current data suggests that this ratio is approximately unity, with no significant rapidity dependence observed to date[117]. This is a surprising feature of the particle source measurement which deserves to be extended over full phase space, to understand how the source geometry is expressed by the faster particles.

One major theoretical concern is the lack of fully-fledged 3D hydrodynamics, which make no assumptions about the boost-invariance of the dynamics. This should be addressed in the next several years[118]. There is also the crucial issue of the proper initial conditions needed to initialize hydrodynamic calculations. To date, hydrodynamical models take the dynamical evolution before a time of 0.6 fm/$c$ as a given, tuning the initial state entropy and energy density to match the final state spectra[109]. It is possible that Color Glass Condensate calculations, which predict the gluon density over the full phase space using existing structure functions and the principle of parton saturation, may be useful in describing the existing RHIC data[119]. It is also possible that a suitably initialized Landau initial condition (full stopping of incoming energy, possibly partial stopping of the initial baryon number[120]) may also be able to predict the entropy and rapidity distribution from even earlier times (~0.1 fm/$c$)[121] as shown in Figure 35. These questions can only be answered definitively by precise measurements with particle identification at forward rapidities.

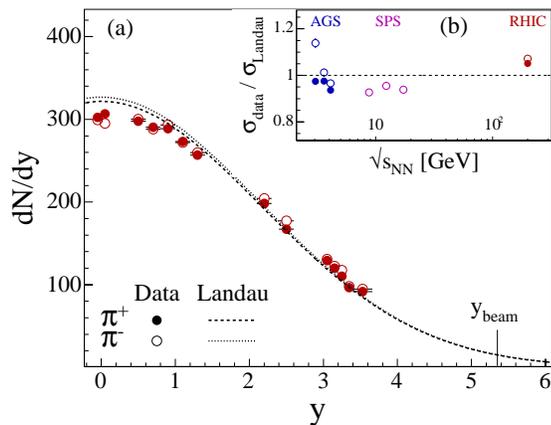

**Figure 35:** Rapidity distribution of produced pions in 200 GeV Au+Au collisions from BRAHMS. A fit to a single Gaussian is shown and the width compared to the Landau predictions.

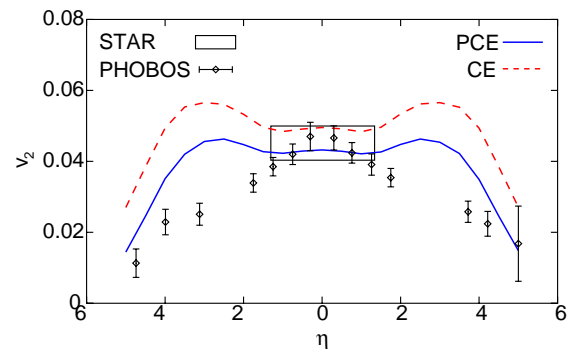

**Figure 36:** Pseudorapidity dependence of $v_2$ compared with hydrodynamic calculations by T. Hirano.

## 5.3 Dynamical Correlations

In the grand-canonical ensemble, a fully equilibrated system of fermions or bosons should experience fluctuations that are predictable using standard statistical methods. However, while such patterns would be an excellent signal to characterize the degree of thermalization achieved by the system, additional dynamical fluctuations may also be seen, due to non-trivial effects like Bose-condensation, hydrodynamic flow, and hard-scatterings of partons in the initial state. However, due to the high-multiplicities in a typical heavy ion collision, it is generally found that a systematic study of multiparticle (2-particle, 3-particle, etc.) correlations provides the cleanest means to discover and quantify dynamical correlations in the full system, rather than, *e.g.*, attempting to isolate entire jets.

Multiple sources of correlation have been discussed in the context of proton-proton collisions and A+A collisions. Studies have shown that hydrodynamic evolution generates a strong correlation between low-momentum particles in the transverse direction which experience a common flow velocity field. This manifests itself as a significant $v_2$ signal even when considering two-particle correlations but is especially clear for higher-order (4-particle) effects. In the context of forward physics, one might also expect some correlation between separated rapidities due to longitudinal hydrodynamic expansion. This may be particularly strong in the case of a Landau initial condition, where the initial state was a region compressed to a fraction of a hadron radius, or due to asymmetric parton-parton collisions, as discussed above in the context of forward-midrapidity azimuthal correlations.

The QCD fragmentation process, by which a fast parton radiates soft gluons, also should generate correlations stretching from the *x* of the parent parton to the lowest *x* reachable by the system. This can be considered in the context of DGLAP evolution or in the CGC by coherent gluon emission. Clearly correlations are seen as jets, which can be sampled by two-particle correlation measures, or fully measured by cone algorithms. STAR and PHENIX have both established strong correlations of high-$p_T$ particles close in azimuth, and used these correlations to establish the presence of jet-like structures in p+p, d+A and Au+Au collisions[122]. These studies have been challenging due to the additional pres-ence of a strong elliptic flow signal, which grows with increasing $p_T$ up to 2 GeV/$c$, where it saturates. This indicates that moderate $p_T$ particles are sensitive to multiple effects. Going out to larger η, where the elliptic flow signal gets substantially weaker, may be a crucial means for disentangling the factors which determine the strength of these correlations. Already, preliminary STAR data taken in a relatively small range in η (η < 1) indicates that central events lead not just to a suppression of away-side correlations, but to a noticeable broadening of the jet-like correlations in pseudorapidity on the near-side[123], as shown in Figure 37. This is for particles below 2 GeV/$c$ – studies out to higher $p_T$ already promise greater insight. This is a relatively new field in heavy ion physics, which would substantially benefit from a forward detector, in order to study these correlations throughout the particle source, as well as to completely avoid spurious effects due to detector biases.

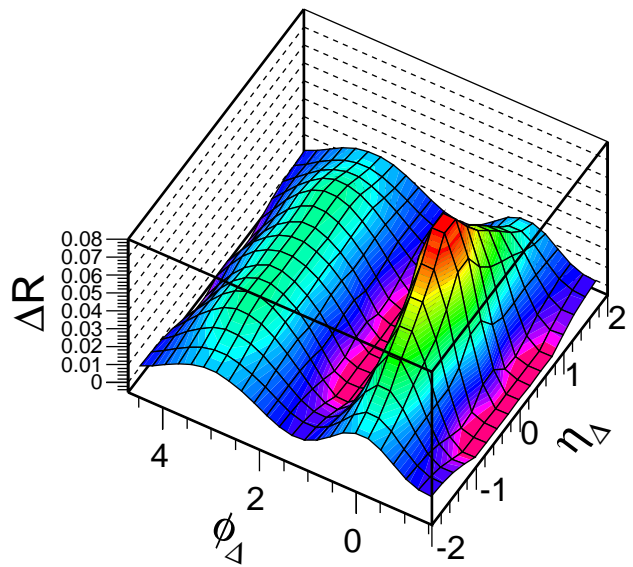

**Figure 37:** Two particle number correlations measured by STAR in 200 GeV Au+Au collisions. A disappearance of the away side correlation is observed along with a broadening in pseudorapidty of the near-side correlations.

It should be noted that even the thermal approach provides for additional fluctuations given the presence of a phase transition. This has been studied by NA49 by looking at dynamical momentum fluctuations as a function of beam energy, which can be mapped into varying the initial state baryon density[124]. They find a sharp maximum in strangeness production, but also find increased fluctuations in the K/π ratio at lower-energies, although a peak is not seen there[125]. As discussed above, looking in the forward direction may well be comparable to studying this beam energy dependence, through the phenomenon of limiting fragmentation. Thus, the study of dynamical fluctuations of particle number, momentum and flavor in the forward rapidities at RHIC will provide insight into the general interplay between particle number and baryon density.

## 5.4 Identified Particle Yields out to the Kinematic Limit

The current state of forward physics at RHIC mainly involves the measurement of inclusive quantities. For instance, the PHOBOS detector has measured inclusive charged particle production as a function of pseudorapidity (η) over the range |η| < 5.4 in p+p, d+Au, and Au+Au[126] reactions, *e.g.* as shown in Figure 29. The BRAHMS experiment has measured identified hadrons out to pseudorapidity η = 3.5[127]. These have facilitated a series of interesting observations about global features of particle production in heavy ion collisions, especially pertaining to the shape of the overall rapidity distributions. One striking phenomenon, which speaks strongly for the need for a forward physics program, is that of "limiting fragmentation" shown by the PHOBOS experiment in Figure 38. This is the independence of many quantities on the beam energy when observed at fixed rapidity interval from the beam rapidity. These quantities include both inclusive particle yields and even the elliptic flow vs. pseudorapidity[128] (shown in Figure 39, and to be discussed below). This phenomenon suggests that midrapidity physics at low energy is effectively forward physics at higher energy. Testing and deepening our understanding of this apparent universality of particle production is a major goal of a new RHIC detector.

A full acceptance forward spectrometer with particle ID would be an excellent means to collect a full data set of identified yields out to high-$p_T$, essentially augmenting the existing BRAHMS program. This would provide an enormous amount of high-statistics information about the detailed modification of particle spectra at large rapidities out to the kinematic bounds observed in p+p collisions (where the maximum $p_T \sim 1/\sinh y$) – and perhaps beyond, due to the multiple collisions experienced by nucleons in the context of a nuclear collision.

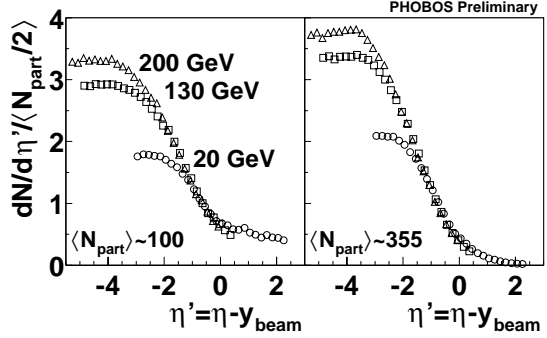
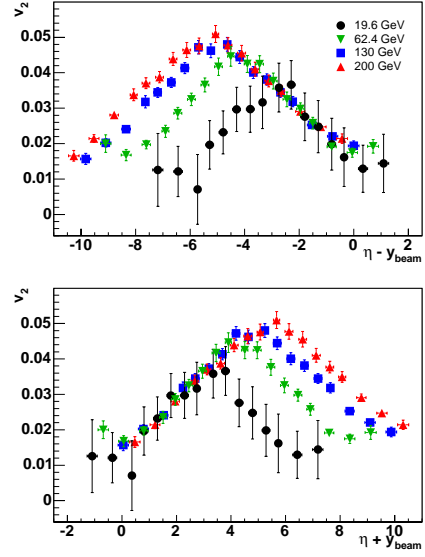

**Figure 38:** Limiting fragmentation of inclusive charged particles in RHIC Au-Au collisions.

**Figure 39:** "Limiting fragmentation" of elliptic flow in Au-Au collisions for 4 RHIC energies.

This high precision data over a large rapidity range may also be able to provide insight into the production of strangeness in a system with large baryon density. By varying the rapidity of the measured particles, BRAHMS[129] has already shown that various particle ratios appear to be connected by the baryochemical potential inferred by the antiproton/proton ratio. This has led to some discussion that the various rapidity ranges are causally "separated" from others, suggesting that measurements in limited rapidity ranges are sufficient to extract the properties of the "relevant" source (*e.g.* at $y = 0$) while $4\pi$ yields measure an inappropriately-averaged source. This may explain why $4\pi$ integrated yields consistently point to an overall 20-25% strangeness suppression while mid-rapidity yields give essentially none[130].

Better understanding of this "factorization" of freeze-out properties may help us understand the time scales in which strangeness and baryon number information is transferred to the final state system at freeze-out. It may also shed some light on the sharp maximum seen by NA49 at moderate SPS energies[131], shown in Figure 40, which may well correspond to large rapidities at RHIC both with respect to particle density and approximately to baryon density.

## 5.5 Baryon Dynamics

One of the more mysterious aspects of strong interactions, in both p+p and A+A collisions, is exactly how energy gets transferred from the incoming projectiles into the final state particles. This is clearly related to the energy available for particle production, and thus in some sense may be considered to be the process which sets up the subsequent dynamical evolution, presumably by the ideal relativistic hydrodynamics described above.

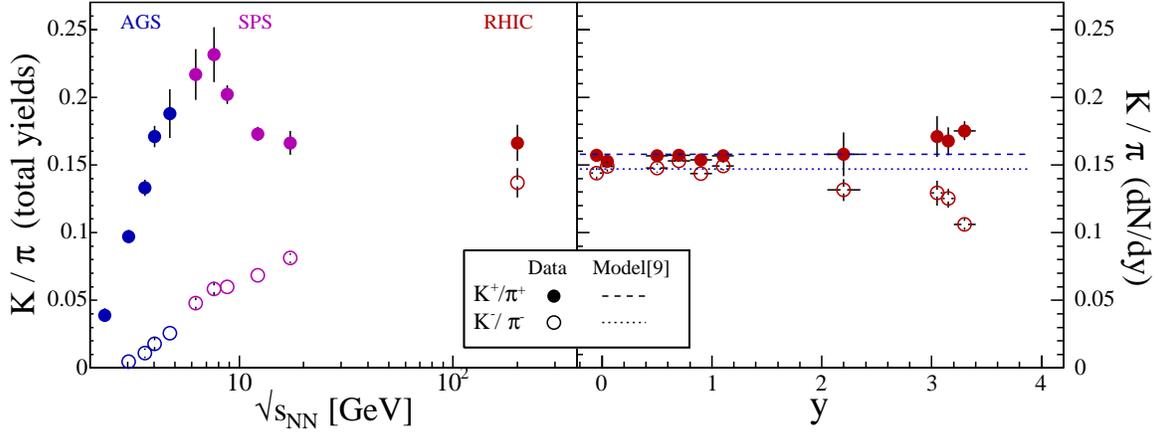

**Figure 40:** $K^+/\pi^+$ and $K^-/\pi^-$ ratios shown as a function of CMS energy (left) and rapidity at 200 GeV.

Baryon number is conserved in all experimentally known interactions. Proton decay experiments have limited any possible decay channel to have a half-life of $10^{33}$ years[132]. Experiments using p+p and p+A interactions have tried for years to elucidate the mechanism of inelastic processes, which degrade the proton energy and "pionize" the released kinetic energy[133]. Interesting phenomena that have been measured here are the "leading protons" in p+p collisions, which appear to be distributed uniformly in the $x_F$ variable ($2p_z/\sqrt{s}$) and directly correlate with the particle multiplicity[134], as well as moderately high-$p_T$ baryon and meson production even at 90-degrees. Proton-nucleus collisions showed a dramatic slowing of the proton, which seemed to saturate at about 2 units of rapidity for a wide range of centralities[135], despite the tendency for the total multiplicity to scale linearly with the number of participants. This makes the direct connection between the proton energy and the multiplicity somewhat non-trivial.

Nucleus-nucleus collisions are even more difficult to understand since despite the large increase in particle density, the net baryon number only seems to stop by only about 2 units of rapidity, as suggested by recent BRAHMS data[136], shown in Figure 41. And yet, despite the large amount of energy in these "leading baryons", heavy ion collisions seem to saturate the limit of the multiplicity found in $e^+e^-$ reactions[137]. Of course, it is well-known that jets in $e^+e^-$ and p+p reactions also have "leading particles" which carry information about the initial state quantum numbers. These also take substantial fractions of the initial energy (typically 25%[138]) so we may be dealing with a possibly generic phenomenology, even if they arise from different physical mechanisms.

Clearly, only a dedicated study of baryon dynamics in p+p, p+A and A+A reactions with full phase space coverage and meson/baryon separation at large rapidities could hope to fully elucidate this mechanism, which is so crucial for understanding the very initial state of the collision. One fascinating possibility would be to study the correlation of leading baryons with multiplicity, and especially direct correlations with slower particles, perhaps going as far back as mid-rapidity and possibly even particles in the opposite hemisphere. The Landau initial condition provides a scenario with a system in full thermal contact in the very initial state, while Bjorken initial conditions envision dynamics where only short range correlations persist due to the extreme longitudinal expansion and finite particle formation time. Systematic studies such as these may offer insight relevant to very old puzzles in the strong interaction.

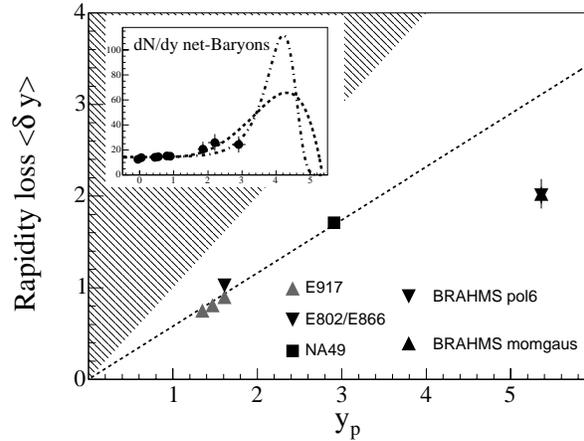

**Figure 41:** Evolution of baryon stopping as a function of energy. Inset is shown the projected net baryon density at forward rapidities.

## 5.6 Observables for a New RHIC II Detector

### 5.6.1 Everything counts: identified particle spectra in full azimuth over a wide rapidity range.

Crucial to elucidating the baryon dynamics of the early stage and the details of the system evolution until freeze-out, a full acceptance detector at RHIC should have the following features:

- Full tracking out to the highest rapidities, with reasonable resolution out to 100 GeV/$c$.
- Reasonable particle identification, at least to separate mesons from baryons, up to 50 GeV/$c$.
- Electromagnetic and charged-particle detectors, to contain as much of the emitted energy and particle number as possible.
- Full-acceptance forward calorimetry for the study of Mueller-Navelet jets in p+p and p+A.

The BRAHMS experiment at RHIC is currently the only one that can do particle identification over a broad range of rapidity and $p_T$. However, this functionality has come at the cost of several crucial features needed for disentangling the dynamics of RHIC collisions. Building a limited-aperture spectrometer has necessitated extremely long running times, since only one track per event makes it through the spectrometer. Moreover reaching the full rapidity coverage requires several positions for the spectrometer relative to the beam pipe. Finally, the spectrometer covers only the forward hemisphere of asymmetric reactions, *e.g.* requiring double the running time to explore the full phase space for d+A collisions.

Having a full-acceptance multiparticle spectrometer will allow a wealth of measurements not possible with the current RHIC detectors:
- Total yields to study the initial state entropy. Local and global fluctuations as a probe of phase transition dynamics
- Chemical equilibrium: local or global in rapidity? Flavor fluctuations as a probe of thermalization
- Validity of the hydrodynamic description away from midrapidity
- Centrality and species dependence of limiting fragmentation
- Leading particles in p+p, p+A and A+A

### 5.6.2 Pushing into the saturation regime: large acceptance in *x* and $Q^2$

One of the limitations of the current large RHIC detectors is the limited acceptance in *x* and $Q^2$ of the partonic subprocesses. In the saturation picture, one makes the approximation $x = (2p_T/\sqrt{s})e^{-y}$. Assuming $Q \sim p_T$, this leads to the simple formula $y = \log(1/x)+\log(Q)$ such that contours of constant *y* cut diagonally through the $\log(1/x)$ and $\log(Q)$ plane (see Figure 9). The effect of changing the beam energy is to move the contours up and down by $\log(\sqrt{s}/m)$. Overlaying these contours of equal *y* on the parton saturation "phase diagram" shown in Figure 31, one sees two important features:

- At mid-rapidity, DGLAP evolution becomes relevant at a substantially lower-$p_T$ than at larger rapidities
- At forward rapidities, most of the accessible particle production is confined both to the CGC region (low-*x*) and the "Color Quantum Fluid" (low-*x*, high-$p_T$).

This points to the need for much larger forward coverage than typically assumed at RHIC. In fact, far-forward coverage can reach physics relevant to lower-*x* than most measurements at HERA. This is essential for minimizing ambiguities between physics relevant to DGLAP evolution and those relevant to CGC evolution. Thus, it is also crucial to have sufficient acceptance and rates to reach the highest-$p_T$ possible at these forward rapidities to push as far into the perturbative regime as possible even at forward rapidities. This will require the high luminosities at RHIC II and the large azimuthal acceptance of a new RHIC II detector.

### 5.6.3 Multiparticle azimuthal correlations: directed and elliptic flow

Provided that nearly-full azimuthal acceptance can be achieved, this detector will be the *ultimate* tool for exploring angular asymmetries in the produced particles over the full rapidity range available at RHIC.

Individual tracking of particles will allow various methods, already employed at RHIC, like reaction plane estimation and 2-, 3-, 4-, and 6-particle cumulants to study the effect of non-flow correlations over a wide range of rapidity and $p_T$. This will offer the opportunity to dial in various amounts of net baryon density as a function of rapidity, which should allow interesting tests of quark recombination. Also, the limiting fragmentation of $v_2$ can now be tested as a function of particle species and $p_T$, to see exactly which components contribute. The particle ID will also allow a study of the fine structure at higher rapidities, providing stringent tests of hydrodynamic models.

### 5.6.4 Return of pQCD dynamics: high-$p_T$ physics in the forward direction

One of the more striking results at RHIC has been the phenomenon of high-$p_T$ suppression in heavy ion collisions over a wide range of energies. This phenomenon has been very striking at RHIC, but less so at CERN, and yet limiting fragmentation suggests that forward physics at RHIC may be quite similar to that at midrapidity at lower energies. This clearly points to achieving high-statistics particle spectra (ideally identified) in the forward direction to see which factors control the behavior of the suppression phenomena. Although saturation models have tried to predict more novel forms of high-$p_T$ suppression, studies made so far at midrapidity have not revealed any measurables particularly suggestive of CGC dynamics.

Beyond the inclusive spectra measured out to the highest-possible $p_T$ values, correlations of high-$p_T$ particles have provided deeper insight into the nature of the suppression phenomena, strongly suggesting that the partons lose energy directly in the medium. CGC models would predict that monojets would be created in the forward regions, as collisions of two gluons in the dense "color liquid" phase would fuse to create a single jet at the $x_F$ value determined by the two colliding partons ($x_1$ and $x_2$). This could be studied by the back-to-back correlations of high-momentum particles in the forward direction within ranges of rapidities comparable to STAR. As discussed above, diphoton correlations may be used to study HBT correlations pertaining to the earliest times, being direct radiation from the primordial quarks and antiquarks in the source.

### 5.6.5 Direct probes of the CGC: Drell-Yan and heavy flavor

The transition from DGLAP to CGC evolution should have some observable effects on probes sensitive to low-x partons intrinsic to the nucleon or nucleus. Both Drell-Yan dileptons, an electromagnetic probe sensitive to the quark and antiquark distributions, and heavy flavor, which is typically sensitive to the gluon distribution, should provide direct access to this physics via leptonic channels.

With precise tracking in the forward direction, it may well be possible to identify charmed particles, both open and closed. Charmonia may be detectable if one can identify the forward-going muons, which should be straightforward out to η~3, but more difficult in the far-forward arm. PHENIX data already shows non-trivial effects in the rapidity dependence of *α*, which characterizes the scaling with nuclear geometry. In 4π, these measurements should be able to address the issues of shadowing vs. anti-shadowing in d+Au data and thus determine the level to which saturation models can be used to understand the very initial state gluon distributions. Open charm may be detectable by displaced vertices (provided sufficient tracking is available close to the interaction point). This can be used to test $k_T$ factorization results that lead to harder charm spectra than one might expect by NLO pQCD[139].

Drell-Yan has no topological structure in its decay, nor does it evince any resonance peaks. Furthermore, the dilepton continuum at high CMS energies has substantial contribution from correlated leptons from charm and bottom decays. Still, by careful subtraction of known sources (onia, open heavy flavor) and high luminosities to push out in pair $p_T$, it may be possible to extract forward dileptons to look for modifications of pQCD expectations.

# 6   The Structure and Dynamics inside the Proton

The spin physics program at RHIC II will focus on rare processes and its success is predicated on attaining good beam polarization (70%) and high luminosity $2 \cdot 10^{32}$ ($8 \cdot 10^{32}$) cm$^{-2}$s$^{-1}$ at $\sqrt{s}$ = 200 GeV (500 GeV) in polarized proton-proton interactions. In order to measure the requisite, extremely small asymmetries (~$10^{-3}$) further improvements in the RHIC II collider performance are required to control bunch-to-bunch variations and limit false asymmetries. These include the use of simultaneous polarization reversals of the beams stored in both RHIC rings using AC-dipole magnets and beam re-cogging, and precise luminosity monitoring at high rate at the experiment.

There is exciting physics in polarized proton-proton collisions that forms the basis of a world class spin physics program in a new comprehensive detector at RHIC II. This physics can be divided into four areas: the flavor decomposition of the contributions from the QCD sea to the polarization in a proton, probes of gluon polarization through the spin-dependent production of heavy quarks, transversity densities of quarks and anti-quarks in the proton and physics beyond the Standard Model. Furthermore, the large acceptance electromagnetic and hadronic calorimetry of the new comprehensive detector coupled with the increased luminosities at RHIC II make available detailed study of the gluon polarization in the proton through jet measurements at large transverse energy. All of these physics topics have been discussed previously and form the basis of a RHIC II spin program[17].

Determination of the contribution of gluons to the polarization of the proton is already underway with the existing RHIC detectors. Determination of the sea polarization, a part of the original RHIC spin program, requires optimal operation of RHIC at $\sqrt{s}$ = 500 GeV for p+p and further upgrades of the existing RHIC detectors. Further probes of gluon polarization in the proton can be made by measurements of charm and bottom production, production of heavy quarkonia, and an understanding of the unpolarized production of heavy quarkonium states. This can be accomplished by measuring various leptonic decay channels over a large kinematic range in the new RHIC II detector. Initial measurements of the transversity densities of quarks and anti-quarks are already underway with the existing detectors at RHIC. The increased luminosities at RHIC II and the range and capabilities of the new RHIC II detector will allow measurements of the transversity in the proton at larger transverse momenta. The existence of physics beyond the Standard Model (SM) is generally exciting and will be investigated using spin measurements at RHIC II. These include new parity-violating interactions, to investigate possible quark compositeness, where small deviations from the SM are predicted with increased jet transverse energy. Such measurements are extremely difficult and require the highest possible luminosities and data rates, and large coverage for jet measurements. There are other interesting measurements, but these have been selected to provide a flavor for the range of exciting spin physics made available with a new comprehensive detector at a high luminosity RHIC II. The strength of a new RHIC II polarized proton-proton program will rely to a large extent on the design of an optimized comprehensive proton-proton collider detector similar to the dedicated collider detectors CDF and D0 programs at the Tevatron. This will be discussed in more detail in the following sections.

## 6.1 Polarization of Quarks and Anti-quarks in the Proton

Initial determination of the contribution of gluons to the polarization of the proton will be accomplished with the existing RHIC detectors in the current RHIC spin program. Although determination of the sea polarization is part of the original RHIC spin program, this measurement is still significantly far away. It requires $\sqrt{s}$ = 500 GeV p+p operation of RHIC with optimal polarization (~70%) and luminosity, plus further upgrades of the existing RHIC detectors. The measurement of the parity-violating longitudinal single-spin asymmetry $A_L$ in W production will determine the underlying polarized quark and anti-quark distributions[19] and further the understanding of the QCD sea. A Feynman graph of the underlying process is shown in Figure 42.

The measurement of $u + \bar{d}(d + \bar{u}) \to W^{+(-)}$ through the electron/muon decay channels at high $p_T$ provides a clean separation of the underlying quark distributions if one can observe the final-state leptons in the forward direction, *i.e.* in a kinematic region where one samples asymmetric collisions. Results obtained in polarized deep inelastic scattering (DIS) experiments suggest that the QCD sea is significantly polarized.[140] It is crucial for our understanding of the nature of the QCD sea to explore W production in polarized proton-proton collisions to determine if the polarization of the QCD sea is flavor-dependent. The difference of $\bar{d} - \bar{u}$ unpolarized anti-quark distributions has been found to be significantly different from zero in particular in the region of small Bjorken-$x$ values.[141] This strong breaking of SU(2) symmetry has an explanation among several non-perturbative QCD mechanisms. An approach by Dressler *et al.*[21] based on the Chiral Soliton Model provides a prediction of the difference $\Delta\bar{u} - \Delta\bar{d}$ that is larger than the difference of the unpolarized quark distributions. The large acceptance of a new comprehensive RHIC II detector for muon and hadron identification in a large magnetic field and full calorimetry coverage will enable high quality measurements of $W^{+(-)}$ decays with large statistics. Such a detector system similar to CDF and D0 in coverage would allow measurement of the missing energy in the unobserved final-state neutrino and aid in the reconstruction of the underlying kinematics. This program clearly requires running RHIC II at the highest possible center-of-mass energy at high luminosity and maximum polarization.

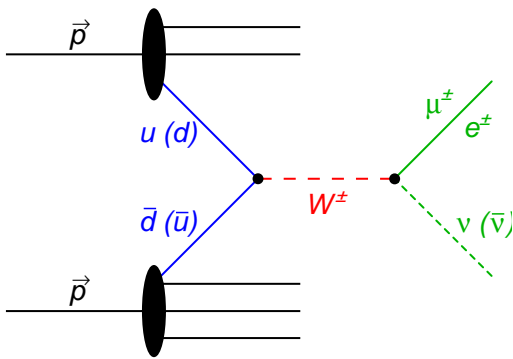

**Figure 42:** Feynman graph of $u + \bar{d}(d + \bar{u}) \to W^{+(-)}$ production in polarized proton-proton collisions.

A long-standing question in spin physics is the contribution of strange quarks and anti-quarks to the QCD sea in the proton. The polarization of strange quarks cannot be accessed via parity violation for inclusive W production at RHIC because $u,d$ quarks provide the dominant contribution to the cross section. But, it should be possible at RHIC II to access the strange quark polarization by looking at parity violation in charm-tagged W production. Polarized intrinsic $s$ quarks in the polarized proton can couple to $\bar{c}$ from the second proton (produced by the splitting of a gluon from that proton into $\bar{c} + c$) and result in a final state W. The $c$ quark from the gluon splitting can fragment into a D-meson and then be used as a tag that the W production resulted from polarized strange quarks. This experiment is ambitious and requires the highest pos-sible luminosity at RHIC II and also a very large acceptance detector to detect the charged lepton from *W*-decay in coincidence with an identified D-meson. To date, the only other method to directly access strange quark polarization is through flavor-tagged semi-inclusive deep inelastic scattering.[18]

## 6.2 Gluon Polarization Accessed via Heavy-Flavor Production

The production of heavy quarks in p+p collisions is dominated by gluon-gluon fusion. The underlying leading-order Feynman graph is shown in Figure 43. The contribution of competing production mechanisms for heavy flavors through quark-antiquark annihilation is small at RHIC, since that requires an anti-quark in the initial state. It has been shown[142] in a leading order approximation that heavy-quark production in polarized proton-proton collisions provides a means to constrain the underlying gluon polarization. Complete next-to-leading-order (NLO) QCD corrections to the polarized hadron-production of heavy flavors are now available.[143] The importance of next-to-leading order (NLO) corrections to provide reliable quantitative predictions has been pointed out.[144]

Furthermore, charm and bottom production probe the gluon density in the proton at different momentum fractions (Bjorken-*x*) and scales ($Q^2$). Measurements of the production of heavy quarkonia also probes gluons in the proton and will only be possible at the higher RHIC II luminosities. [Understanding the unpolarized production of heavy quarkonium states is therefore also important to the spin physics program as well as the heavy ion program.] Heavy-flavor production can be selected through various leptonic decay channels over a large kinematic range in the proposed detector. In addition, $B \to J/\psi + X$ tagging the $J/\psi$ through displaced electron and muon vertices provides a means of identifying open beauty production.

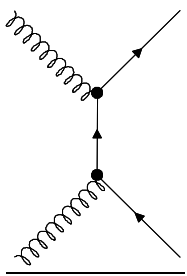

**Figure 43:** Leading order Feynman graph for heavy quark production in gluon-gluon fusion.

It was shown in Ref. 8 that NLO calculations have considerably less uncertainty in terms of the unphysical renormalization and factorization scales. It was demonstrated based on the current PHENIX acceptance that the gluon polarization can be constrained by heavy flavor production. Charm and bottom production will access different regions of Bjorken-*x*.

Heavy-flavor production can be selected through various leptonic decay channels. Examples are $pp \to e^{\pm} + X$, $pp \to \mu^{\pm} + X$, $pp \to e^+e^- + X$, $pp \to \mu^+\mu^- + X$ and $pp \to e^{+(-)}\mu^{-(+)} + X$. In addition, $B \to J/\psi + X$ tagging $J/\psi$ through displaced electron vertices provides a means of identifying open beauty production and thus a probe to access the gluon polarization.

Besides the potential of constraining the gluon polarization in heavy flavor production, the unpolarized aspect of hadron production of heavy flavors has attracted considerable attention in recent years. This is related to the fact that open charm production is reasonably well understood, however beauty production has exhibited large differences between data and theory[22]. Discrepancies between data and theory have been reported in *ep* collisions at HERA[23] and even in γγ collisions at LEP[24]. This has led to an extensive discussion of physics beyond the Standard Model (SM) which could provide an explanation for this discrepancy. In fact, in the minimal supersymmetric extension of the SM, gluinos can decay into a standard model bottom quark and a lighter supersymmetric sbottom quark. This would increase the yield in bottom production and thus would provide a mechanism to explain the apparent discrepancy between data and theory. RHIC could play an important role in understanding this discrepancy through its ability to investigate energy- and spin-dependent charm and bottom production.

## 6.3 Transversity of Quarks and Anti-Quarks in the Proton and other Transverse Spin Effects

The transversity density of quarks and anti-quarks will be accessible at RHIC, but can be most cleanly studied at RHIC II. Although transversity is a leading twist distribution function, equally important to our understanding of the spin structure of the proton as the helicity asymmetry distribution, it cannot be probed in inclusive deep-inelastic scattering because the chiral-odd character of the distribution function is incompatible with the fact that electroweak (and QCD) vertices preserve quark chirality. Measurements of semi-inclusive deep inelastic scattering that probe transversity through the use of a spin-dependent, chiral-odd fragmentation function (the Collins function) are underway by the HERMES and COMPASS collaborations, and are proposed at Jefferson Laboratory. Similarly, initial measurements of transverse spin asymmetries through the Collins mechanism are also possible with the existing RHIC detectors. Single spin asymmetries in polarized proton collisions can probe transversity through a spin- and transverse momentum dependent fragmentation function (Collins function), but these spin effects are expected to decrease with increasing $p_T$. Furthermore, effects from transversity have to be distinguished from other possible dynamics that can also give rise to non-zero single-spin effects. The increased luminosities at RHIC II and the range and capabilities of the proposed detector will allow measurements of the transversity densities of quarks and anti-quarks in the proton with increased sensitivies at larger transverse momenta. Double transverse-spin asymmetry ($A_{TT}$) measurements in Drell-Yan production[145] provide direct sensitivity to the transversity distribution function, but require non-zero transverse polarization of antiquarks within a transversely polarized proton. Another method is to measure $A_{TT}$ for jet production[146] at very high $p_T$ (>40 GeV/$c$) to minimize the contributions from gluons in the initial state that serve to dilute the two-spin asymmetry. Alternatively, di-jet production can be studied, and the two-spin asymmetry can be measured for the linear combination of the charge sign of pions that suppress contributions from gluon fragmentation.

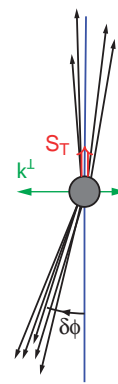

The first results from the RHIC spin program have revealed that large transverse single spin asymmetries, first observed in lower energy polarized proton collisions, persist to RHIC energies. One of the proposed mechanisms for the large analyzing power in polarized proton collisions is the Sivers effect. The Sivers function is a correlation of the form $S_T \cdot (P \times K^\perp)$ between the original proton transverse polarization vector ($S_T$), its momentum (P), and the transverse parton momentum relative to the proton direction ($K^\perp$). In collisions of unpolarized protons with transversely polarized protons the single transverse-spin asymmetry in the relative intra-jet azimuthal angle (*i.e.* non-collinearity depicted by δφ in Figure 44) of di-jets or alternatively the measurement of the relative azimuthal angle between photons and away-side jets are sensitive to the Sivers function.[147]

**Figure 44**: A "beams-eye" view, along the direction of the colliding beams, of the di-jet final state with finite relative azimuthal angle (δφ). $S_T$ is the original proton transverse polarization vector and $K^\perp$ the transverse parton momentum relative to the proton direction[147].

Recently it was realized that the time-reversal odd Sivers function can be non-zero because of subtleties of gluon exchange between the quark, involved in either Drell-Yan production or that absorbs the virtual photon in semi-inclusive DIS, and target spectator system.[148] Because of these subtle interactions, transverse spin effects associated with the Sivers function in semi-inclusive deep inelastic scattering and the analyzing power for Drell-Yan production in polarized proton collisions are expected to be opposite in sign. It is important to quantify the role played by the Sivers function in single semi-inclusive deep inelastic scattering and polarized proton collisions because it is non-zero from orbital motion of the partons within the proton.

## 6.4 Physics Beyond the Standard Model (SM)

There are predictions for physics beyond the Standard Model (SM) whose existence may be discovered using spin measurements at RHIC II[26]. These include new parity-violating interactions that would lead to significant modifications of SM predictions. Here parity violation arises within the SM for quark-quark scattering through the interference of gluon- and $Z^0$-exchange. Figure 45 displays the single spin asymmetry for inclusive jet production as a function of jet transverse energy. The standard model prediction[27] is shown (labeled SM) along with predictions for quark compositeness (solid curves labeled $\varepsilon\eta = \pm 1$) and leptophobic model predictions for an extra heavy vector boson that couples directly to quarks (dashed and dotted curves labeled by mass of the new vector boson). Observation of a parity-violating single-spin asymmetry in inclusive single-jet production at RHIC would signify quark compositeness, as depicted by the solid curves labeled $\varepsilon\eta = \pm 1$ in Figure 45. Since the magnitude of deviations from the SM prediction (solid curve labeled SM in Figure 45) is extremely small and increases with transverse jet energy, such measurements are very difficult requiring the highest possible luminosities and data rates, and large coverage for jet measurements to the highest possible transverse energies.

Adding polarization information to W production should lead to interesting tests of the Standard Model. This is particularly accessible with large detector acceptance. A search for parity violation in di-jet or di-hadron events separated by a large rapidity interval should open the possibility of searches for new physics. This was pointed out initially by Bjorken.[149]

It has been pointed out that a polarized RHIC program would allow one to place constraints on several models by selection of a specific region in phase space which is unconstrained by current and future experimental efforts at other collider facilities such as the Tevatron. RHIC would therefore be in a unique position to explore a certain aspect of physics beyond the SM. This program clearly requires running RHIC at high luminosity and at the highest possible center-of-mass energy. This requirement was already raised in the previous section. The capabilities of a new comprehensive (hermetic) detector system certainly benefit the study of physics beyond the standard model at RHIC II.

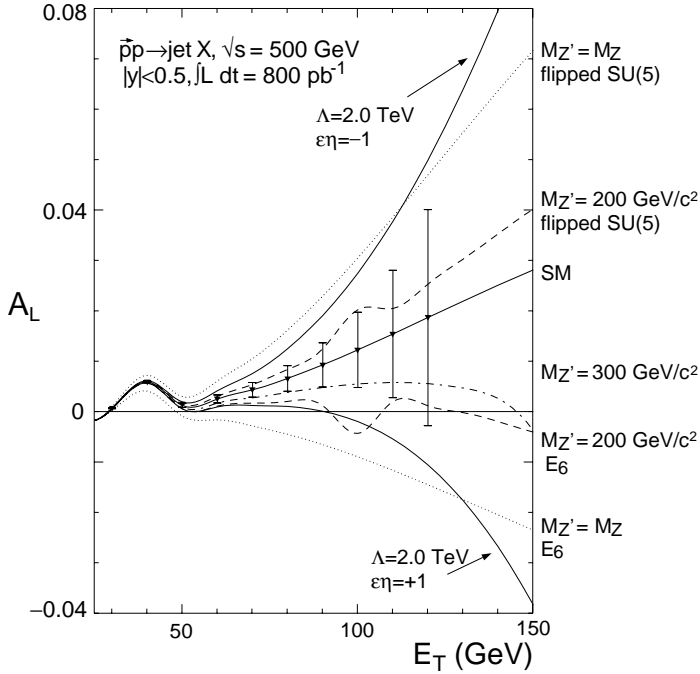

**Figure 45:** $A_L$ for one-jet inclusive jet production in polarized proton-proton collisions[27]. The solid line refers to the Standard Model prediction. The dashed and dotted curves are predictions from different leptophobic models. The lower solid curve refers to a prediction involving contact interactions.

## 6.5 Gluon Polarization in the Proton from Jet Measurements at Large Transverse Energy

Initial determination of the contribution of gluons to the polarization of the proton will be accomplished with the existing RHIC detectors in the current RHIC spin program. The production of jets in polarized proton-proton interactions has been widely recognized as a means to probe the gluon polarization. The underlying partonic reactions are shown in Figure 46 for the quark-gluon Compton process ("prompt photon production"), gluon-gluon scattering and for gluon-quark scattering leading to the production of jets. Jet measurements are a central part of the spin program in STAR and are reconstructed within STAR using a combination of TPC tracking and EMC cluster information. The addition of hadronic energy information, increased magnetic field strength and fast triggering and data-acquisition in a new RHIC II detector would improve jet measurements significantly. A polarized proton-proton program with increased luminosity at RHIC II extends the transverse energy range for jet measurements to ~ 50 GeV.

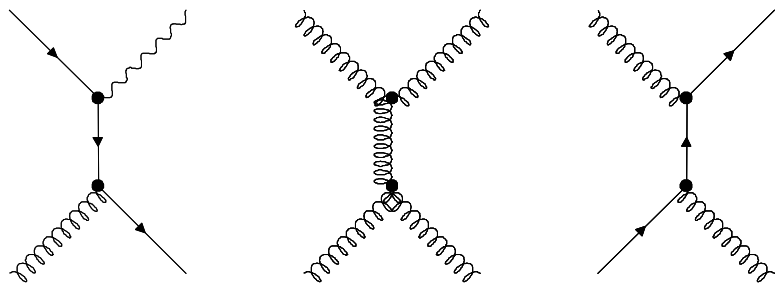

**Figure 46:** Leading order Feynman graphs for gluon-initiated process in proton-proton collisions resulting in various final states involving jets.

## 6.6 Detector Requirements Based on Spin Physics Program

The above spin physics topics – in summary - require the following detector capabilities:

- Heavy Quark Production: electron and muon detection
- QCD (especially jet physics): jet reconstruction, single-photon detection ($\gamma/\pi^0$ separation), $b/c$-tagging
- Electroweak Physics: electron and muon detection and missing energy measurement
- Physics beyond the Standard Model (SM): electron and muon detection, jet reconstruction, $b/c$-tagging and missing energy measurement

This translates into the following basic detector requirements, which must still be refined and strengthened by detailed detector simulations:

Full acceptance in the barrel and forward/backward region is required up to at least $|\eta| = 3.5$ with specialized calorimeter and tracking system beyond $|\eta|\sim 4$ to extend acceptance in the very forward acceptance region to probe the kinematic region of small Bjorken-$x$ values and to provide missing energy information.

Electromagnetic and hadronic energy measurements in the barrel and forward/backward region are essential, as are transverse and longitudinal tower segmentation (for jet reconstruction and electron/hadron separation).

Full tracking coverage is required in the barrel and forward/backward region. Also necessary are a precision inner vertex detector system based on inner barrel layers and forward wheels with a combination of pixel, silicon and GEM-type tracking detectors to provide reliable secondary vertex reconstruction, momentum resolution and at the limits of resolution a charge-sign determination in the forward/backward region. A muon detection system in the barrel and forward/backward region would aid the reconstruction of heavy quarks and heavy-bosons.

An axial magnetic field of at least 1–2 T is preferred. A precise relative luminosity measurement in a high-rate environment is essential, as are local polarimeter information to tune spin rotator magnets and track the local beam polarization, and an absolute luminosity measurement.

Required are a high rate DAQ system, and a system of several layers of triggering to provide triggers on rare processes and a secondary-vertex trigger as is now employed at CDF.

# 7   Detector Concepts

The requirements for a new RHIC II detector are well defined by the previous chapters. The primary requirements are:

- Excellent momentum resolution out to $p_T$ = 40 GeV/$c$ for charged particles in the central rapidity region.
- Complete hadronic and electromagnetic calorimetry over a large phase space (~4$\pi$).
- Particle identification out to very high $p_T$ ($p \sim$ 20–30 GeV/$c$) including hadron ($\pi$, K, p) and lepton (e/h, $\mu$/h) separation in the central and forward region.
- High rate detectors, data acquisition, and trigger capabilities.

These requirements are quite stringent. Because of the focus on hard probes the requirement on multiple scattering in material for low-$p_T$ particles is only a minor concern.

It is our intention to utilize, where possible, detector components from other collider experiments that are decommissioned, or will be in the near future. In particular, it should be possible to identify and procure an existing high field magnet and a large amount of electromagnetic and/or hadronic calorimetry depending upon the eventual geometry. As a proof of principle, we will use the SLD magnet. It was originally planned as a superconducting magnet but due to funding limitations was constructed with warm coils resulting in a field strength B = 0.6 T. However, the iron is sufficiently thick to contain considerably larger flux. After discussion with SLD experts it was concluded that the magnet iron as constructed would allow replacement of the warm coils with superconducting coils yielding a field strength B = 1.3 T. The inner radius of the magnet is 2.8 m, with an inside length of 6 m (overall outer length of 7 m). This would provide a bending power of 2.2 T-m over the tracking volume of radius 1.7 m [this is more than 2.5 times that of STAR]. SLD management is very interested in the re-use of the SLD magnet, support structure, and other components. A possible layout for a RHIC II detector using the SLD magnet is shown in Figure 47.

Large components of other high energy experiments such as CDF, D0, HERA-B and CLEO are expected to become available before the end of the decade to join a growing number of already decommissioned experiments such as SLD, DELPHI, ALEPH, etc. In particular, D0 is of interest with its novel thin magnetic coil design with calorimetry plus muon chambers. CLEO has a high field magnet plus crystal calorimetry that could be utilized.

Figure 47 shows a detector layout based on the SLD magnet. The proposed layout is very similar to a classical high energy experiment, but, in addition, it retains particular features necessary for heavy ion physics, such as extensive particle identification and very fine granularity calorimetry in the forward direction. From the inside out the central ($|\eta| < 1.2$) detector consists of a small radius, very thin Beryllium beam pipe, a Silicon microvertex tracker, a solid-state or gas main tracker, a RICH, a ToF/Aerogel combination, an electromagnetic calorimeter (EMCal), and the SLD magnet with its embedded hadronic calorimeter (HCal) and muon chamber sections. The SLD magnetic is hermetic out to $|\eta| < 3.5$ and the detector layout in the magnet endcap direction (1.2 < $|\eta|$ < 3.5) will mimic the central layout, *i.e.* a forward disk tracker, followed by the PID components and the calorimetry and muon chambers. Beyond the 'doors' of the SLD magnet (at $|\eta|$ >3.5) we propose to build a dedicated forward spectrometer consisting of a silicon disk tracker in its own magnetic field, followed by a RICH, an EMCal and HCal. The forward spectrometer is limited in the forward direction by the RHIC DX magnets, which means the maximum achievable spectrometer range will be η = 4.8. Details are given in section 7.7. Beyond the DX magnet we are considering a Roman pot configuration with Silicon readout.

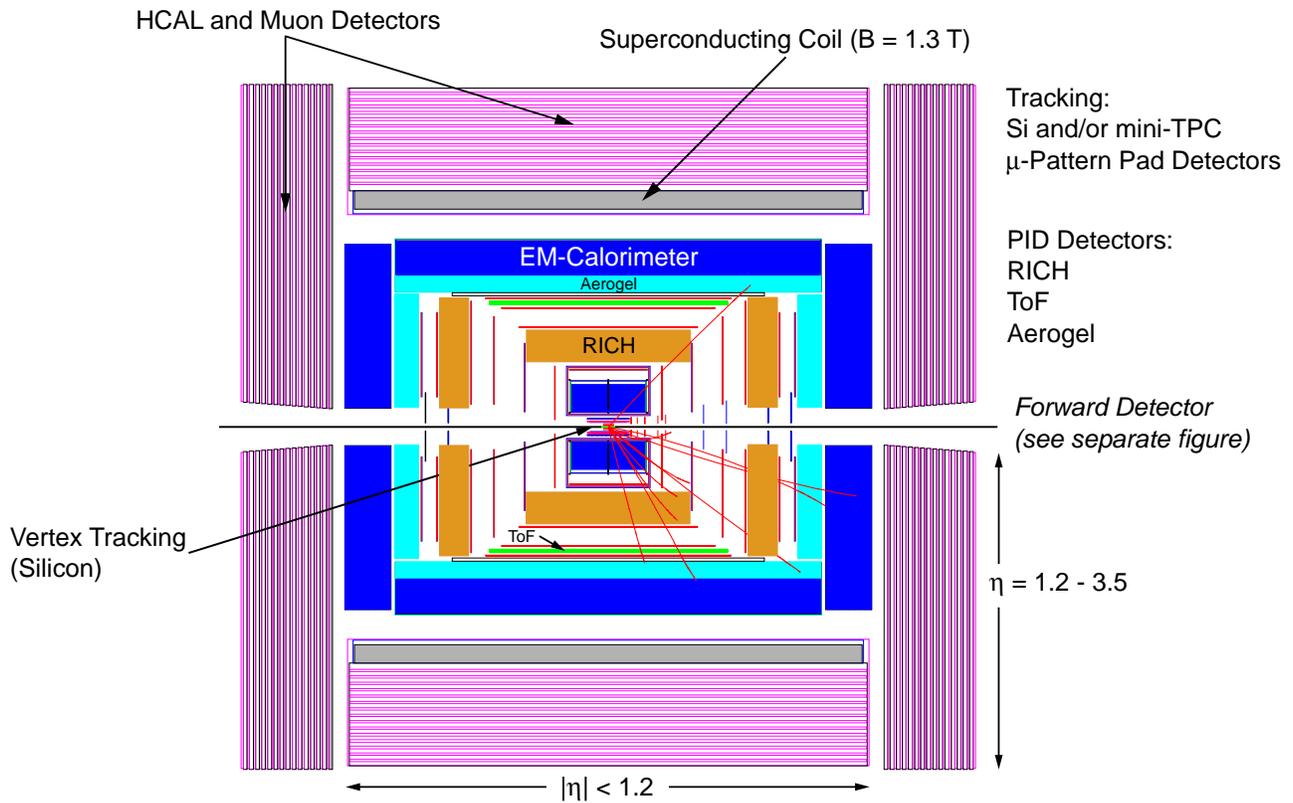

**Figure 47:** Conceptual design of the SLD magnet instrumented for a RHIC II detector.

In the following we will briefly discuss the various detector components:

## 7.1 Magnet

Good momentum resolution at high $p_T$ requires a large bending power $\int$ B·dl. A large diameter solenoid with a high field also provides sufficient volume to add detectors for optimal tracking, particle identification (PID) and calorimetry. As discussed above, the SLD magnet would be an excellent choice that combines all these necessary features. The SLD magnet and support structure consists of four basic elements: the superconducting coil, the exoskeleton support frame, the barrel iron (flux return yoke), and the Fe endcaps (or doors).

The superconducting coil, to be constructed, will have a 5.8 m diameter and will be 6.2 m long. It will provide a solenoidal magnetic field of 1.3 T in a volume of 162 m³. The existing iron structure (in the barrel and endcaps) consists of 15 layers of 5 cm thick plates, separated by 3.2 cm gaps instrumented with plastic limited streamer tube chambers. The iron serves as a flux return system, a Warm Iron Calorimeter (WIC), and muon detector. This type of detector uses *x* and *y* strip (~8 mm width) and pad (~3×3 cm²) readout, the signals are comfortably large, the tubes and FEEs are robust and rather easy to produce. The detector has perfect efficiency for minimum ionizing particles, and it is reasonably inexpensive. Because of aging effects the available streamer tubes might have to be replaced in which case modern IC FEEs could be implemented. One 100-ton or two 40-ton handling capacity cranes are required for magnet and detector assembly.

The requirements for the magnet and the tracking system are set by the momentum resolution at high $p_T$. With our high $p_T$ tracker (see below) embedded in the SLD magnet we expect to reach resolutions of $\Delta p_T/p_T \sim$ 1% at 20 GeV/$c$ and ~ 3% at 40 GeV/$c$ at mid-rapidity.

## 7.2  Vertex Detector

The proposed high resolution tracking detector needs to be complemented with a very high resolution vertex detector as a pointing and track seed device. Therefore the small thickness Be beam-pipe (radius ~2 cm) will be surrounded by four layers of Silicon detectors (radii from 4 to 16 cm), necessary to generate reliable seeds for the tracking pattern recognition. The detectors must be fast and must have very good position resolution. The most reliable technology presently is the hybrid pixel detectors, as used in all LHC experiments. We believe that resolutions of less than 20 μm are achievable in detectors with a minimum radiation length of about 0.3% per layer. With respect to alternate technologies it remains to be seen whether recent R&D on CCD's and active pixel sensors (APS) will lead to radiation hard devices of sufficient readout speed. In that case both CCD and APS would be superior to hybrid pixels in terms of resolution and radiation length. We are therefore closely following the R&D effort by the STAR collaboration for their APS microvertex detector. One can also conceive of a mixed technology detector such as the inner tracking system in ALICE in order to further lower the cost without degrading the resolution capabilities. In such a device the very thin, high resolution inner layers could be followed by less expensive Silicon strip for the outer two layers of vertex detector. Detailed simulations will have to determine whether such a device could yield the necessary track pointing resolution. A possible configuration of the vertex detector is shown in Figure 47 and detailed in Table 6 in the next section as part of the overall tracking.

The main physics requirement for the vertex detector, besides the unambiguous track seed determination for the main tracker pattern recognition, will be the tagging of *b*- and *c*-jets. With the proposed setup we expect around 70% *b*-tagging efficiency with 95% purity and about 50% *c*-tagging efficiency with 80% purity.

## 7.3  Tracking Detectors

The large volume inside the uniform magnetic field enables us to "construct" a high quality tracking system to cover |η| < 3.5. In our tracking approach we distinguish between a primary tracking device and a high $p_T$ tracker just inside the magnet coil.

### 7.3.1  Main central tracker

Two variants of the primary tracking device are being considered: a miniTPC or a multilayer (6-7) Silicon strip detector (radii from 25 to 80 cm), followed up with three to four layers of Gas Pad de-tectors (micropattern technology) in front of the EMC (high $p_T$ tracker). Such a combination of tracking components will be very powerful for track finding, to distinguish primary particles from secondary vertices (tagging), and to guarantee high quality momentum reconstruction starting from low $p_T$ (~500 MeV/$c$) and up to ~40 GeV/$c$. To reach lower $p_T$ we will either utilize the vertex detector as stand alone tracker or lower the field for certain runs.

A fast, compact TPC (anode-cathode distance 40-45 cm, drift speed ~10 cm/μs, 16 identical modules with 35 pad-rows (triple GEM foils, pad size 0.2×0.8 cm²)) is convenient for track reconstruction and dE/dx measurements. The specific proposed design would be based on a short drift distance, fast, low diffusion working gas and micro-pattern detector readout. Hit reconstruction (space) precision should be 100–120 μm in r·ϕ and ~200 μm in $z$ (drift direction).

An additional option is to install into the same gas volume a gas pad detector with a CsI top to be sensitive to UV-light produced by charged particles inside the TPC gas. The combination of particle track measurements, dE/dx and Cherenkov light will provide a powerful $e^{+/-}$ identification capability for lepton energies from 0.05 GeV to 5 GeV. Although the detector design is attractive, electronics must still be developed for demonstrated operation in a continuous readout mode to attain the high rates at RHIC II. This means the miniTPC will be useful only if it can be demonstrated that it can operate without a gating-grid, and that the ~4 μs maximum drift time is acceptable.

A solid-state tracker has the advantages of superior resolution and fast readout speed. However detailed simulations are required to determine the number of layers of Silicon needed such that pattern recognition capabilities are sufficient to resolve all tracks in the active volume. Too many layers might make the device prohibitively expensive. Track-seeding in the vertex detector will certainly help to lower the number of tracking layers, and presently the cheapest off-the-shelf Silicon technology, namely double-sided Silicon strip detectors, could be ideal to develop a large solid state tracker that is cost competitive. Besides the single-point resolution, the two-track resolution of any tracker is important in order to resolve particles within a single jet. From state-of-the-art Silicon strip detectors we expect single point resolutions down to 20 μm and two-track resolution of less than 500 μm for a reasonably segmented device.

### 7.3.2 High $p_T$ tracker

Independent of the main tracker technology, a so-called high $p_T$ tracker will be added radially outward in order to provide fast high momentum tracking. This device is envisioned as a three to four layer micro-pattern pad detector in front of the calorimeters (and magnet coil) in the central and endcap positions. Even with a fast Silicon main tracker this device will be necessary in order to provide additional points between tracker and calorimeter for precise energy flow measurements. The proposed pad detector would use micro-pattern detector technology (GEM, MicroMeGas) which can achieve 100 μm resolution in the r·ϕ direction and fast (~10 ns) response. Pad sizes can be varied from 0.04×4 cm² to 0.04×10 cm². Such detectors can be mass-produced and are reasonably inexpensive and thus a convenient approach to construct large surface, low mass, fast, and high precision tracking systems.

### 7.3.3 Central tracker layout

One possible layout for the tracking detectors is displayed and described in Figure 47. For the present set of simulations the component positions, segmentation in radius (r) and azimuthal angle ($\phi$), and thicknesses are listed in Table 6.

| Detector | Radius (cm) | Halflength (cm) | Sigma r-phi (cm) | Sigma z (cm) | Thickness (cm) |
|---|---|---|---|---|---|
| **Vertex** | 2.8 | 9.6 | 0.001 | 0.001 | 0.01 |
| (APS or | 4.3 | 12 | | | |
| Hybrid pixels) | 6.5 | 21 | | | |
| | 10.5 | 27 | | | |
| **Main Si-strip** | 19 | 39 | 0.003 | 0.03 | 0.03 |
| | 24.5 | 42 | | | |
| | 31 | 45 | | | |
| | 38.5 | 51 | | | |
| | 46 | 57 | | | |
| | 56 | 60 | | | |
| or | 22.5-60 | 55 | 0.012 | 0.035 | 0.2 |
| **Main mTPC** | | | | | (mylar+gas) |
| **High $p_T$ track** | 70 | 76 | 0.17 | 0.17 | |
| micropattern | 115 | 110 | 0.01 | 0.9 | 0.3 G10 + |
| | 135 | 130 | 0.01 | 1.2 | 1.0 Gas + |
| | 170 | 165 | 0.01 | 1.4 | 0.05 Mylar |

**Table 6:** Position, segmentation in radius (r) and azimuthal angle ($\phi$), and thicknesses of the various central tracking detectors. For details see text.

### 7.3.4 Endcap tracking

With respect to bulk matter measurements in the forward direction ($1.2 < |\eta| < 3.5$) reasonably precise tracking will be required to look for any modification in spectral shape with increasing rapidity. In our simulations we used the layout in Table 7, although the sizes and number of layers still needs to be optimized and evaluated in terms of cost-performance tradeoffs. Generally a $\delta p/p$ of about 10% can be expected.

| Layer | Technology | Z position (cm) | $R_{min}$ (cm) | $R_{max}$ (cm) |
|---|---|---|---|---|
| 1 | Micro-pattern | 70. | 10. | 80. |
| 2 | Micro-pattern | 110. | 20. | 110. |
| 3 | Micro-pattern | 150. | 30. | 145. |
| 4 | Micro-pattern | 180. | 35. | 165. |
| 5 | Micro-pattern | 225. | 40. | 150. |
| 6 | Micro-pattern | 245. | 45 | 150. |

**Table 7:** A proposed layout and dimensions for an endcap disk tracker.

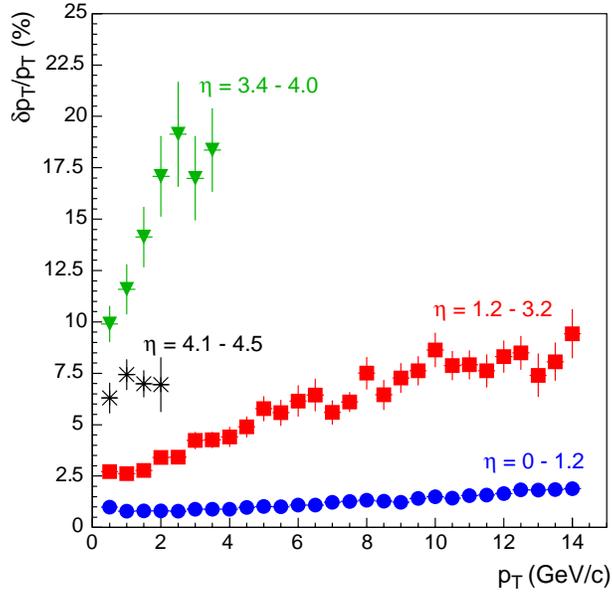

**Figure 48:** Momentum resolutions based on the tracking devices of Figure 47 for the different regions of pseudo-rapidity as described in this section and the forward spectrometer section.

### 7.3.5 Tracking performance simulation

The main physics requirements for the tracker are excellent momentum and two-track resolution out to the highest $p_T$ in order to measure all charged particles in a jet. We estimate the momentum resolution to be below 1% out to about 20 GeV/$c$ and less than 3% out to 40 GeV/$c$. For the tracking detectors shown in Figure 47, a complete simulation of reconstruction and tracking capabilities resulted in the momentum resolutions shown in Figure 48 for various regions of pseudo-rapidity. The two-track resolution for any charged-particle pair did not exceed 500 μm. In addition, the above mentioned energy flow measurements, *i.e.* the tagging of a tracked high momentum particle to a particular part of the calorimeter system, will be strongly enhanced through excellent point-back features, *i.e.* excellent single point resolution in all components of the integrated tracker. By combining a higher resolution primary tracker with a Silicon-based vertex detector and a large radius pad detector we expect to extend the excellent single point resolution out to distances near the calorimeter.

### 7.4 Particle Identification

The requirement to identify all hadrons in a high $p_T$ jet requires good hadron identification up to momenta of at least 20 GeV/$c$. Lepton particle identification will be achieved through the *e/h* capabilities in the calorimeters and the muon chambers. Hadron and lepton particle identification will be achieved through a combination of *dE/dx* in the tracking ($p_T$ < 1 GeV/$c$), a time-of-flight device ($p_T$ < 3 GeV/$c$), and a combination of various Aerogel Cherenkov-threshold counters and a RICH detector with gas radiator ($p_T$ up to 20–30 GeV/$c$). The time-of-flight device should be based on resistive plate chambers, since otherwise it would be prohibitively expensive and its use in such a high magnetic field questionable. One possible configuration using time-of-flight detectors, two Aerogel Cherenkov-threshold detectors with $n$ = 1.05 and $n$ = 1.01, and a $C_5F_{12}$ gaseous RICH detector with $n$ = 1.00175 is shown in Figure 47. The hadron particle identification capabilities for this configuration are shown in Figure 49. The eventual hadron particle identification scheme will need to be optimized through further simulations. If particle identification is needed for momenta higher than 20 GeV/$c$, then another RICH gas, such as $C_4F_{10}$, could be used with further optimization of the entire particle identification detector scheme. We are presently evaluating the possibility of covering the complete pseudo-rapidity range inside the magnet (–3.5 < |η| < 3.5) with RICH/Aerogel devices.

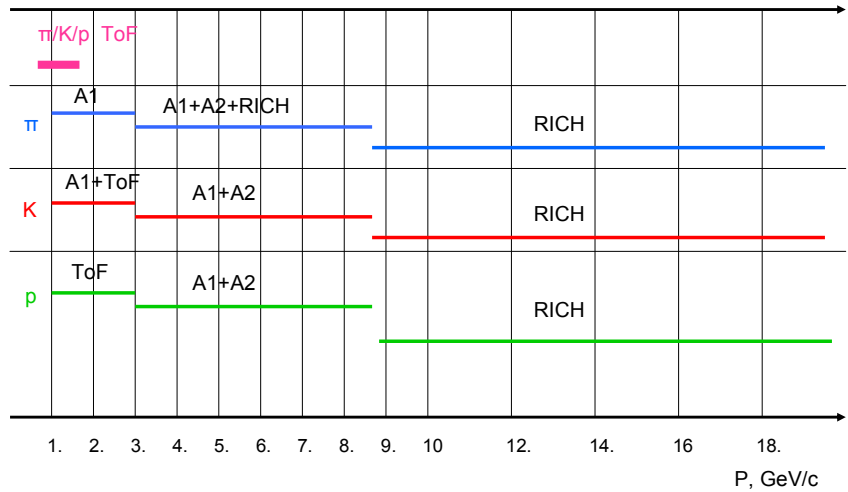

**Figure 49:** Charged hadron particle identification as a function of momentum using the ToF (time-of-flight), Aerogel 1 ($n$ =1.01), Aerogel 2 ($n$ =1.05), and gas-RICH ($n$=1.00175) detectors. The horizontal lines indicate where each particle (listed in the left column) can be identified based on combinations of signals in the detectors listed above the lines.

## 7.5  Hadronic Calorimetry and Muon Detection

The layout of the SLD Magnet enables us to install the EMCal and the 1st section of a HCal inside the magnet, and to instrument the return yoke as a "Tail Catcher Hadron Calorimeter" and a Muon Detector.

None of the existing RHIC detectors features a hadronic calorimeter, nor are any HCal upgrades foreseen to either STAR or PHENIX. The main arguments to add effective hadron calorimetry to the new detector are as follows:

- It allows isolation cuts to be applied in pp-collisions so that a γ-jet can be distinguished from background fragmentation photons[150].
- It improves the jet energy resolution by measuring the neutral particle energy component.
- It removes the trigger bias of the electromagnetic calorimeter.

In addition a hermetic (4π) hadronic calorimeter will allow us to measure the missing energy in W production thus enabling us to measure the W-decay into di-jets.

The primary candidate for barrel calorimetry is the SLD LAr calorimeter. This calorimeter consists of two sections: the electromagnetic section, see below, made of 22 radiation lengths of lead plates of 2 mm thickness with 2.75 mm gaps between the plates (resolution ~11%/$\sqrt{E}$), and the hadronic section made of 6 mm thickness lead plates with 2.75 mm gaps (~2.8 absorption lengths, resolution ~60%/$\sqrt{E}$). Required modifications include: a.) installation of very small thickness read out plates with the pad structure on the "ground side" of each gap (G10, Kapton) , comparable to the H1 calorimeter design, and b.) development of frontend electronics with ~130 ns shaping time, comparable to D0 electronics. These are needed to be compatible with the RHIC-II luminosity and the high rate DAQ and trigger requirements.

For the endcaps (1.2 < |η| < 3.5) we are considering a Fe(Pb)+Scintillator (> 3.5 absorption lengths) design for the Hadron Calorimeter, because the SLD LAr calorimeter would not be capable of withstanding the high occupancy in central Au+Au event.

The SLD magnet comes with Tail Catcher hadronic calorimetry and muon detection embedded in the iron covering |η| < 3.5. The system consists of Fe+Wire plastic tubes (streamer mode, pad-readout). Hits are measured in 15 planes yielding a reasonable hadron energy resolution of ~80 %/$\sqrt{E}$ with good muon identification in both the barrel and endcaps. The muon efficiency of the SLD muon detector barrel was found to be 87% taking into account muon identification, reconstruction and track matching with detectors inside the magnet[151]. Analysis of $K^0 \to \pi\pi$ decays was used to check the pion misidentification. For $p$ > 2 GeV/$c$ this was less than 0.3%. In a Monte Carlo simulation the contribution to the muon spectrum from all misidentified particles was found to be ~ 6% for $p$ < 6 GeV/$c$.

## 7.6 Electromagnetic Calorimetry

In a modified SLD approach an EMCal would be installed directly in front of the SLD magnet coil, *i.e.* at 2.8 m radius covering the barrel and in front of the endcap hadron calorimeter. Different technologies are presently foreseen for the EMCal barrel and endcap sections.

If simulations show that the resolution of the 1st section of the SLD Liquid Argon calorimeter is not sufficient, then the barrel EMCal could be constructed of PBWO$_4$ crystals of 8 cm length and 4.5×4.5 m$^2$ cross-section as pre-shower detectors. They also provide moderate timing resolution (~1 ns). The anticipated resolution would below 10%/$\sqrt{E}$. Behind the crystals a Fe+Scintillator accordion type detector with 15 radiation length is being considered. In the present cost estimate we assume that the existing SLD barrel EMCal is sufficient but we will replace the endcap EMCal because for the forward (endcap) EM calorimeter (1.2 < |η| < 3.5) we anticipate the occupancy in central RHIC-II Au-Au events to be too high for the rather coarse granularity of the SLD device. Possible alternate technologies include simple Fe-Scintillator type projective towers or a fine-granularity crystal detector. Building a crystal detector of the proper size and granularity is prohibitively expensive, but we are considering the use of the existing CLEO calorimeter crystals as an option for the endcap or the very forward EMCal (see next section). The CLEO calorimeter has an energy resolution better than 6%/$\sqrt{E}$.

### 7.6.1 Calorimeter resolution simulations

The low $p_T$ cutoff for photon-tagging of photon-jet events depends critically on the performance of the EMC and, based on our simulations, requires an energy resolution of 10%/$\sqrt{E}$ and fine granularity. We expect to resolve photons over the soft background above ~ 7 GeV/$c$ momentum.

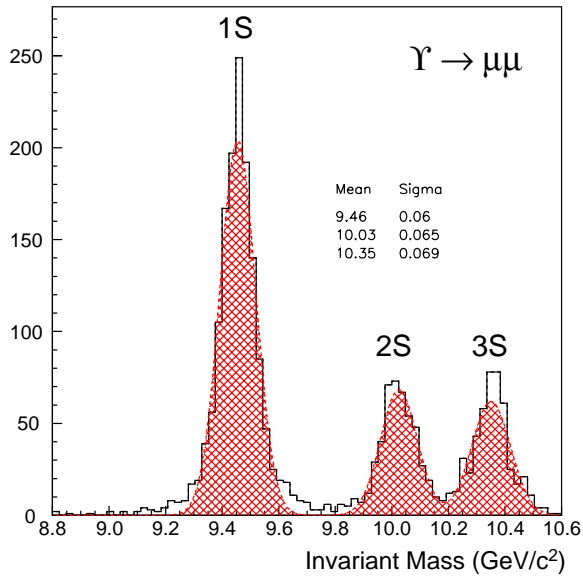

**Figure 50:** Mass resolutions for ϒ via reconstructed tracks from simulation. Backgrounds are not realistic for nucleus-nucleus case.

With the anticipated high luminosity and fast detectors we expect to be able to have sufficient statistics for photon-tagged jets out to a photon $p_T$ of ~ 20 GeV/$c$.

Our recent simulation studies have shown that an energy resolution of better than $10\%/\sqrt{E}$ is required in order to resolve the ϒ states with calorimeter information alone. This clearly shows that the quarkonia physics can only be fully realized when using the PID information from the EMC in combination with high resolution tracking and a muon chamber, as is done in CDF[152]. Shown in Figure 50 is the mass resolution from simulations expected for the ϒ states.

## 7.7 Very Forward Detectors

Beyond η = 3.5, *i.e.* outside the doors of the SLD magnet, we are proposing to build a dedicated forward spectrometer, consisting of a tracking detector, a PID detector, and a set of calorimeters followed by a muon chamber. The schematic layout is shown in Figure 51.

The rather small opening in the SLD magnet in the forward direction (*r* = 20 cm) and the limiting RHIC DX magnet downstream in the RHIC tunnel defines the layout of the spectrometer and limits its pseudo-rapidity range to 3.5 < |η| < 4.8. This forward spectrometer would cover a z-range between 4.5 m and 8 m from the interaction point. Regarding the tracking detectors we are evaluating the usefulness of a two stage detector, with its first nine disk layers inside the main magnet, and five more Silicon strip disks with small disk radii located outside of the main magnet (*z* = 450–610 cm), but surrounded by a dedicated forward spectrometer magnet. The layout presently featured in our simulation package is described in Table 8.

This tracker could either be preceded or followed up by a dedicated forward RICH of sufficient depth. That means the RICH is either located inside the 'door' of the SLD endcap calorimeter or resides behind the final tracking layer (*z* = 6.1 m) and in front of a high precision 'plug' calorimeter consisting of a crystal based electromagnetic section and a digital hadronic calorimeter based on Si-W. Regarding the crystal calorimeter, only the existing CLEO crystals (5 by 5 cm, 30 cm long) have a sufficiently small Moliere radius to handle the anticipated occupancy in the pseudo-rapidity range from 3.5 < |η| < 4.8.

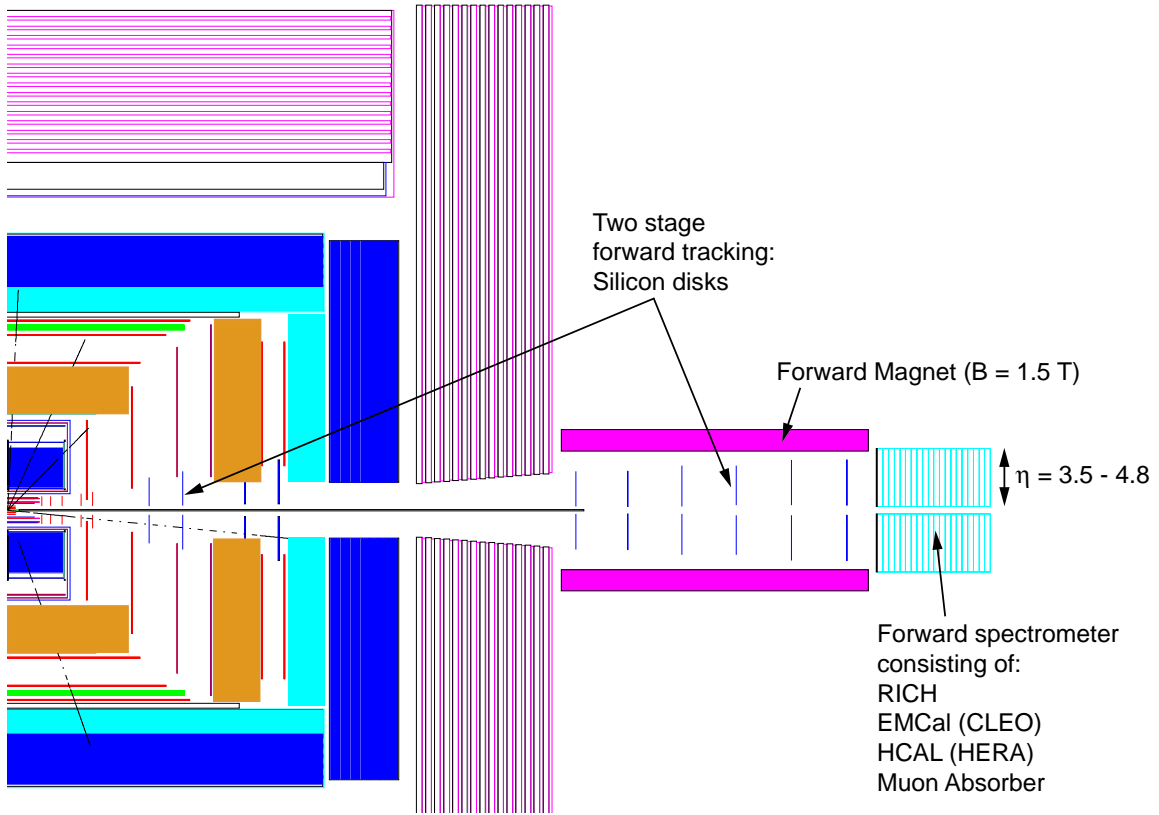

**Figure 51:** Conceptual design of a forward spectrometer for a RHIC II detector based on the SLD magnet.

| Layer | Technology | Z position (cm) | $R_{min}$ (cm) | $R_{max}$ (cm) |
|---|---|---|---|---|
| 1 | Si pixel | 35 | 4.4 | 8.7 |
| 2 | Si pixel | 45 | 4.45 | 10.45 |
| 3 | Si strip | 55 | 4.5 | 12.2 |
| 4 | Si strip | 65 | 4.5 | 13.9 |
| 5 | Si strip | 75 | 4.55 | 15.5 |
| 6 | Micro-pattern | 125 | 4.6 | 28. |
| 7 | Micro-pattern | 155 | 4.7 | 33.5 |
| 8 | Micro-pattern | 200 | 6. | 35. |
| 9 | Micro-pattern | 240 | 6. | 40. |
| 10 | Si strip | 450 | 10. | 20. |
| 11 | Si strip | 490 | 10. | 25. |
| 12 | Si strip | 530 | 10. | 30. |
| 13 | Si strip | 570 | 10. | 35. |
| 14 | Si strip | 610 | 10. | 40. |

**Table 8:** Layout and dimensions of a two-stage forward tracker.

The hadronic section could be constructed from components of the existing HERA-B hadronic calorimeter. Finally we are evaluating a set of forward hadron detectors based on Roman pots instrumented with scintillating fiber spectrometers at very large distances (30–50 m) from the interaction vertex. These detectors will allow us to measure the forward protons necessary to trigger on diffractive processes (*e.g.* pomeron exchange, rapidity gap measurements).

Alternative guidance in the design of the far-forward region may be taken from work done for the FELIX detector at the LHC[153]. The concept is based on a series of compact stations consisting of silicon trackers and electromagnetic calorimeters inside strong dipole fields and just surrounding the beam pipe. A judiciously chosen setup can then cover the full range in a modular way, such that a staged design is possible, allowing desired regions to be covered as funding permits.

The basic tracker design would consist of stations of 4 planes of 50 mm × 300 mm pads. The calorimeter design would be a compact design based possibly on tungsten with silicon strip readout. A design like this is being used by the PHENIX collaboration for their Nose Cone calorimeter to achieve large forward rapidity coverage with a small detector.

A forward detector has clear benefits for the physics program at RHIC-II. However, achieving coverage in the far forward region involves tracking particles of 50 GeV or more. This may require close collaboration with the RHIC accelerator group to find a design that achieves the physics aims without compromising the machine stability or performance.

In order to facilitate tracking in the far forward region, it may be necessary to reconsider the beam optics and beam pipe geometry. Clear priorities are: overall luminosity for the high-$p_T$ program, minimizing backgrounds and good momentum resolution. It is not clear that these priorities are all consistent with one another. Forward tracking may require moderately strong dipole fields in the forward direction which will deflect the primary gold or proton beams, requiring delicate compensating fields to ensure the stability of the collider. High luminosity may necessitate putting focusing quadrupoles close to the IR, creating enormous backgrounds in the far forward region. The bottom line is that a detector system like this will need to be designed in a tightly coupled way to the upgraded RHIC-II machine.

One consequence of a far forward spectrometer at RHIC-II is that it would have comparable rapidity reach to experiments currently being performed at the SPS. Thus, if a fixed target program would be feasible at RHIC, *e.g.* by installation of a gas-jet target, then this detector system would already be suitable for studies of particle production from lowest SPS energies to nearly the top energy. This would provide important cross checks of older results and consistency with the much higher-energy results from the RHIC collider mode.

# 8   Preliminary Budget and In-Kind Contributions

Table 9 shows the present preliminary budget. An overall contingency of 30% is applied to the project. Only the DAQ/TRG project carries a higher contingency (see below). The additional cost for civil engineering of a new hall for this detector is not shown, but early engineering estimates set the cost to about $5 Million.

| Detector Component | Cost (incl. contingency) |
|---|---|
| **Central Detector** | **$69 M** |
| Magnet | $2 M + in kind (SLD) |
| Coil (replace warm with superconducting) | $10 M |
| Tracking central (Silicon or mTPC) | $15 M |
| Tracking endcap (Silicon) | $2 M |
| Vertex (APS or Si-pixel) | $5 M |
| HCAL | $2 M + in kind (SLD) |
| Muon chambers | in kind (part of SLD magnet) |
| EMCal central | $1 M + in kind (SLD,D0) |
| EMCal endcap | $10 M |
| PID RICH (central) | $8 M |
| PID RICH (endcap) | $4 M |
| PID ToF | $8 M |
| PID Aerogel | $2 M |
| **Forward Detector** | **$11 M** |
| Magnet | $1 M + in kind (D0) |
| Tracking | $5M |
| RICH | $2M |
| Hcal | $2M + in kind (HERA) |
| EMCal | $1M + in kind (CLEO) |
| **Combined DAQ/TRG** | **$15 M** |
| **Grand Total** | **$95 M** |

**Table 9:** Preliminary cost estimate for the new RHIC-II detector.

Several of the detector components that have to be constructed are based on existing RHIC-I detector technology or ongoing RHIC-II R&D. For example, the active pixel sensor (APS) and the ToF are part of the STAR upgrade plan. The Aerogel detector and the mini-TPC, which is one of the options for the central tracking detector, are part of the PHENIX upgrade plan. Otherwise standard 'off-the-shelf' technologies have been used, such as Silicon strip detectors for the central tracker and Pb-Scintillator calorimeters for the EMCs. Also, in this cost estimate we are only considering a single forward spectrometer arm. Certainly there can be physics arguments to have a symmetric forward coverage on both sides of the central detector. We estimate the cost of the forward detector to increase by around 50% if one would choose to build two arms.

In the following we detail some of resources used to generate this cost estimate.

## 8.1 Magnets

We are presently considering the re-commissioning of the existing SLD magnet for our detector. After deliberations with the SLD management we came to the conclusion that dismantling, shipping and re-erecting the existing device will cost on the order of $2 Million (including contingency). We are considering replacing the existing warm coil with a superconducting coil in order to double the maximum achievable magnetic field. The cost estimate for the coil replacement is based on the comparative coil study done by STAR in the early 90's. Preliminary discussion with a coil vendor in order to obtain a more recent estimate will commence before the end of the year.

For the very forward spectrometer we anticipate to re-use the D0 central magnet as is. Therefore only shipping and installation cost are considered.

## 8.2 Vertex Detector

The vertex detector cost is based on two separate ongoing efforts: a.) the APS microvertex detector for STAR and b.) the Hybrid-Silicon pixel device for ALICE. The ALICE effort is almost completed whereas the STAR effort is in its proposal stage. The detector that we are proposing is considerably larger than the STAR microvertex detector, but the technology used would be identical. Part of the STAR proposal is to build the APS in two stages in order to address the slow readout problem separately. If a solution to this issue can be found, then an APS would be our first technology for our vertex detector. If not, the fallback option is a hybrid Silicon pixel detector. This technology has been used in ATLAS, CMS, and ALICE and therefore is readily available. Based on the ALICE and STAR cost estimates we anticipate the cost of these two technologies to be roughly comparable.

## 8.3 Tracking Detectors

The full tracking detector package adds up to $22 Million in this proposal. It contains a central main tracker ($|\eta| < 1.2$), a forward tracker ($1.2 < |\eta| < 3.5$) and a very forward tracker ($3.5 < \eta < 4.8$).

The main tracker consists of either a mini-TPC or a multi-layer Silicon detector followed up by a four layer, large area micro-pattern detector right in front of the magnet coil.

In the forward direction ($1.2 < |\eta| < 3.5$) we propose a 6-layer Silicon disk device based on micro-pattern technology (see detector concept chapter), and in the very forward direction ($3.5 < \eta < 4.8$) we propose a two-stage multi-layer detector with one stage inside the main magnet plus an additional small five layer Silicon strip disk device inside the forward spectrometer. The cost estimates are mostly based on a similar exercise for a large solid state tracker for the next linear collider (NLC) project, which can be found on the hep-archive[154]. One of our collaborators was involved in costing the Silicon effort for this project. We do not anticipate a big cost difference between the TPC or Silicon option in the main tracker, therefore we are simply applying the Silicon to all components costed here. Because we anticipate to use our fast tracking detectors in the trigger layout, some of the electronics cost can be found in the trigger cost. In general the on-detector frontend electronics are part of the detector cost, the readout electronics are part of the trigger cost. The cost for the forward trackers is in good agreement with the NLC estimates. The cost for the central tracker can also be derived from similar Silicon efforts in ATLAS and CMS. We are proposing to use only mass-produced 'off the shelf' Silicon strip detectors for our central tracker. The necessary number of Silicon wafers requires us to use the simplest and most common Silicon detectors, *e.g.* double sided strip detectors. Companies, such as Hamamatsu and Micron, have shown that they can produce the necessary number of wafers, and the prices charged by these companies to the LHC projects were used as guidance for the Silicon cost. The mass produced electronics chain is estimated to cost considerably less than $1 per channel. As guidance for the central tracker we estimate the cost of the silicon to account for about 25% of the total cost, the electronics for about 35% and the mechanical support for about 15%.

## 8.4 Calorimeters

The cost estimate for the calorimeter effort relies heavily on recycling available high energy components and is therefore most susceptible to change. Certainly many different high energy experiments are scheduled to stop operation in this decade, but, we are well aware that none of these detectors come for free and that there might be a.) competition in obtaining the device, and b.) considerable cost in dismantling, shipping, and retro-fitting the device. We have applied these costs to the best of our ability and we were generous in the contingency we have applied to the necessary operations. Our deliberations with the SLD management included not only obtaining their magnet but also the connected components, *i.e.* the muon detectors, the EMCal and the HCal. We have evaluated these components and concluded in some of the cases that the performance or the readout might not be optimized for our requirements. Therefore we have proposed alternate options for these components. In particular:

- The hadronic (central+endcap) calorimeter is embedded in the magnet and will therefore be used by us. Based on our conversations with the SLD management, we are estimating $2 M for replacing the streamer tube readout. No additional retro-fit or shipping cost is necessary. The shipping cost is part of the magnet shipping cost.
- The central electromagnetic calorimeter could also be provided by SLD via their 1$^{st}$ LAr section. An alternative would be the D0 LAr detector which will also be available by 2008. An alternative new detector based on Russian pre-shower PbWO4 and a Fe-Scintillator accordion type EMC is being investigated but not costed here. Our cost estimate assumes only a moderate shipping and retro-fit cost of $1 M in order to use either the SLD or D0 detector. We estimate that problems with occupancy or resolution in the EMCal will only occur in the endcap region of the central detector.
- The endcap EMCal ($1.2 < |\eta| < 3.5$) will have to handle considerably higher occupancies at RHIC than at SLAC and therefore we propose to build a new endcap EMCal based on Fe-scintillator or Pb-scintillator technology. The cost for an accordion type calorimeter is based on the cost estimates by STAR for their central barrel and endcap calorimeters.
- In the very forward direction ($3.5 < |\eta| < 4.8$) we propose to retro-fit small, but very sophisticated existing calorimeters to our device, namely the HCal from HERA and the crystal EMCal from CLEO. We have started discussions with the CLEO group and we estimate to retro-fits for both calorimeters to cost about $3M including shipping and labor. The retro-fit of the CLEO crystals includes cutting the projective crystals and re-wrapping them.

## 8.5 Muon Detectors

The iron structure of the SLD magnet works as a muon absorber and is instrumented with the proper readout. Therefore we are presently applying no extra cost to the muon capabilities in our system. Questions of whether this detector is too shallow for the momentum range to be covered at RHIC-II are under investigation.

## 8.6 PID Detectors

We propose three kinds of PID detectors, a $2\pi$ central ToF, RICH detectors in the central, endcap and very forward regions and Aerogel Cerenkov counters in the central and endcap regions. All three components have already been successfully employed by existing RHIC-I detectors or are advanced detector upgrades for the existing RHIC detectors. Therefore the funding for these components is rather well understood and documented.

The ToF is a copy of the $2\pi$ ToF for STAR, which is already partially completed. The technology, layout and readout of the STAR device will be adopted. If the STAR device would be available, our proposal would be to continue to build STAR modules until the phase space of our detector is filled. Therefore we estimate that availability would offset at least 50% of the estimated cost. Although our device is slightly larger than the STAR device we estimate the same overall cost simply by assuming that production lines and equipment developed for STAR could be used in the future by the same groups.

The Aerogel Cerenkov extension of the ToF device is identical to the PHENIX upgrade proposal. The RICH detectors total $14 Million and cover the full phase space of the proposed detector including a dedicated RICH in very forward direction. The RICH cost is scaled from the construction cost for the ALICE RICH. Although we might use different gases in different sections of our RICH layout, the detector design and the electronics will be identical and therefore we cost all RICH components the same.

## 8.7 DAQ and Triggers

This would be a very challenging and exciting detector from the TRG/DAQ standpoint, because it would be possible to make the whole chain a digital pipeline. The assumption that we can run gaseous tracking detectors without a gating grid means that we can simply digitize the signal for each time unit, something we can accomplish using currently available ADCs. This means that the trigger can be continuously more sophisticated as we work our way through the digital information. Selection based on calorimetry and hits can quickly be augmented by tracking to allow PID based triggers. Farms of CPUs can gather and build events easily with a token-based bookkeeping architecture. We imagine that by data taking time (~ 2014) the storage will be optical devices capable of GB/s rates and PB volumes. Most of the "analysis" can be done in the DAQ environment, leading to streams equivalent to current μDST streams.

For the DAQ and trigger cost we have adopted a hybrid model between the original RHIC-I detector costing, which did not include any electronics cost in the DAQ/TRG estimate and the estimates by ALICE and the NLC detectors, which folded part of the readout electronics chain into the DAQ/TRG cost. For detectors than can and will provide trigger capabilities (ToF, trackers, calorimeters) we have included the late stage of the electronics into the TRG cost, whereas for non-triggering detectors (vertex, RICH) we have decided to keep the full readout chain cost in that subproject. Our estimate will depend strongly on new technologies for data acquisition and triggering that are to be developed in the next decade and therefore this estimate carries a higher contingency (40%). However it is also success oriented in its estimate of the improvements in DAQ throughput through technology advances. The main guidance for the present cost estimate is based on the subsystem channel count shown in Table 10.

| Subsystem | Channel Count | Comments |
|---|---|---|
| Vertex Tracker | 388,000,000 | 40 by 40 μm pixels |
| Main Barrel Tracker | 3,000,000 | 100 μm pitch strips (double-sided) |
| High pt Barrel Tracker | 2,200,000 | pad sizes (0.04 by 4, 0.04 by 10 cm) |
| Endcap Tracker | 800,000 | pad sizes (0.04 by 10 cm) |
| Forward Tracker | 800,000 | various technologies (see text) |
| Barrel EMCal | 34,400 | based on SLD layout |
| Endcap EMCal | 9,600 | based on SLD layout |
| Forward EMCal | 200 | based on CLEO crystals |
| Barrel HCal | 9,800 | based on SLD layout (towers) |
| Endcap HCal | 110,800 | based on SLD layout (strips) |
| Barrel TOF | 50,000 | based on STAR layout |
| Barrel RICH | 260,000 | 6 by 6 mm pads |

**Table 10:** preliminary channel count based on the proposed technologies for the main detector components.

# References

[1] See RIKEN BNL Research Center Workshop *"New Discoveries at RHIC - The Strongly Interactive QGP"* at http://www.bnl.gov/riken/May14-152004 workshop.htm; also Proceedings of the RIKEN BNL Research Center Workshop on "New Discoveries at RHIC", Formal report BNL-72391-2004.

[2] T. Matsui and H. Satz, Phys. Lett. B178, 416 (1986).

[3] X.N. Wang, Phys. Rev. Lett. 77 (1996) 231; E. Wang and X.N. Wang, Phys. Rev. Lett. 89 (2002) 162301.

[4] I. Vitev and M. Gyulassy, Phys. Rev. Lett. 89 (2002) 252301.

[5] C. Bourrely and J. Soffer, Phys. Rev. D68 (2003) 014003.

[6] P. Kolb, J. Sollfrank and U. Heinz Phys Rev. C62, 054909 (2000); ibid, Phys. Lett. B459, 667 (1999).

[7] Y.L.Dokshitzer and D.E.Kharzeev, Phys. Lett. B519 (2001) 199, [hep-ph/0106202].

[8] D. Kharzeev and J. Sandweiss, private communication (2004).

[9] From B. Mueller, Proceedings of the RIKEN Workshop on "New Discoveries at RHIC", BNL-72391-2004, p. 125.

[10] B. Aubert *et al.* (BABAR Collaboration), hep-ex/0304021; D. Besson *et al.* (CLEO Collaboration), hep-ex/0305100.

[11] M. Harada, M. Rho and C. Sasaki, hep-ph/0312182; M. Nowak, M. Rho and I. Zahed, hep-ph/0307102.

[12] I. Arsene *et al*. (BRAHMS collaboration), preprint submitted to Phys. Rev. Lett., nucl-ex/0403005.

[13] L. McLerran, Acta Phys.Polon. B34 (2003) 3029 ; D. Kharzeev and E. Levin, Phys. Lett. B523 (2001) 79 ; D. Kharzeev, E. Levin, L. McLerran, Phys. Lett. B561 (2003) 93.

[14] R. Vogt, hep-ph/0405060.

[15] V. Guzey, W. Vogelsang and M. Strikman, hep-ph/0407201.

[16] D. Kharzeev private communication (2003).

[17] G. Bunce *et al.*, Ann. Rev. Nucl. Part. Sci. 50 (2000) 525.

[18] A. Airapetian *et al*. (HERMES collaboration), hep-ex/0407032.

[19] P.M. Nadolsky and C.-P. Yuan, Nucl. Phys. B666 (2003) 31.

[20] E.A. Hawker *et al*., Phys. Rev. Lett. 80 (1998) 3715.

[21] B. Dressler *et al*., *hep-ph*/9809487.

[22] E. Asakawa *et al*., hep-ph/9912373; J.I. Illana, hep-ph/9912467.

[23] K. Klimek, hep-ph/0305266.

[24] M. Acciarri *et al*. (L3 Collaboration), Phys. Lett. B503 (2001) 10.

[25] D. Boer and W. Vogelsang, hep-ph/0312320.

[26] T. Gehrmann, D. Maitre and D. Wyler, hep-ph/0406222.

[27] P. Taxil and J.M. Virey, Phys. Lett. B364 (1995) 181; P.Taxil and J.M. Viery, Phys. Rev. D55 (1997) 4480.

[28] W.Vogelsang and M.R. Whalley, J. Phys. G. 23 (1997) a1.

[29] B.A.Kniehl, G.Kramer, B.Poetter, hep-ph/0011155.

[30] C.Bourrely and J.Soffer, hep-ph/0305070.

[31] K.Ackerstaff *et al.*, OPAL, Europ.Phys.J. C8 (1999) 241 and hep-ex/9805025.

[32] K. Abe *et al.* (SLD), hep-ex/0106063.

[33] M.Gyulassy *et al.*, nucl-th/0302077.

[34] P. Abreu *et al*. (DELPHI), hep-ex/0106063.

[35] Y.L.Dokshitzer and D.E.Kharzeev, Phys. Lett. B519, 199 (2001) [hep-ph/0106202].

[36] J.Ashman *et al*., EMC collaboration, Z.Phys.C52 (1991) 1.

[37] A.Airepatian *et al*., HERMES, EPJ C20 (2001) 479, V.Muccifora *et al.*hep-ex/0106088.

[38] A.Accardi and H.J.Pirner, nucl-th/0205008.

[39] B. Kopeliovich *et al*., nucl-th/9607036.

[40] X.N.Wang & E.Wang, hep-ph/0202105.

[41] C.Gale *et al*., nucl-th/0403054.

[42] C.Stewart *et al*., Phys. Rev. D42, 1385 (1990).

[43] D.Jaffe *et al*., Phys. Rev. D40, 2777(1989) .

[44] K.Streets *et al*., FERMILAB report 89/42-E.

[45] I.P.Lokhtin and A.M.Snigirev, hep-ph/0303121.

[46] P.Abreu *et al*., Phys. Lett. B416 (1998) 247.

[47] E.M. Aitala *et al*., Phys. Lett. B495 (2000) 42 and references therein.

[48] G.A.Alves *et al*., Phys. Rev. Lett. 77 (1996) 2392 and references therein.

[49] K.P. Das and R.C. Hwa, Phys. Lett. B68 (1977) 459.
[50] C.H. Chang et al., hep-ph/0301253.
[51] E.L. Berger et al., Phys. Rev. D23 (1981) 99.
[52] R. Hwa et al., nucl-th/0407081.
[53] See, e.g., F. Karsch, Lect. Notes Phys. 583 (2002) 209.
[54] R. Rapp and J. Wambach, Adv. Nucl. Phys. 25 (2000) 1.
[55] G.E. Brown and M. Rho, Phys. Rep. 363 (2002) 85.
[56] R. Rapp, Nucl. Phys. A725 (2003) 254.
[57] D. Kharzeev, R.D. Pisarski, and M.H.G. Tytgat, Phys. Rev. Lett. 81 (1998) 512.
[58] D. Kharzeev, hep-ph/0406125 and S. Voloshin, hep-ph/0406311.
[59] D. Kharzeev, private communication.
[60] T. Matsui and H. Satz, Phys. Lett. B178 (1986) 416.
[61] D. Kharzeev and H. Satz, Phys. Lett. B366 (1996) 316.
[62] Z. Lin and C.M. Ko, Phys. Lett. B503 (2001) 104-112.
[63] J.F. Gunion and R. Vogt, Nucl. Phys. B492 (1997) 301
[64] S. Digal, P. Petreczky and H. Satz, Phys. Rev. D64 (2001) 094015.
[65] S. Digal, P. Petreczky and H. Satz, Phys. Lett. B514 (2001) 57-62.
[66] F. Karsch at QM2004, hep-lat/0403016, to be published in J. Phys. G.
[67] M. Asakawa and T. Hatsuda, hep-lat/0308034.
[68] S. Datta et al., Phys. Rev. D69 (2004) 094507.
[69] D. Kharzeev, private communication.
[70] P. Braun-Munzinger and J. Stachel, Phys. Lett. B490 (2000) 196 and Nucl. Phys. A690 (2001) 119.
[71] L. Grandchamp and R. Rapp, Phys. Lett. B523 (2001) 60.
[72] C.M. Ko et al., Phys,. Lett. B444 (1998) 237.
[73] R.L. Thews, M. Schroedter and J. Rafelski, Phys. Rev. C63 (2001) 054905.
[74] G.T. Bodwin, E. Braaten and G.P. Lepage, Phys. Rev. D51 (1995) 1125.
[75] B.L. Ioffe and D. Kharzeev, Phys. Rev. C68 (2003) 061902.
[76] L. Grandchamp and R. Rapp, Nucl. Phys. A715 (2003) 545-548.
[77] S. Gavin and J. Milana, Phys. Rev. Lett. 68 (1992) 1834.
[78] M.B. Johnson et al., Phys. Rev. C65, 025203 (2002).
[79] Y.L. Dokshitzer and D.E. Kharzeev, Phys. Lett. B519 (2001) 199.
[80] R. Baier, D. Schiff and B.G. Zakharov, Ann. Rev. Nucl. Part. Sci. 50 (2000) 37.
[81] A.L.S. Angelis et al. Eur. Phys. J. C5 (1998) 63-75.
[82] M. C. Abreu et al., Phys. Lett. B466 (1999) 408.
[83] M. C. Abreu et al., Phys. Lett. B477 (2000) 28.
[84] A. Baldit et al., CERN-SPSC-2000-010; SPSC-P-316. (2000) 1-64.
[85] G. Borges (NA50), QM2004 Proceedings (2004) to be published in J. Phys. G.
[86] M. J. Leitch et al., Phys. Rev. Lett. 84 (2000) 3256.
[87] I. Abt et al. (HERA-B), Phys. Lett. B561 (2003) 61-72.
[88] P.M. Dinh et al., Nuc. Phys. A698 (2002) 579c.
[89] M. Bishai (FNAL), Talk at the 2004 RHIC & AGS User Meeting, BNL (2004)
https://www.bnl.gov/rhic_ags/users_meeting/Workshops/R3a/bishai.pdf
[90] R.D. Field, Phys. Rev. D 65 (2002) 094006.
[91] R. Gavai et al., Int. J. Mod. Phys. A10 (1995) 3043-3070.
[92] E866, Phys. Rev. Lett. 84 (2000) 3256.
[93] E772, Phys. Rev. Lett. 66 (1991) 2285.
[94] T.H. Chang (E866) Ph.D. thesis, hep-ex/0012034.
[95] R. Vogt, Nucl. Phys. A700 (2002) 539.
[96] D. Kharzeev and E. Levin, Phys. Lett. B523 (2001) 79.
[97] D. Kharzeev, E. Levin, and L. McLerran, Phys. Lett. B561 (2003) 93.
[98] M.J. Leitch et al., Phys. Rev. Lett. 84 (2000) 3256.
[99] K.J. Eskola et al., Nucl. Phys. B535 (1998) 351.
[100] Y. Dokshitzer, V. Khoze, S.I. Troian, and A.H. Mueller, Rev. Mod. Phys. 60 (1988) 373.
[101] I. Arsene et al., Submitted to PRL, nucl-ex/0403005.
[102] Y. Kovchegov and L. McLerran, Phys. Rev. C63 (2001) 024903.

[103] D.Kharzeev, private communication (2004).
[104] M.M. Aggarwal *et al.,* Phys. Rev. Lett. 93 (2004) 022301.
[105] J. Sandweiss and A. Chikanian, private communication.
[106] A.H.Mueller and H. Navelet, Nucl. Phys. B282 (1987) 727.
[107] BRAHMS, PHENIX, PHOBOS, STAR, Phys. Rev. Lett.91 (2003).
[108] K.Ikatura, Y.Kovchegov, L.McLerran, and D. Teaney, Nucl. Phys. A730 (2004) 160.
[109] U. Heinz and P. Kolb, "Quark-Gluon Plasma 3", Eds. R.C. Hwa and X.N. Wang, World Scientific, 2003.
[110] K.H.Ackermann *et al.*, Phys. Rev. Lett. 86 (2001) 402.
[111] C. Adler *et al.*, Phys. Rev. Lett.87 (2001) 182301.
[112] J. Adams *et al.*, Phys. Rev. Lett 92 (2004) 052302.
[113] A.M.Poskanzer and S.Voloshin, Phys. Lett. B474 (2000) 27.
[114] B.B.Back *et al.*, Phys. Rev. Lett. 89 (2002) 222301.
[115] PHENIX and STAR, preliminary data.
[116] C.M. Hung and E.V.Shuryak, Phys. Rev. Lett.75 (1995) 4003.
[117] B.Holzman *et al.*, QM2004, nucl-ex/0406027.
[118] T. Hirano, Phys. Rev.C65 (2002) 011901.
[119] T.Hirano and Y.Nara, nucl-th/0404039.
[120] L.Landau and S.Z.Belenkij, Nuovo Cim.Suppl.3 (1956) S10.
[121] P. Steinberg, nucl-ex/0405022.
[122] C.Adler *et al.*, Phys. Rev. Lett. 90 (2003) 082302.
[123] T. Trainor, 20$^{th}$ Winter Workshop on Nuclear Dynamics, Jamaica, 2004.
[124] T. Anticic *et al.*, hep-ex/0311009.
[125] C. Roland *et al.*, QM2004, nucl-ex/0403035.
[126] B.B. Back *et al.*, Phys. Rev. Lett 91(2003).
[127] M. Murray *et al.*, QM2004, nucl-ex/0404007.
[128] B.B. Back *et al.*, nucl-ex/0406021, 2004.
[129] D. Ouerdane *et al.*, QM2004, nucl-ex/0403049.
[130] F. Becattini *et al.*, Phys. Rev. C69 (2004) 024905.
[131] S.V.Afanasiev *et al.*, Phys. Rev. C66 (2002) 054902.
[132] A. Rubbia, hep-ph/0407297.
[133] W.Busza and A.Goldhaber, Phys. Lett. B139 (1984) 235.
[134] M.Basile *et al.*, Phys. Lett. B95 (1980) 311.
[135] J.E.Elias *et al.*, Phys. Rev. Lett.41 (1987) 285.
[136] I.G.Bearden *et al.*, nucl-ex/0312023.
[137] B.B.Back *et al.*, nucl-ex/0301017.
[138] K.Abe *et al*., Phys. Rev. D69 (2004) 072003.
[139] D. Kharzeev and K. Tuchin, Nucl. Phys. A735 (2004) 248.
[140] D. Adams *et al*., Phys. Rev. D56 (1997) 5330.
[141] FNAL E866/NuSea Collaboration (J.C. Peng *et al.*), Phys. Rev. D58 (1998) 092004.
[142] A. P. Contogouris, S. Papadopoulos and B. Kamal, Phys. Lett. B246 (1990) 523.
[143] I. Bojak and M. Stratmann, Phys.Rev. D67 (2003) 034010.
[144] M. Karliner and R.W. Robinett, Phys. Lett*.* B324 (1994) 209.
[145] J. Ralston and D.E. Soper, Nucl. Phys. B152 (1979) 109.
[146] J. Soffer, M. Stratmann and W. Vogelsang, Phys. Rev. D65 (2002) 114024.
[147] D. Boer and W. Vogelsang, hep-ph/0312320.
[148] S.J. Brodsky, D.S. Hwang and I. Schmidt, Phys. Lett. B530 (2002) 99; John C. Collins, Phys. Lett. B536 (2002) 43.
[149] J.D. Bjorken, Phys. Rev. D 47, 101 (1993).
[150] Glueck *et al*., Phys. Rev. Lett. 73, 388
[151] SLD Collaboration, SLAC Pub-8516 (IHEP 2000, Osaka, Japan, 2000).
[152] CDF, Phys. Rev. Lett. 75 (1995) 4358.
[153] FELIX LOI, http://felix.we.cern.ch/FELIX/Loi/loi.html
[154] T. Abe *et al*., Linear Collider Resource Book, Part IV, hep-ex/0106058.